\documentclass[tradiabstract]{aa}
\usepackage{graphicx}
\usepackage{txfonts}
\usepackage{amsmath}
\usepackage{amsfonts}
\usepackage{amssymb}
\usepackage{booktabs}
\usepackage{setspace}
\usepackage[colorlinks,citecolor=blue]{hyperref}
\usepackage{blindtext}
\usepackage[english]{babel}
\usepackage{natbib}
\usepackage{tabularx,multirow,booktabs,blindtext}
\usepackage{subfigure}
\usepackage{lipsum}
\usepackage{adjustbox}
\usepackage{longtable}
\usepackage{rotating} 
\usepackage{lscape}
\usepackage{footnote}
\usepackage{threeparttable}
\usepackage{multicol}
\usepackage{graphicx}
\usepackage{caption}
\usepackage{tablefootnote}

\bibpunct{(}{)}{;}{a}{}{,} 
\begin{document}

		\title{First eROSITA study of nearby M dwarfs \\ and 
	the rotation-activity relation in combination with TESS}
	\titlerunning{A first eROSITA view of nearby M dwarfs}
	\authorrunning{Magaudda et al.}
	\author{E. Magaudda$^1$, B. Stelzer$^{1,2},$ St. Raetz$^1$ and A. Klutsch$^1$}
	\institute{ Institut f\"{u}r Astronomie und Astrophysik, Eberhard-Karls Universit\"{a}t T\"{u}bingen, Sand 1, D-72076 T\"{u}bingen, Germany
		\and INAF- Osservatorio Astronomico di Palermo, Piazza Parlamento 1, I-90134 Palermo, Italy}
	\abstract
	{We present the first results with the {\em ROentgen Survey with an Imaging Telescope Array (eROSITA)} on board the Russian Spektrum-Roentgen-Gamma mission (SRG), and we combine the new X-ray data with observations with the {\em Transiting Exoplanet Survey Satellite (TESS)}. We use the {\sc superblink} proper motion catalog of nearby M dwarfs as input sample to search for {\em eROSITA} and {\em TESS} data. We have extracted {\em Gaia}\,DR2 data for the full M dwarf catalog that comprises $\sim9000$ stars, and we calculated the stellar parameters from empirical relations with optical/IR colors. Then we cross-matched this catalog with the {\em eROSITA Final Equatorial Depth Survey} (eFEDS) and the first {\em eROSITA} all-sky survey (eRASS1). 
	After a meticulous source identification in which we associate the closest {\em Gaia} source to the {\em eROSITA} X-ray detections, our sample of M dwarfs is defined by $704$ stars with SpT = K5..M7 ($690$ from eRASS1 and $14$ from eFEDS). While for eRASS1 we used the data from the source catalog provided by the eROSITA\_DE consortium, for the much smaller eFEDS sample we performed the data extraction and we analyzed the X-ray spectra and light curves. This unprecedented data base for X-ray emitting M dwarfs allowed us to put a quantitative constraint on the mass dependence of the X-ray luminosity, and to determine the change in the activity level with respect to pre-main-sequence stars. {\em TESS} observations are available for $501$ of $704$ X-ray detected M dwarfs, and applying standard period search methods we could determine the rotation period for $180$ X-ray detected M dwarfs, about one forth of the X-ray sample. With the joint {\em eROSITA} and {\em TESS} sample, and combining it with our compilation of historical X-ray and rotation data for M dwarfs, we examine the mass dependence of the `saturated' regime of the rotation-activity relation. A first comparison of {\em eROSITA} hardness ratios and spectra shows that 65\,\%  of the X-ray detected M dwarfs have coronal temperatures of $\sim 0.5$\,keV. We performed a  statistical investigation of the long-term X-ray variability of M dwarfs comparing the {\em eROSITA} measurements to those obtained $\sim 30$\,years earlier during the {\em ROSAT} all-sky survey (RASS). Evidence for X-ray flares are found in various parts of our analysis: directly from an inspection of the eFEDS light curves, in the relation between RASS and eRASS1 X-ray luminosities, and in a subset of stars that displays X-ray emission hotter than the bulk of the sample according to the hardness ratios. Finally, we point out the need to obtain X-ray spectroscopy for more M dwarfs to study the coronal temperature-luminosity relation that is not well constrained by our eFEDS results.}
\maketitle
\section{Introduction}\label{sect:intro}

M dwarfs are the most numerous stars in the Galaxy. Long overlooked because of their relative faintness, they have lately become of central interest for astronomy motivated by their importance as hosts of habitable planets \citep{Tarter2007}. Understanding the evolution of planets around M dwarfs and their potential for hosting life requires good knowledge of stellar magnetic activity because planetary atmospheres react sensitively to both short-wavelength (UV, extreme ultraviolet, and X-ray) radiation and stellar winds.

The X-ray emission of M dwarfs is also of paramount importance for several unresolved problems in stellar astrophysics. Being a manifestation of magnetic heating, the UV and X-ray emissions of late-type stars are proxies for the efficiency of stellar dynamos. In analogy to the Sun, standard ($\alpha\Omega$) stellar dynamos are thought to be driven by convection and rotation, and located in the tachocline connecting radiative interior and convective envelope \citep{Parker1993}. As a consequence, the X-ray emission of M dwarfs can be expected to undergo drastic changes at the transition where stellar interiors become fully convective (SpT $\sim$ M3). Early studies have given controversial results, with the occurrence of a qualitative change of the X-ray emission across this boundary being debated \citep[e.g.][]{Rosner1985,Fleming2003}. More recently, based on an improved mass function, an exceptionally large spread of the X-ray emission level and rotation rates was observed for spectral types M3...M4 \citep{Reiners2012,Stelzer2013}. 
This spread is  likely a signature of ongoing spin-down and an associated decay of dynamo efficiency, but may also indicate a transition related to the fact that stellar interiors becoming fully convective.

Through their link with the dynamo, the secular evolution of a stars’ high-energy radiative output and its angular momentum should occur in parallel. This evolution likely depends on the initial conditions which differ from star to star. Depending on the initial rotation after the disk phase, it takes a G-type star from a few tens to a few hundreds of Myr to spin down to $\rm \approx 1 - 10$ times the solar rotation rate \citep{Johnstone15.0,Tu2015}. On the contrary, M dwarfs stay in the saturation regime for much longer; even for a $\rm 0.5\,M_{\odot}$ star (SpT$\sim$M1/M2), saturation may last as long as 1\,Gyr for half of the objects \citep{Johnstone15.0,Magaudda2020}, resulting in the prolonged irradiation mentioned above. The large spread of rotation rates in mid-M type stars is likely a major responsible for their large spread in X-ray luminosities mentioned above. 

The {\em Einstein} and {\em ROSAT} satellites have  provided the first significant numbers of X-ray detections from M dwarfs \citep{Fleming88.0,Fleming98.0,Schmitt2004}.  However, \cite{Stelzer2013} showed that about $40$\,\% of the closest M dwarfs, those in a volume of $10$\,pc around the Sun, have remained below the detection threshold of the {\em ROSAT} all-sky survey (RASS).
RASS observations have also been the major resource for seminal studies of the rotation-activity relation, e.g.  \cite{Pizzolato2003} and \cite{Wright2011}. Contrary to the first studies of the link between stellar rotation and magnetic activity \citep{Pallavicini1981}, these works made use of photometric rotation measurements that avoid the ambiguity due to the generally unknown inclination angle that affects studies based on spectroscopic $v \sin{i}$ measurements.

\cite{Magaudda2020} have presented a comprehensive study of the relation between rotation, X-ray activity and age for M dwarfs. Therein we have updated and homogenized data from the literature and added in new very sensitive observations from dedicated observations with the X-ray satellites {\em XMM-Newton} and {\em Chandra} and the photometry mission {\em K2} from which we  derived rotation periods. 
Among the new results was a significantly steeper slope in the non-saturated regime for stars beyond the fully convective transition as compared to early-M dwarfs. We confirmed that the X-ray emission level of fast-rotating stars (i.e. those in the saturated regime) is non-constant,
as was previously mentioned by \cite{Reiners2014}.  Moreover, we calculated the evolution of the X-ray emission for M dwarfs older than $\sim 600$\,Myr, by combining the results from the empirical $L_{\rm x}-P_{\rm rot}$ relation with the evolution of $P_{\rm rot}$ predicted by the angular momentum evolution model of \citet{Matt_2015}.

All previous observational work on X-ray activity - rotation relations is based on data that has been collected over decades with different telescopes and instruments, and with a focus on different regions of the parameter  space, introducing various biases. 
New space missions with an  all-sky observing strategy are now available 
that allow to acquire X-ray and rotation data for statistical samples with well characterized stellar parameters that are biased only by a relatively uniform sensitivity limit. This offers new prospects for systematic studies of the X-ray emission of M dwarfs and, in particular their   rotation-activity relation. 
We present here the first results from a combined study using the {\it extended ROentgen survey with an Imaging Telescope Array} \citep[eROSITA;][]{Predehl2021} on the Russian Spektrum-Roentgen-Gamma (SRG) mission to measure X-ray luminosities and the {\it Transiting Exoplanet Survey Satellite} \citep[TESS;][]{Ricker14.0} to obtain  rotation periods. We search the sample of M dwarfs compiled from the {\sc superblink} proper motion survey by \cite{L_pine_2011} for {\em eROSITA} and {\em TESS} data, and we homogeneously  characterize the stars using {\em Gaia} data. 
More details on our samples are given in Sect.~\ref{sect:database}, and how we construct our input catalog of M dwarfs with additional information from {\em Gaia} is described in Sect.~\ref{sect:sample}.
The {\em eROSITA} and {\em TESS} data analysis is described in separate sections (Sect.~\ref{sect:analysis_efeds} and Sect.~\ref{sect:analysis_erass1}) for the two subsamples examined in our work that are described in Sect.~\ref{sect:database}.
This is owed to the different, complementary scientific goals we pursue with the two samples. 
The presentation and interpretation of our results are found in Sect.~\ref{sect:results}, where we also put our findings in context to previous work on the rotation-activity relation.
In Sect.~\ref{sect:conclusions} we summarize our conclusions and give an outlook to future studies in this field.

\section{Database}\label{sect:database}

 This work is based on the {\sc superblink} proper motion catalog of nearby M dwarfs from \cite{L_pine_2011} (henceforth referred to as LG11). The LG11 catalog is an all-sky list of 8889 M dwarfs  (photometric spectral types K7 to M6) brighter than $J=10$\,mag and within $100$\,pc.  
  
In this work we study the X-ray emission of M dwarfs from LG11 in two data sets, the {\em eROSITA} Final Equatorial-Depth Survey (eFEDS) and the first {\em eROSITA} All-Sky survey (eRASS1). 
eFEDS corresponds to a $\sim 140$ sq.deg large area in the southern sky that was observed in four individual field scans during the calibration and performance verification (CalPV) phase \citep{Predehl2021} of {\em eROSITA} (see Brunner et al., A\&A subm). For the sake of simplicity we refer to each field scan observation as field.
As eRASS1 database we use the catalog produced at MPE in its version 201008 that is based on the c946 data processing. While an official eFEDS X-ray source list will be made public in the data release related to this A\&A special issue, eRASS1 data will be officially released  thereafter. 

The X-ray samples from eFEDS and eRASS1 and the way we treat them in this work are complementary. The eFEDS fields comprise a relatively small number of stars for which we present a detailed X-ray study including {\em eROSITA} light curves and spectra.
The part of our study that makes use of eRASS1 data is focused  on global properties, taking advantage of the large number of targets provided by the all-sky survey. In particular, we study  hardness ratios as a proxy for the coronal temperature and the long-term variability of the X-ray luminosity of M dwarfs in comparison to eFEDS and {\em ROSAT} data, and the relation between X-ray emission and rotation periods derived from {\em TESS} light curves. 
A complete discussion of the X-ray properties of the M dwarf sample based on spectral and temporal analysis of eRASS1 data for individual stars is beyond the scope of this work. 
Similarly, for the M dwarfs in eFEDS, we provide an exhaustive analysis of {\em TESS} data using both $2$-min and $29$-min cadences, while our analysis is restricted to the $2$-min light curves for the much larger eRASS1 sample. 

In Table~\ref{tab:sample_def} we anticipate the number of targets in the various catalogs studied throughout this paper. The definitions of the samples are provided in the subsequent sections. 

\begin{table}
	\begin{center}
 		\begin{threeparttable}[b]
 		\caption{Number of stars in the different samples of main-sequence M dwarfs (see Sects.~\ref{sect:sample}, \ref{subsect:analysis_efeds_TESS} and~\ref{subsect:analysis_erass1_TESS} for the definitions).}
		\label{tab:sample_def} 

			\begin{tabular}{lrr}\hline
				Sample name & `full' & `validated' \\ \hline
				LG11-{\em Gaia} & 8229 & 7319 \\
				LG11-{\em Gaia}/eFEDS & 14 & 13 \\
				LG11-{\em Gaia}/eRASS1 & $690$ & $593$ \\ 
				LG11-{\em Gaia}/eFEDS/TESS\tnote{1} & $3$ & $3$ \\
				LG11-{\em Gaia}/eRASS1/TESS\tnote{1} & $177$ & $138$\\ 
				\hline
			\end{tabular}
			\begin{tablenotes}
				\item[1] Only stars with reliable {\em TESS} rotation period are included in these samples. 
			\end{tablenotes}
		\end{threeparttable}
	\end{center}
\end{table}

\section{Preparation of the M dwarf catalog}\label{sect:sample}

 To thoroughly characterize the M dwarf sample we exploited {\em Gaia} data and published empirical calibrations for stellar parameters based on photometry. 
 
 We started by matching the LG11 catalog with the second data release of the \textit{Gaia} mission \citep[\textit{Gaia} DR2,][]{2018A&A...616A...1G}. We used the proper motions (P.M.)  given in LG11 to correct the epoch 2000 coordinates provided in the LG11 catalog to the \textit{Gaia} epoch (J2015.5). Then we performed a multicone search with a search radius of $3^{\prime\prime}$ around each target in LG11 with \begin{scriptsize}TOPCAT\end{scriptsize} \citep{2005ASPC..347...29T}. In that step we found 9638 \textit{Gaia} entries for the 8889 LG11 stars of which 736 have multiple matches ($723$ doubles and $13$ triples). 
To identify reliable
\textit{Gaia} counterparts and to avoid sources in the search radius that do not belong to our targets (i.e. that are no companions to our targets) we used two additional criteria. 

\begin{figure}[t]
	\begin{center}
		\includegraphics[width=0.5\textwidth]{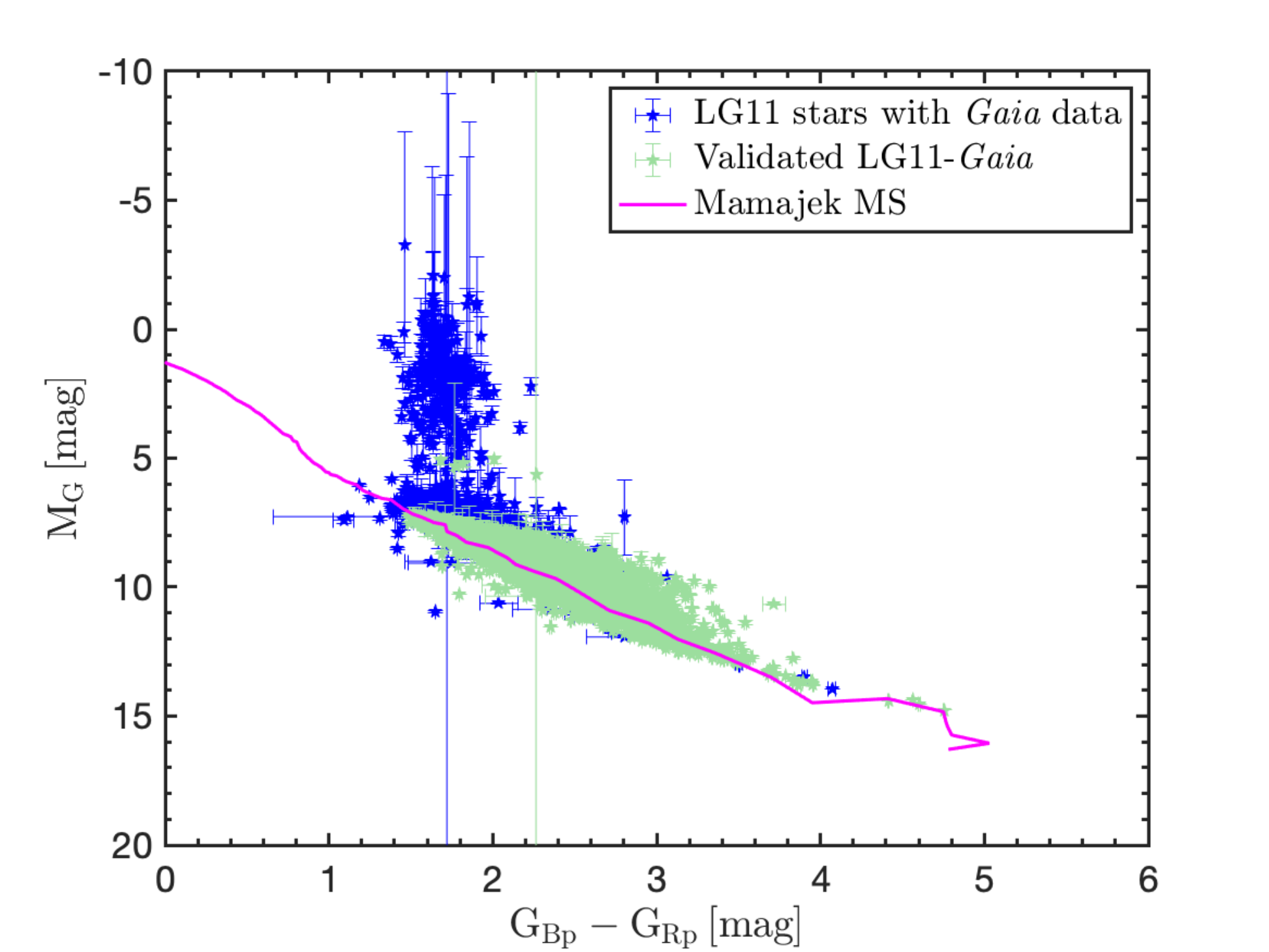}
		\caption{{\em Gaia} color-magnitude diagram based on DR2 data for the  LG11 sample with {\em Gaia} distance and photometry. The pink line represents the main-sequence by E.Mamajek (see footnote~\ref{note1}). A residual contamination by giant stars is present. Stars with $M_{\rm G} > 5$\,mag and $M_{\rm K_s} > 4.6$ (the magnitude above  which the polynomial relations between photometry and stellar parameters of \cite{Mann2015} are valid) are highlighted in green.}
\label{fig:gaia_cmd_fullLG11}
\end{center}
\end{figure}

Firstly, we calculated for all {\em Gaia} entries in our matched catalog the total \textit{Gaia} P.M. by taking the square root of the quadratic sum of the proper motions in right ascension and in declination. 
Then we computed the difference between the total LG11 proper motion and the total {\em Gaia} proper motion. 
The histogram of this proper motion difference shows a Gaussian shape with a sharp edge at $\pm 0.2^{\prime\prime}/{\rm yr}$ when only the stars with a single \textit{Gaia} match are considered. Therefore, we removed all {\em Gaia} counterparts from the full list that show a proper motion difference outside this range. This criterion reduced our catalog to $8779$ entries of which $440$ show multiple {\it Gaia} matches. During that step, all stars without proper motions given in {\em Gaia}\,DR2 were also removed.
As a second step we converted the \textit{Gaia} magnitudes to $J$-band magnitudes (that we call $J_{\rm Gaia}$) using the relation given on the ESA webpage\footnote{\url{https://gea.esac.esa.int/archive/documentation/GDR2/Data_processing/chap_cu5pho/sec_cu5pho_calibr/ssec_cu5pho_PhotTransf.html}}. 
Then we calculated the difference between the Two Micron All-Sky Survey \citep[2MASS;][]{2006AJ....131.1163S} $J$ band magnitude listed in the LG11 catalog  and $J_{\rm Gaia}$.
Significant differences between these magnitudes can arise in cases where two adjacent {\em Gaia} sources are not resolved in 2MASS. Often the two {\em Gaia} sources form a common proper motion (CPM) pair. Therefore, to keep in our catalog the CPM pairs with a moderate $J$ band magnitude difference that are potentially composed of two M stars we removed all {\em Gaia} counterparts with $J_{\rm Gaia} - J > \pm 2$\,mag. 
After this removal of wrong identifications and faint companions that are not relevant as a potential X-ray source the catalog results in $8489$ entries of which $169$ still show multiple {\em Gaia} matches. We note that in this step  all {\em Gaia} counterparts  
 without magnitudes in \textit{Gaia}\,DR2 have also been removed. The 169 multiple matches were checked through by-eye-inspection using ESASky\footnote{\url{sky.esa.int}} \citep{Merin2017}
 and $150$ of them were found to be co-moving pairs.

After the application of our selection criteria $571$ LG11 stars were found to have no \textit{Gaia} counterpart, either because their {\em Gaia} data are missing or incomplete, i.e. no proper motions and/or no magnitudes are available in DR2.  These stars have been removed during the cleaning process described above. 
To recover the \textit{Gaia} IDs for the targets with incomplete data our initial $3^{\prime\prime}$ multicone match was repeated for these $571$ stars. We found that $414$ of them have {\em Gaia} matches (albeit with incomplete photometric and/or astrometric information), and $33$ of these have multiple {\em Gaia} sources in the search radius. The multiple matches were again checked with  by-eye-inspection, and we found that $30$ of them are CPM pairs.  
As a final check for our match we used the cross-identification function of the SIMBAD astronomical database \citep{2000A&AS..143....9W}. We uploaded the list of target names from LG11 and converted them into \textit{Gaia}\,DR2 {\sc source\_id}. The results of our match procedure explained in the previous paragraphs and the output from SIMBAD were compared and if not consistent by-eye-inspection was applied. (SIMBAD does not always provide the correct \textit{Gaia} ID for a given target which is why we have performed the match procedure described above).  

As final result we found $8917$ \textit{Gaia} counterparts for $8736$ targets from the LG11 catalog.
{\em Gaia} data are not available for the remaining $153$ LG11 targets. The excess of Gaia counterparts
represents the $180$ stars for which we found common proper motion
companions of $\left| \Delta  J\right|\leq 2$\, mag, of which one is a triple system. These objects add to the binary stars that are listed as resolved pairs in the original LG11 catalog ($104$ binary pairs). 
Since the CPM companions discovered through our match of the LG11 catalog with {\em Gaia}\,DR2 have been added to our table the total number of objects in our target list is $9070$
($8917$ stars with {\em Gaia} counterparts including the CPM companions and $153$ stars without {\em Gaia} counterpart). 

We matched our final catalog of $9070$ stars with \cite{BailerJones2018} (hereafter BJ\,18) to obtain {\em Gaia}\,DR2 distances, $d_{\rm BJ18}$. We find that $531$ entries from our catalog do not have data in BJ\,18, including the $153$ stars without any {\em Gaia} counterpart.
Among the stars for which we have a distance from BJ18 there are $20$ without {\em Gaia} photometry. We removed these latter ones because we aim at a well-characterized sample, and we calculated the spectral types (SpTs) from the $G_{\rm BP}-G_{\rm RP}$ color with the values provided for the main sequence by E.~Mamajek.\footnote{\label{note1} The table {\em A Modern Mean Dwarf Stellar Color and Effective Temperature Sequence} is maintained by E.~Mamajek at  \url{http://www.pas.rochester.edu/~emamajek/EEM_dwarf_UBVIJHK_colors_Teff.txt.}} 

\begin{figure*}[t]
\begin{center}
\parbox{18cm}{
\parbox{9cm}{
\includegraphics[width=0.5\textwidth]{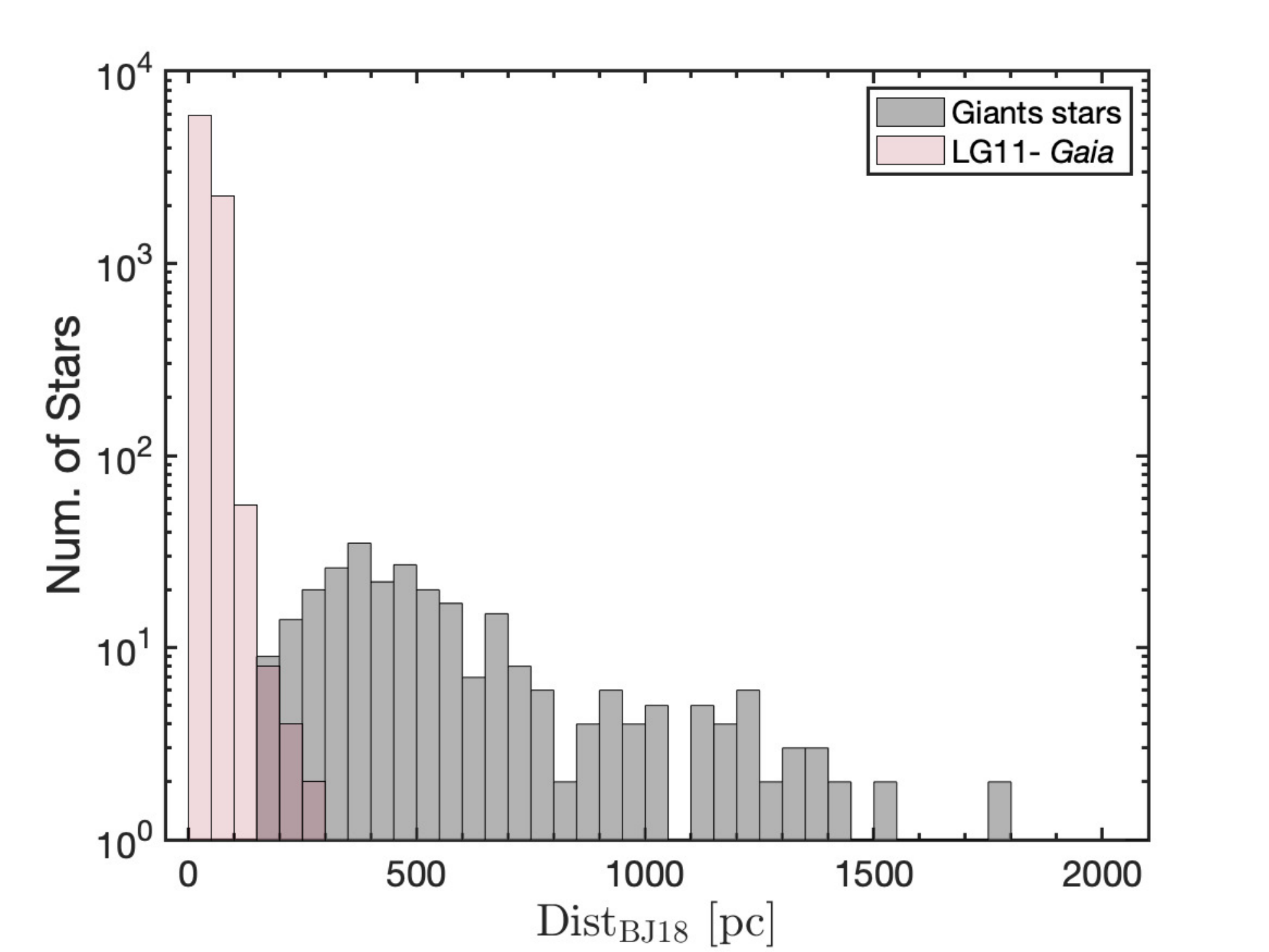}}
\parbox{9cm}{
\includegraphics[width=0.5\textwidth]{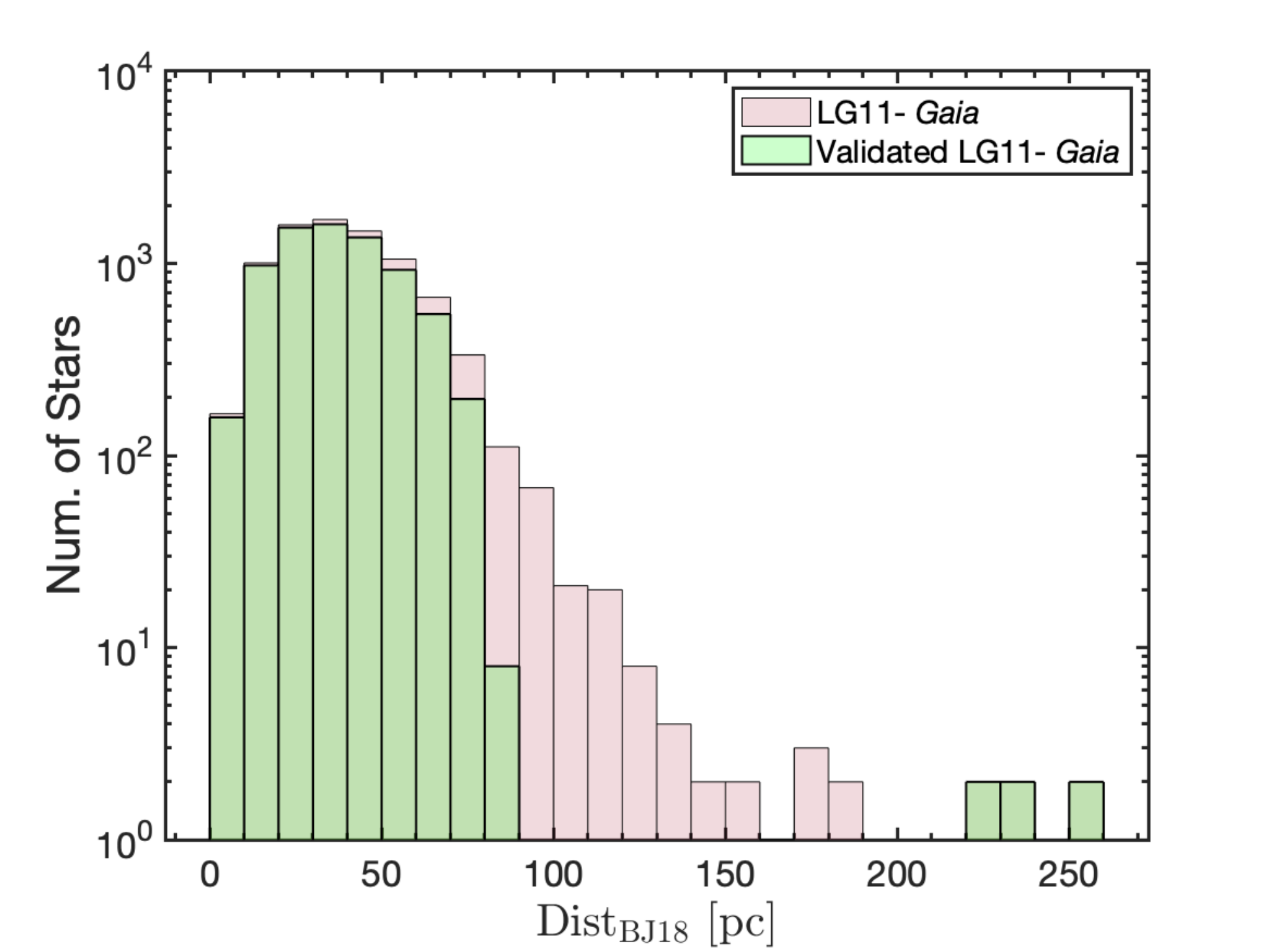}}
}
\caption{Distance distributions ($d_{\rm BJ18}$) for various sub-samples of the LG11 catalog according to our match with {\em Gaia}\,DR2. {\em left} - giant stars (in gray) and MS stars (in red); {\em right} - zoom into the MS sample (red) and subsample of `validated' MS stars (green). See text in Sect.~\ref{sect:sample}.}
\label{fig:dist_spt_fullLG11}
\end{center}
\end{figure*}

In  Fig.~\ref{fig:gaia_cmd_fullLG11} we show the {\em Gaia} color-magnitude diagram (CMD) for the $8519$ stars with {\em Gaia} distance and photometry.
Note that this sample includes the companions in CPM binaries.
The {\em Gaia} CMD shows two distinct  populations, the main sequence and a cluster of stars above it centered at $G_{\rm BP} - G_{\rm RP} \sim 1.7$  (corresponding to late-K SpT  according to the Mamajek scale). We take $M_{\rm G} = 5$\,mag as a rough dividing line between the two populations, and show the distance distributions of these two groups in the left panel of Fig.~\ref{fig:dist_spt_fullLG11}.

Here it is evident that all main-sequence stars are within $300$\,pc with a strong peak around $\sim 50$\,pc, while the remaining stars, which make $\sim 3.4$\,\% of the whole sample cover a wide range of distances from $\sim 100$\,pc to $\sim 1.5$\,kpc and include a few outliers with distance up to $5$\,kpc that are not shown in the figure.  Based on their position in Fig.~\ref{fig:gaia_cmd_fullLG11} and
their large distances these stars are probably giants that `contaminate' the LG11 dwarf star catalog. 
LG11 discussed the compromise in their catalog between the aim of catching as many M dwarfs as possible, including those with small sky motion, and reducing the contamination with M giants. They argued that the majority of red giants have proper motions lower than their cutoff, $\mu = 40\,{\rm mas/yr}$, and they applied additional cuts in absolute magnitude, reduced proper motion and colors. Nevertheless, an over-density of optically bright stars at low Galactic latitude discussed by LG11 suggests the presence of unrecognized giants, consistent with our finding.  

Our study is focused on dwarf stars, and therefore in the following we concentrate on the main-sequence stars ($M_{\rm G} > 5$\,mag).
Henceforth this sample of $8229$ stars is called the `LG11-{\em Gaia} sample' (see Table~\ref{tab:sample_def}). 
To calculate their stellar parameters we applied the empirical relations from \cite{Mann2015} which these authors calibrated on spectroscopically confirmed M dwarfs. 
Specifically, \cite{Mann2015} obtained stellar masses ($M_{\star}$) from the absolute magnitude in the 2MASS $K_{\rm s}$ band ($M_{\rm K_{\rm s}}$), the bolometric correction ($BC_{\rm K_{s}}$) from $V-J$, and the bolometric luminosity ($L_{\rm bol}$) from $ BC_{K_{\rm s}}$. 
For the application of these relations to our LG11-{\em Gaia} sample we calculated the $M_{\rm Ks}$ values from $d_{\rm BJ18}$ and the apparent $K_{\rm s}$ magnitude\footnote{For the stars for which BJ18 report no {\em Gaia} distance, we considered adopting the photometric distances calculated as described by \citet{Magaudda2020} from $M_{\rm Ks}$ obtained from an empirical relation with $V-J$. However, when we revisited the Magaudda et al. sample we noticed that it includes a small number of stars with FGK spectral types for which  the $M_{\rm K_s}$ values derived from the photometric distances have yielded a mass in the M-type regime. To avoid such a contamination, we therefore decided to limit the sample studied in this work to stars with {\em Gaia} distance and photometry, for which we can determine reliable SpT and stellar parameters.\label{Footnote:4}}. The \cite{Mann2015,Mann_2016} relations have been calibrated for the range $4.6 < M_{\rm K_s} < 9.8$.
Considering this criterion reduces our sample to a total of $7319$ stars.
This subsample that fulfills the validity range of \cite{Mann2015} is highlighted in Fig.~\ref{fig:gaia_cmd_fullLG11} in green color and is henceforth referred to as the `validated LG11-{\em Gaia} sample' (see Table~\ref{tab:sample_def}).

In the right panel of Fig.~\ref{fig:dist_spt_fullLG11} 
the full LG11-{\em Gaia} sample of main-sequence stars (red) and the subset of the `validated' main-sequence (green) stars are compared in terms of their distance distribution. Not unexpectedly, the validated sample -- which is defined by a magnitude cut -- comprises (with few exceptions) the more nearby stars.
Our work is based on these two samples of main sequence M dwarfs.

Specifically, this article is focused on two subsamples of the LG11-{\em Gaia} stars, those that are located within the eFEDS fields (Sect.~\ref{subsect:sample_eFEDS}) and those that are detected in eRASS1 (Sect.~\ref{subsect:sample_eRASS1}).
In those parts of our analysis that rely on the stellar mass we restrict the sample to the validated stars. 

\subsection{M dwarfs in the eFEDS fields}
\label{subsect:sample_eFEDS}

\begin{figure*}[ht]
	\begin{center}
	\includegraphics[width=1.0\textwidth]{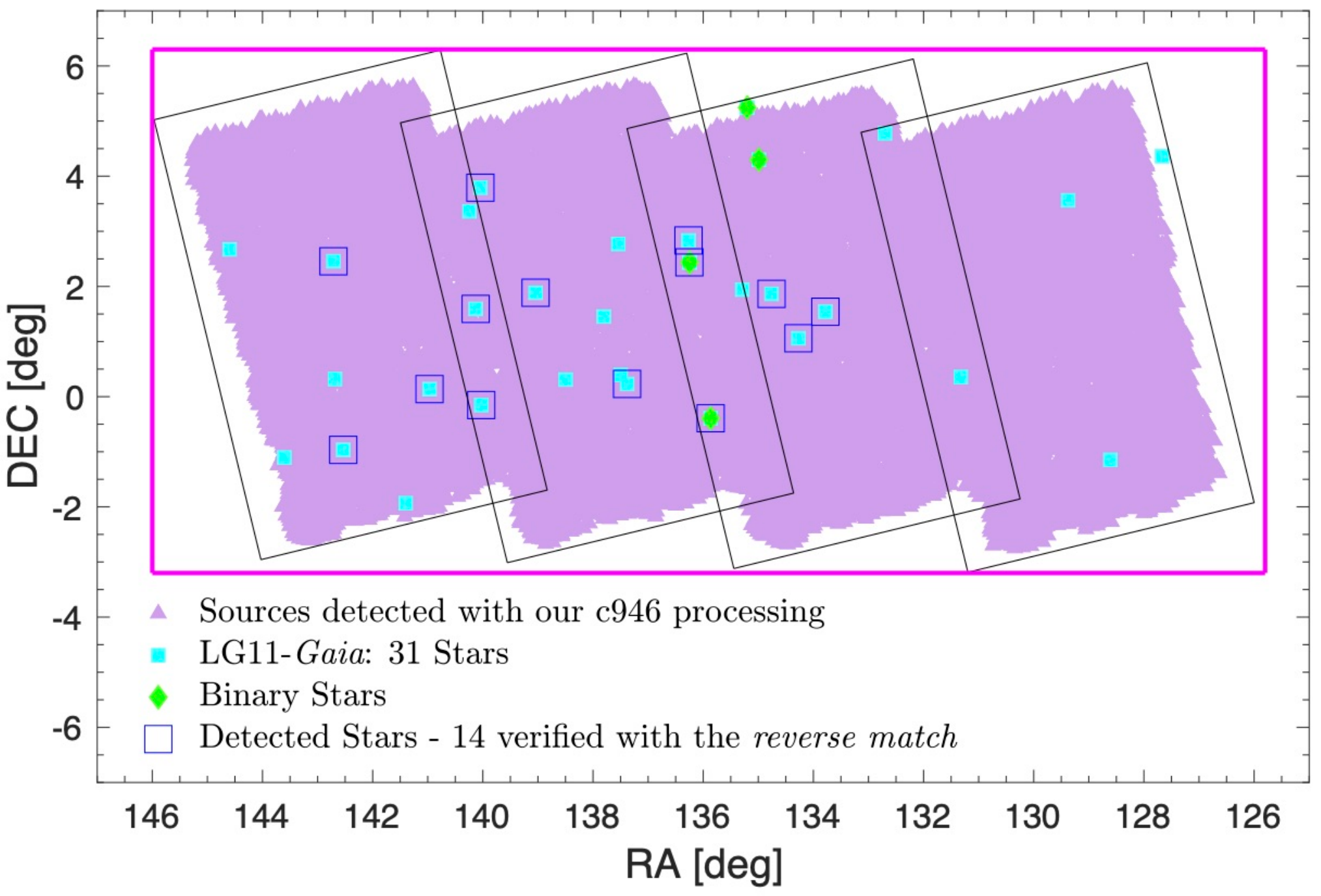}
	\caption{ Cross-match between the X-ray sources in the eFEDS fields and the LG11-{\em Gaia} catalog. We show the search area we used to match LG11 with the survey area (purple rectangle), the eFEDS event file boundaries (four open black rectangles), and the new X-ray sources provided in this work from the  c946 processing of the data (filled purple triangles). Stars from the LG11 catalog are represented with cyan filled squares, and overlapping green filled diamonds represent the two binary systems in our sample. The large open blue squares mark those LG11-{\em Gaia} stars that are detected in eFEDS.}
	\label{eFEDS_field}
	\end{center}
\end{figure*}

An official X-ray source catalog for the eFEDS fields was produced and is presented by Brunner et al. (A\&A subm). For our study of M dwarfs in the eFEDS we have performed our own {\em eROSITA} data analysis, which is described in Sect.~\ref{subsubsect:analysis_efeds_xrays_extraction}. Here we describe how we identified the M dwarfs in the eFEDS sky area. Our method is visualized in Fig.~\ref{eFEDS_field}.

To determine the boundaries of the eFEDS fields we used the event files that we extracted with the eSASS software (Brunner et al., A\&A subm.) from the c946 processing of the eFEDS data. We visually inspected the events and identified the edges as that point beyond which no photons were detected.
These boundaries are shown in Fig.~\ref{eFEDS_field} as black rectangles. This area is somewhat larger than the one defined by the catalog obtained from our source detection (24376 X-ray sources shown in purple) because the density of the registered events decreases towards the boundary of the fields.
Details on the data extraction are given in Sect.~\ref{subsubsect:analysis_efeds_xrays_extraction}. 

Since the four eFEDS fields are overlapping and they delineate a complex shape on the sky, for the down-selection of the M dwarf sample we first defined a rectangular box on the sky that comprises all four eFEDS fields (purple in Fig.~\ref{eFEDS_field}).
We matched our LG11-{\em Gaia} catalog with this rectangular sky area.
Subsequently we removed through visual inspection of the X-ray images the objects that are inside the purple rectangular box but not inside the eFEDS fields boundaries (black rectangles). This way we found that $31$ M dwarfs are covered by the eFEDS observation (filled cyan squares in Fig.~\ref{eFEDS_field}, except for the two co-moving binary systems that are shown as filled green diamonds). 
Note that one of the two CPM pairs located inside the black rectangle but outside the area defined by the purple sources is indeed undetected but still inside the eFEDS fields.

We note, here, that our work on the eFEDS data was performed in parallel with that of the construction of the official eFEDS source catalog by Brunner et al., (A\&A subm.).
To validate our detections we compared  our X-ray source catalog with the official one (eFEDS\_c001\_V4\_main). First, we found a discrepancy between the X-ray coordinates of our catalog and those in eFEDS\_c001\_V4\_main. 
We suspected that this is due to an astrometric correction that was used in eFEDS\_c001\_V4\_main to correct for the mean linear offset between the X-ray sources and the {\em Gaia} positions of objects in the {\em Gaia}-unWISE catalog of candidate Active Galactic Nuclei (AGN) by \cite{Shu2019}. This offset is different for each of the four eFEDS fields and is given in Table~1 of Brunner et al. (A\&A subm.). We  applied these corrections to the X-ray coordinates of our catalog and we verified that this removed the offset between the X-ray positions in our catalog and  eFEDS\_c001\_V4\_main. Finally, we calculated the final absolute coordinates (RA\_CORR, DEC\_CORR) by applying Eq.~1 from Brunner et al. (A\&A subm.) to our X-ray catalog. 
 
We then used our catalog with the corrected X-ray coordinates to search for X-ray detections in the LG11-{\em Gaia} sample.
We base our match between optical and X-ray position on {\em Gaia} coordinates and proper-motions from our LG11-{\em Gaia} catalog.
We first corrected the coordinates of our stars by their P.M. to the eFEDS mean epoch (Nov 5, 2019) and then we matched them with our final X-ray coordinates (RA\_CORR, DEC\_CORR). 
This way we found $15$ matches within $15^{\prime\prime}$\footnote{We performed a cross-check of our detections by matching the P.M.-corrected LG11-{\em Gaia} sample also with the official eFEDS catalog, finding all the $15$ stars detected in our catalog.}. 

While it is quite plausible that the M dwarfs from the LG11 catalog are X-ray emitting, the limited sensitivity of {\em eROSITA} combined with its modest spatial resolution requires a cross-check for other possible optical counterparts to the {\em eROSITA} sources, i.e. we have to verify the associations between our target stars and the detected X-ray sources. We pursue here a conservative approach, i.e. we aim at keeping only those M dwarfs in our sample that we consider `safe' counterparts to the X-ray sources, and we base this assessment on the separation between the optical and X-ray position with respect to that of other {\em Gaia} objects in the vicinity.

To find the alternative possible {\em Gaia} counterparts for each of the $15$ X-ray sources
we performed a `reverse' match, in which we searched for all {\em Gaia} sources within $15^{\prime\prime}$
of the X-ray coordinates from our catalog ($\rm RA\_CORR$, $\rm DEC\_CORR$). 
This way we found a total of $21$ potential {\em Gaia} counterparts, including $14$ stars from the LG11-{\em Gaia} sample. 
Then we inspected the separations between the X-ray positions and the {\em Gaia} coordinates for the $21$ {\em Gaia} sources (${\rm Sep_{X,opt}}$). 
Hereby, we considered for those {\em Gaia} sources that are identified with a star in our input catalog the P.M. correction to the mean eFEDS observing date.
As a result, the star from LG11-{\em Gaia} is the closest {\em Gaia} object to an X-ray source for all $14$ cases, and these M dwarfs define our list of bonafide eFEDS X-ray emitters. The missing 
one is a high proper motion star, that is not recovered in the reverse match because the P.M. correction can be applied only a posteriori and a search radius of $15^{\prime\prime}$ is too small for this star. Therefore we increased the search radius up to $20^{\prime\prime}$, finding the {\em Gaia} source associated with this M~dwarf which, however, does not turn out to be the closest {\em Gaia} counterpart.
Adhering to our conservative approach, we excluded this star from our LG11-{\em Gaia}/eFEDS sample.
This sample, thus, consists of $14$ M dwarfs. 
All but one of them are also part of our `validated' LG11-{\em Gaia}/eFEDS sample' (see Table~\ref{tab:sample_def}).
Finally, we compared the X-ray optical separation ($\rm Sep_{X,opt}$) with the uncertainties on the X-ray positions from our X-ray source catalog ($\rm RADEC\_ERR$), finding that all $14$ LG11-{\em Gaia} stars in our eFEDS X-ray emitter sample have $\rm Sep_{X,opt} < 3~x~RADEC\_ERR$, confirming that the association of the M dwarfs with the eFEDS X-ray sources is consistent with the positional accuracy of {\em eROSITA}. 
Among these there is one CPM pair that has the same X-ray source associated to each component of the system but that is resolved by {\em Gaia} and 2MASS. We chose to ascribe the X-ray emission to the component that is closest to the X-ray source, and we treat it in the same manner as the single stars. Since the two stars in the CPM pair have similar stellar parameters ($M_{\star}$ and SpT), this approach does not influence our results.
 
\begin{figure*}[t]
 	\parbox{18cm}{
 	\parbox{9cm}{\includegraphics[width=0.5\textwidth]{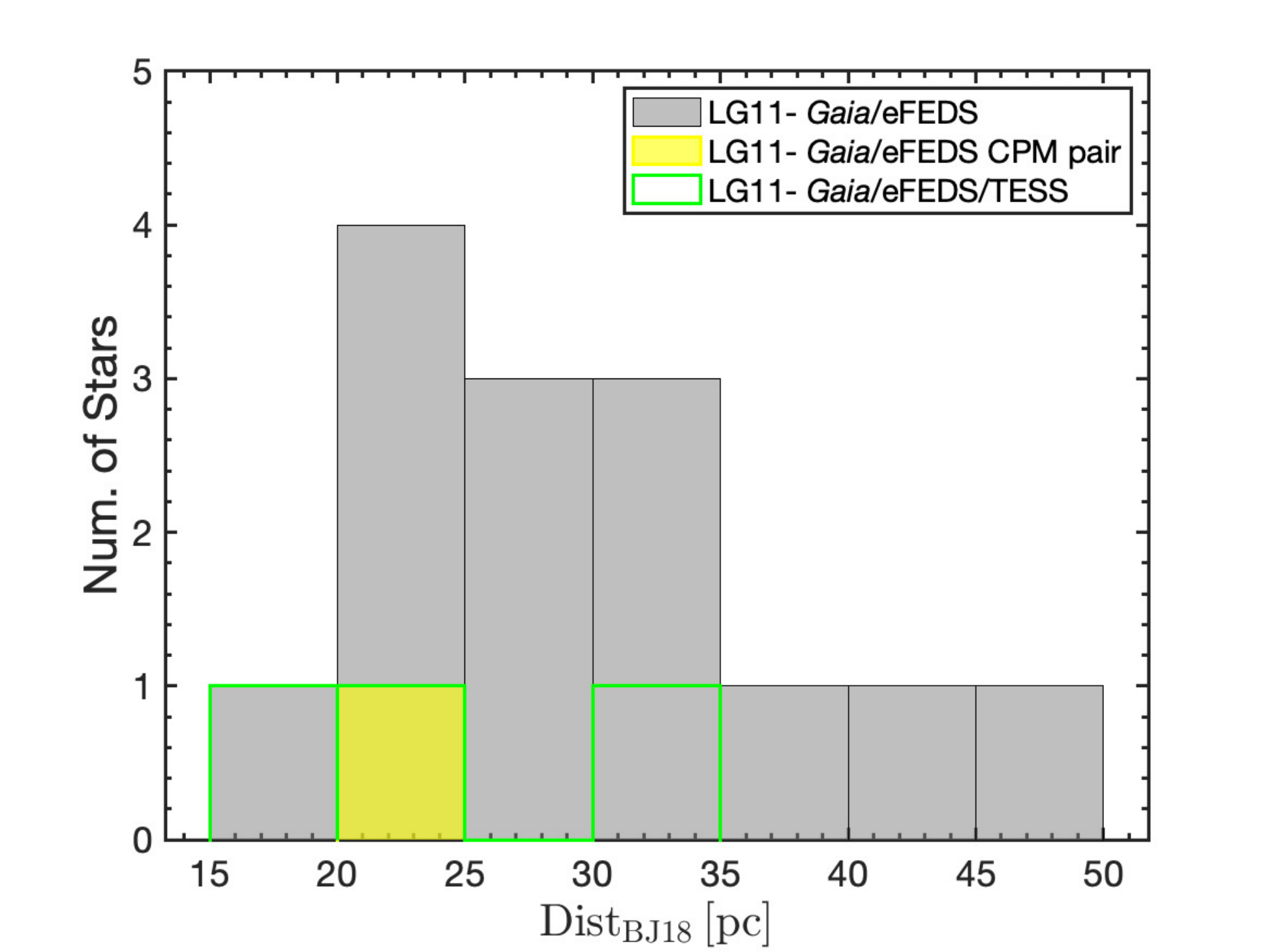}}
 		\parbox{9cm}{\includegraphics[width=0.5\textwidth]{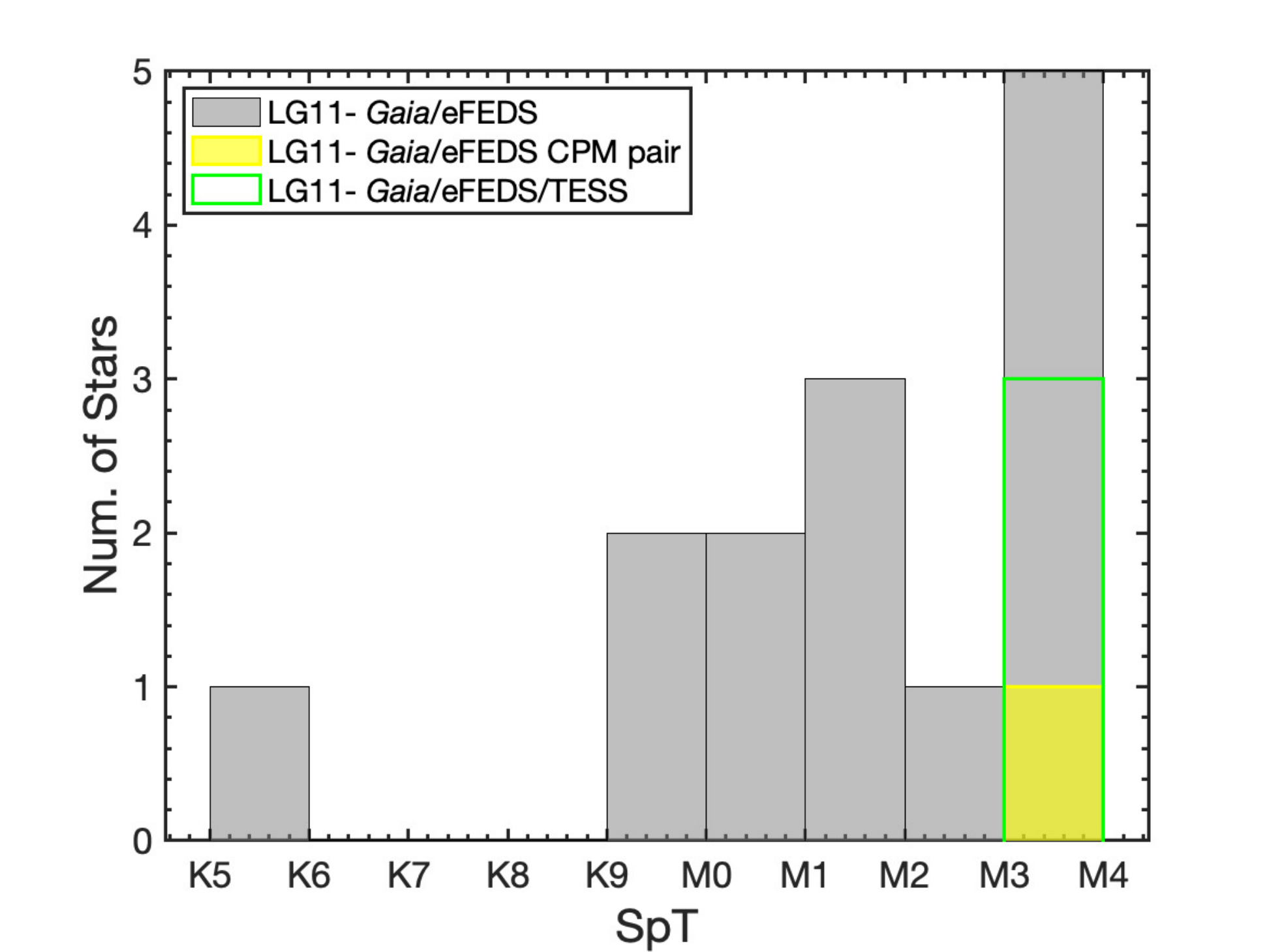}}
 		}
 	\parbox{18cm}{
 	\parbox{9cm}{\includegraphics[width=0.5\textwidth]{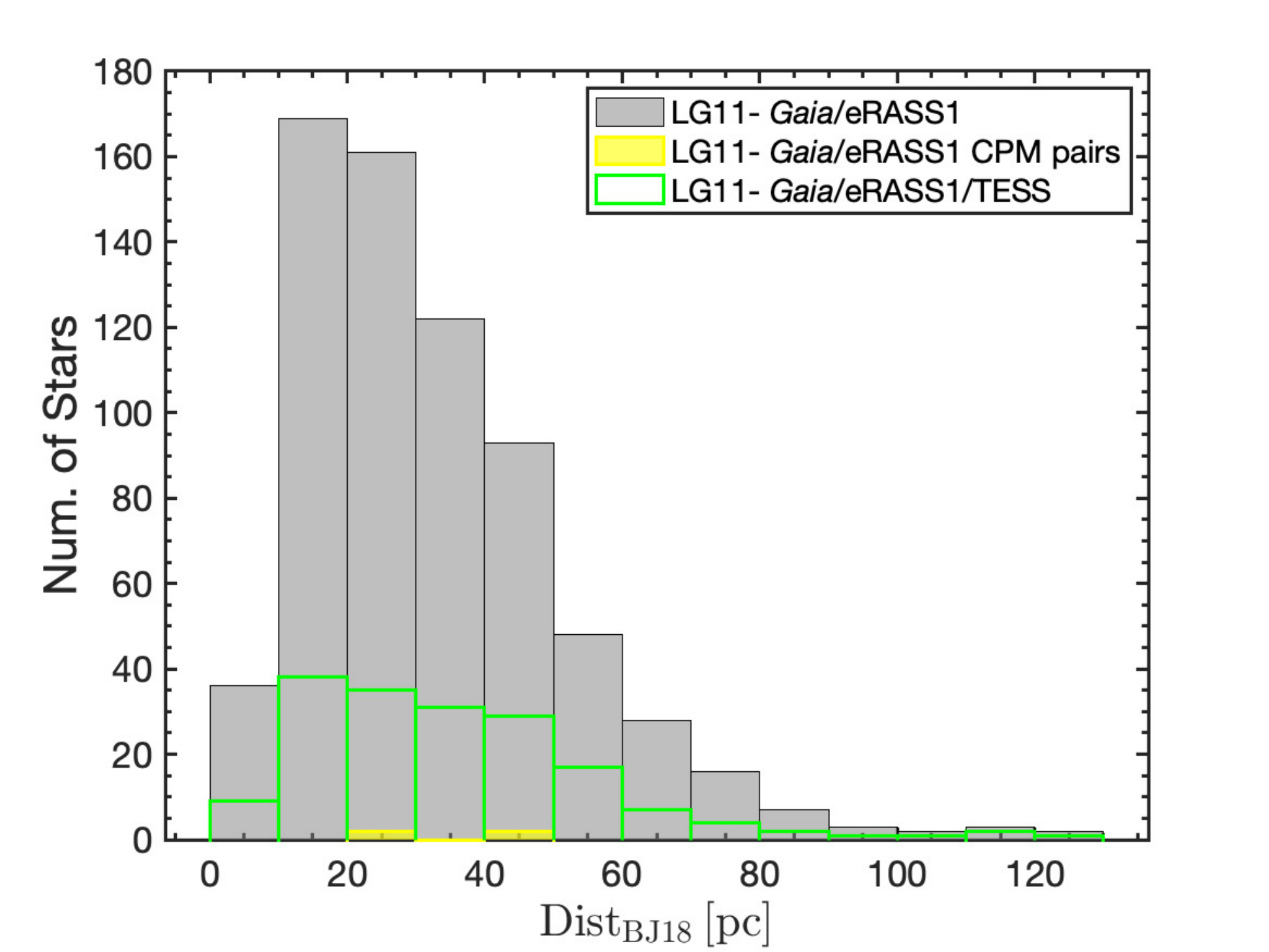}}
 		\parbox{9cm}{\includegraphics[width=0.5\textwidth]{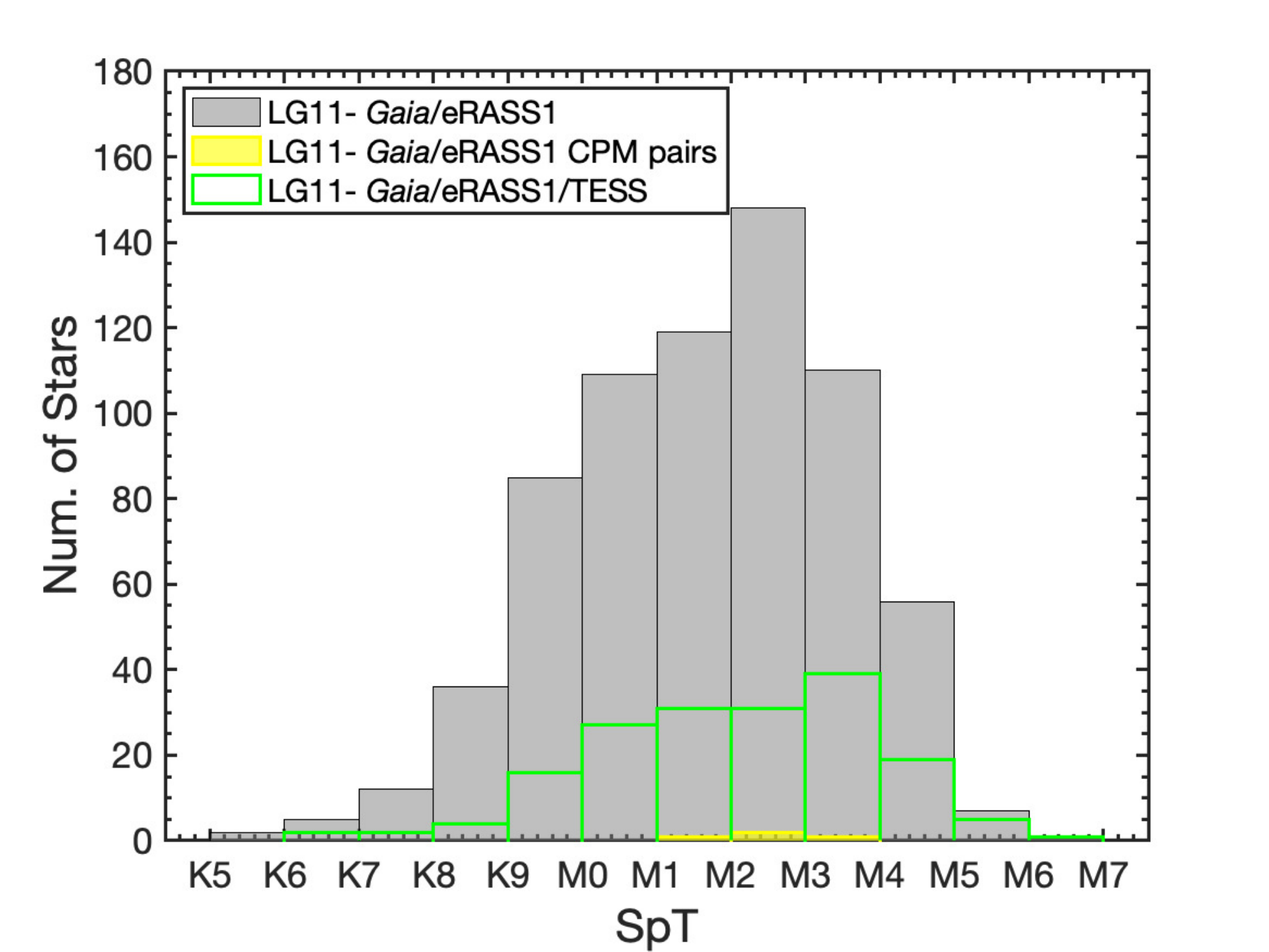}}
 		}
 	\caption{{\bf Top panel:} Distribution of {\em Gaia} distances from \cite{BailerJones2018} and  spectral types of the stars from LG11-{\em Gaia} in the eFEDS fields;  undetected (in gray) and  detected  (in red). {\bf Bottom panel:} Same as for the upper panels but for the LG11-{\em Gaia} stars identified
 	in the eRASS1 catalog v.200108. In all panels the green contours represent the distance and SpT distribution for those stars observed by {\em TESS,} and in yellow we show the CPM pairs.}
 	\label{fig:samples_spt_dist}
 \end{figure*}

In the upper row of Fig.~\ref{fig:samples_spt_dist} we show the distribution of the distances and spectral types for our LG11-{\em Gaia}/eFEDS sample. Fig.~\ref{fig:samples_spt_dist} (top panels) also highlights the one CPM pair of the sample as well as the subsample of stars with {\em TESS} rotation period, i.e. $3$ out of $14$ stars (see Sect.~\ref{subsect:analysis_efeds_TESS}). The Gaia source IDs, stellar parameters and distances for the $14$ LG11/{\em Gaia} stars in our eFEDS sample 
 are listed in Table~\ref{table:stellar_par}. 
 \begin{table*}
    \begin{center}
    \caption{Stellar parameters and distances of the {\em eROSITA} samples of M dwarfs.} 
    \label{table:stellar_par}
        \begin{tabular}{ccccccc}
        \midrule[0.5mm]  
            \multicolumn{1}{l}{LG11 Name} 
				&\multicolumn{1}{c}{Gaia-DR2 designation}  
				&\multicolumn{1}{c}{SpT}  
				&\multicolumn{1}{c}{$Dist_{\rm BJ18}$}
				&\multicolumn{1}{c}{$M_{\rm K_{s}}$}
				&\multicolumn{1}{c}{$M_{\star}$}
				&\multicolumn{1}{c}{binary}\\

				\multicolumn{1}{c}{} 
				&\multicolumn{1}{c}{}  
				&\multicolumn{1}{c}{}  
				&\multicolumn{1}{c}{[pc]} 
				&\multicolumn{1}{c}{[mag]} 
				&\multicolumn{1}{c}{[$\rm M_{\odot}$]}
				&\multicolumn{1}{c}{}\\
            \midrule
            \multicolumn{7}{c}{LG11-{\it Gaia}/eFEDS sample}\\
            \midrule
            PM I08551$+$0132&577602496345490176&K9.4&20.53$\pm$0.02&0.03$\pm$0.03&0.67$\pm$0.01&0\\
            PM I08570$+$0103&576773808175184768&M0.1&48.98$\pm$0.12&0.04$\pm$0.04&0.60$\pm$0.01&0\\
            PM I08590$+$0151&576970105360152192&K5.3&40.89$\pm$0.47&0.14$\pm$0.14&0.78$\pm$0.02&0\\
            $\cdot\cdot$&$\cdot\cdot$&$\cdot\cdot$&$\cdot\cdot$&$\cdot\cdot$&$\cdot\cdot$&$\cdot\cdot$\\
            $\cdot\cdot$&$\cdot\cdot$&$\cdot\cdot$&$\cdot\cdot$&$\cdot\cdot$&$\cdot\cdot$&$\cdot\cdot$\\
            \midrule
            \multicolumn{7}{c}{LG11-{\it Gaia}/eRASS1 sample}\\
            \midrule
            PM I00016$-$7613&4684946035804965632&M2.3&34.50$\pm$0.03&5.91$\pm$0.03&0.48$\pm$0.01&0\\
            PM I00054$-$3721&2306965202564506752&M1.4&4.34$\pm$0.00&6.33$\pm$0.02&0.41$\pm$0.01&0\\
            PM I00082$-$5705&4919497979411495296&M2.9&12.80$\pm$0.01&6.86$\pm$0.02&0.33$\pm$0.01&0\\
            $\cdot\cdot$&$\cdot\cdot$&$\cdot\cdot$&$\cdot\cdot$&$\cdot\cdot$&$\cdot\cdot$&$\cdot\cdot$\\
            $\cdot\cdot$&$\cdot\cdot$&$\cdot\cdot$&$\cdot\cdot$&$\cdot\cdot$&$\cdot\cdot$&$\cdot\cdot$\\
            \bottomrule[0.5mm] 
        \end{tabular}
    \end{center}
 \end{table*}
 
\subsection{M dwarfs detected in eRASS1}\label{subsect:sample_eRASS1}

For the cross-match of our LG11-{\em Gaia} catalog with the eRASS1 catalog (v201008) we used the {\em Gaia} coordinates, corrected by their proper motions to the rough mean observing date of eRASS1 (March 10, 2020). 
We cross-matched these  extrapolated positions of the M dwarfs with the boresight corrected coordinates (col. RA\_CORR, DEC\_CORR) of the eRASS1 catalog within a radius of $25^{\prime\prime}$. At about $10^{\prime\prime}$ the cumulative histogram of identifications flattens out. The cross-matching radius is a compromise between defining a complete sample and avoiding to pick wrong counterparts.  Based on the shape of the cumulative separation distribution we, therefore, consider in the following only the matches within $15^{\prime\prime}$. 
After removing ten stars that are propriety of the Russian {\em eROSITA} consortium the catalog results in $843$ X-ray sources. 
The choice of $15^{\prime\prime}$ as identification radius amounts to only $\sim 2$\,\% less sources than the $25^{\prime\prime}$ match radius and $\sim 5$\,\% more sources than would be in a $10^{\prime\prime}$ radius. 

Analogous to the procedure in Sect.~\ref{subsect:sample_eFEDS}, to uncover the {\em Gaia} sources that are alternative potential counterparts to the X-ray sources we performed a `reverse' match, in which we searched for all {\em Gaia} sources within $15^{\prime\prime}$ of the boresight corrected positions of the $843$ eRASS1 sources. This resulted in a total of $2145$ potential {\em Gaia} counterparts. These multiple optical counterparts should include all $843$ LG11-{\em Gaia} M dwarfs identified in the first match with an X-ray source. In practice we recover only $841$ of them. This is explained by the fact that two stars are not recovered within $15^{\prime\prime}$ because of their high proper motions. 
These cases are similar to the one discussed in Sect.~\ref{subsect:sample_eFEDS} for which a search radius of $15^{\prime\prime}$ was too small. 

We inspected the separations between X-ray position and {\em Gaia} coordinates for all potential {\em Gaia} counterparts  considering for those {\em Gaia} sources that are identified with a target star the P.M. correction to the mean eRASS1 observing date. 
This way we found that  the star from the LG11-{\em Gaia} list is the closest {\em Gaia} object to an X-ray source in $764$ cases.
Specifically to find the two stars from the LG11-{\em Gaia} sample with very high proper motion that are not recovered in the reverse match we increased the search radius to $45^{\prime\prime}$. This leads to $766$ closest {\em Gaia} counterparts identified with a star from our LG11-{\em Gaia} catalog, of which $25$ have a companion.
The components of these binary systems are associated to the same X-ray source, thus special attention is needed. These systems are, by definition of how we identified multiples in the LG11 catalog, resolved by {\em Gaia}. However, $21$ of them are associated with a single 2MASS source. Since for such systems we can not determine reliable stellar parameters we disregard them. For the remaining four CPM pairs that are resolved with {\em Gaia} and 2MASS but not with {\em eROSITA} we adopt the same approach as for the only binary in the LG11-{\em Gaia}/eFEDS sample (see Sect.~\ref{subsect:sample_eFEDS}), i.e. we ascribe the X-ray emission to  the component of the binary that is closest to the X-ray position\footnote{Note that in one of these four CPM pairs resolved with 2MASS one component has no complete {\em Gaia} data, and thus it is removed in the first place from our LG11-{\em Gaia} sample (see Sect.~\ref{sect:sample}). Moreover, this component is not the closest to the X-ray source, and would be removed anyway.\label{footnote:CPM_2MASS}}.
Finally, we removed all stars among the $742$ for which ${\rm Sep_{X,opt}}$ is higher than 3 times the uncertainty on the X-ray position in the eRASS1 catalog ({\sc RADEC\_ERR}).
This leaves $690$ eRASS1 X-ray sources and these define our eRASS1 M dwarf sample, LG11-{\em Gaia}/eRASS1. Among these, $593$ stars are also included in our `validated' LG11-{\em Gaia}/eRASS1 sample (see Table~\ref{tab:sample_def}). 

The {\em Gaia}\,DR2 source IDs, stellar parameters and distances for the LG11-{\em Gaia}/eRASS1 sample
are presented in Table~\ref{table:stellar_par}, and their  histograms of distance and SpT are shown in the bottom panels of Fig.~\ref{fig:samples_spt_dist}. Overlaid is the subsample with {\em TESS} rotation periods that is described in Sect.~\ref{subsect:analysis_erass1_TESS} and the four stars having a co-moving companion.
These distributions are similar to those of the M dwarfs in the eFEDS (displayed in the top panels of the same figure) but they provide a more than $20$-fold higher number statistics. It can also be seen from Fig.~\ref{fig:samples_spt_dist} that the subsample observed by {\em TESS} is a representation 
of the X-ray detected stars which is unbiased in terms of distance and SpT. 

\begin{figure}
    \centering
    \includegraphics[width=0.5\textwidth]{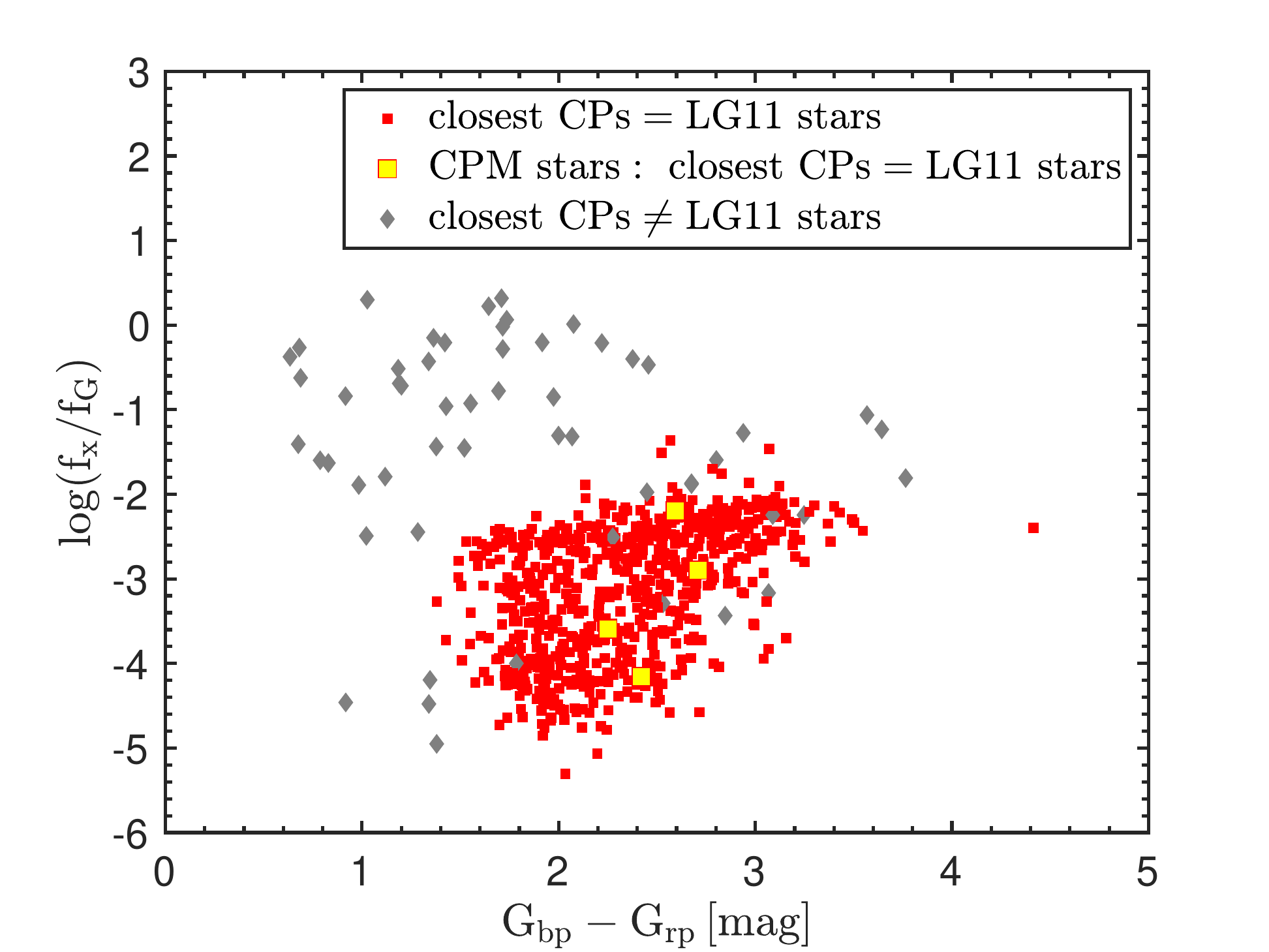}

    \caption{The X-ray-to-optical flux ratio versus {\em Gaia} color for the closest optical counterparts to the X-ray sources selected from the match of LG11-{\em Gaia} with the eRASS1 catalog. Two distinct regions are visible: the `extragalactic' area with high $\log(f_{x}/f_{G})$ values and relatively blue {\em Gaia} color and the `star' region with lower values of $\log(f_{\rm x}/f_{\rm G})$ and redder {\em Gaia} color. The filled red squares are the $690$ that we identified as M dwarfs (see text in Sect.~\ref{subsect:sample_eRASS1}), and the filled yellow squares represent the four stars that have a co-moving companion. We show only those sources with $\rm Sep_{X,opt} < 3~x~RADEC\_ERR$.}
    \label{fig:fxfg_bpminrp}
\end{figure}

The source identification is always a compromise between completeness and avoidance to include wrong counterparts. As explained in Sect.~\ref{subsect:sample_eFEDS} we aim at defining secure X-ray associations with MS stars from the LG11 catalog, at the expense of possibly missing some of them as X-ray emitters. Therefore, we have removed all but those M dwarfs that have been determined through the above analysis to have the smallest separation to the X-ray source. The nature of the remaining closest {\em Gaia} counterparts, i.e. those that are not part of the LG11-{\em Gaia} catalog, is not of interest to our work. However, a quick assessment can be done with help of a diagram that combines X-ray-to-optical flux ratio, $f_{\rm x}/f_{\rm G}$, with {\em Gaia} color. 
On the basis of {\em eROSITA} observations from the eFEDS fields Stelzer et al. (A\&A subm.) show how stars and extragalactic objects separate in this diagram. In fact, in Fig.~\ref{fig:fxfg_bpminrp} the closest {\em Gaia} counterparts of our `reverse' match split into two strongly populated areas. Note that in this figure we consider only the objects with $\rm Sep_{X,opt} < 3~x~RADEC\_ERR$.  The $690$ M dwarfs from LG11-{\em Gaia}/eRASS1 are located in the lower right, including the four stars that have a co-moving companion (highlighted in yellow). Most of the remaining closest {\em Gaia} counterparts that are not objects from our input catalog are located in the upper left of the diagram which defines the extragalactic region. 

\section{Data analysis for the eFEDS fields}
\label{sect:analysis_efeds}

\subsection{eROSITA}\label{subsect:analysis_efeds_xrays}

M dwarf stars are soft X-ray emitters. Indeed, 
in early versions of our data reduction we had noticed that  for the detected stars in our sample no photons have been collected at energies above 5.0\,keV. 
The lowest energy recommended to be used with {\em eROSITA} data is 0.2\,keV  \citep{Predehl2021}.
Therefore, we performed the analysis in the $0.2-5.0$~keV energy band. In the following we explain the details of the data  extraction and analysis regarding the M dwarfs in the eFEDS fields. 

\subsubsection{Data extraction}
\label{subsubsect:analysis_efeds_xrays_extraction}

We analyzed the eFEDS c946 processing data using the eSASSusers\_200602 software release\footnote{We extracted eFEDS data in parallel with the construction of the official catalog, therefore both the processing and the software release we used are not the ones published by the consortium.}. The data processing provides seven events files in the whole {\em eROSITA} energy band ($\rm 0.2-10.0$\,keV), one for each telescope/camera system  on board {\em eROSITA}. We merged the seven files to create one single events and image file filtered for corrupted events in the energy range $\rm 0.2-5.0$\,keV. We calculated the exposure map and the detection mask,
which are needed for the source detection, in the same energy band.
We computed the background map with the {\tt erbackmap} routine, using a smooth fit with a smoothing value of $15$.
Source detection was performed using the {\tt ermldet} pipeline for which we adopted a minimum threshold for the detection maximum likelihood of $6.0$.

We detected a total of $24376$ X-ray sources in the combined four eFEDS fields. The slight difference with respect to the number of sources in the eFEDS\_c001\_V4\_main catalog ($27910$ sources) is most likely to be attributed to different parameters set in the extraction process. These differences do not have any
effect on our study as we have shown in Sect.~\ref{subsect:sample_eFEDS} where we have anticipated our result for the identification of our M dwarf target list with the eFEDS X-ray sources. 

The basic X-ray parameters of the $14$ M dwarfs detected in eFEDS are given in Table~\ref{table:basic_xrayparams}. In particular, we provide the name of the star in LG11 (col.1), 
the X-ray coordinates with their uncertainty  (cols.~2$-$4), the offset between the proper motion corrected optical position and the X-ray coordinates (col.\,5), the $0.2-5.0$\,keV count rate obtained from the source detection procedure (col.\,6), and the detection maximum likelihood in the same energy band (col.\,7). 

\begin{table*}
    \begin{center}
    \caption{Basic X-ray parameters of the LG11-{\em Gaia}/eFEDS and LG11-{\em Gaia}/eRASS1 samples. The data are  from our own eFEDS X-ray source catalog and from the official eRASS1 catalog v201008, and they refer to the $0.2-5.0$\,keV band.} 
    \label{table:basic_xrayparams}
    \begin{tabular}{lcccccc}
        \midrule[0.5mm] 
            \multicolumn{1}{l}{LG11 Name} 
				&\multicolumn{1}{c}{RA\_CORR}  
				&\multicolumn{1}{c}{DEC\_CORR}  
				&\multicolumn{1}{c}{RADEC\_CORR}
				&\multicolumn{1}{c}{$\rm Sep_{X,opt}$}
				&\multicolumn{1}{c}{ML\_RATE\_0}
				&\multicolumn{1}{c}{DET\_ML\_0}\\

				\multicolumn{1}{c}{} 
				&\multicolumn{1}{c}{[deg]}  
				&\multicolumn{1}{c}{[deg]}  
				&\multicolumn{1}{c}{[arcsec]} 
				&\multicolumn{1}{c}{[arcsec]} 
				&\multicolumn{1}{c}{$\cdot 10^{-3}$ [cnt/s]}
				&\multicolumn{1}{c}{}\\
            \midrule
            \multicolumn{7}{c}{LG11-{\it Gaia}/eFEDS sample}\\
            \midrule
            PM I08551$+$0132&133.781820&1.540727&4.00&0.67&31.87$\pm$\phantom{1}6.97&\phantom{1}30.38\\ 
            PM I08570$+$0103&134.270789&1.057574&4.61&5.36&14.96$\pm$\phantom{1}5.06&\phantom{1}10.24\\ 
            PM I08590$+$0151&134.758412&1.864988&2.38&2.01&80.68$\pm$10.13&113.11\\ 
            $\cdot\cdot$&$\cdot\cdot$&$\cdot\cdot$&$\cdot\cdot$&$\cdot\cdot$&$\cdot\cdot$&$\cdot\cdot$\\
            $\cdot\cdot$&$\cdot\cdot$&$\cdot\cdot$&$\cdot\cdot$&$\cdot\cdot$&$\cdot\cdot$&$\cdot\cdot$\\
            \midrule
            \multicolumn{7}{c}{LG11-{\it Gaia}/eRASS1 sample}\\
            \midrule
            PM I00016$-$7613&0.416684&-76.230118&3.09&\phantom{1}2.47&110.61$\pm$31.42&31.66\\ 
            PM I00054$-$3721&1.395104&-37.370675&5.11&10.36&\phantom{1}81.23$\pm$34.24&11.28\\ 
            PM I00082$-$5705&2.070065&-57.096650&4.86&\phantom{1}6.41&\phantom{1}79.60$\pm$30.13&11.15\\ 
            $\cdot\cdot$&$\cdot\cdot$&$\cdot\cdot$&$\cdot\cdot$&$\cdot\cdot$&$\cdot\cdot$&$\cdot\cdot$\\
            $\cdot\cdot$&$\cdot\cdot$&$\cdot\cdot$&$\cdot\cdot$&$\cdot\cdot$&$\cdot\cdot$&$\cdot\cdot$\\
            \bottomrule[0.5mm] 
        \end{tabular}
    \end{center}
 \end{table*}

We also carried out a spectral and temporal analysis for these stars. To this end, we used the {\tt srctool} routine selecting a circular region for the source (with radius of $30\,^{\prime\prime}-40\,^{\prime\prime}$ depending on the source brightness). 
The analysis of the light curves and spectra is explained in the following Sects.~\ref{subsubsect:analysis_efeds_xrays_spectra} and~\ref{subsubsect:analysis_efeds_xrays_lcs}.

\subsubsection{Spectral analysis}
\label{subsubsect:analysis_efeds_xrays_spectra}

\begin{table*}
	\caption{X-ray spectral parameters for the ten M stars in eFEDS with more than 30 counts in the $\rm 0.2-5.0$\,keV band;  1~$\sigma$ uncertainties were computed with the {\sc error} pipeline provided in the XSPEC software package.}            
	\begin{center}
		\begin{tabular}{cccccccc}
			\midrule[0.5mm]                    
			\multicolumn{1}{c}{Name}
			&\multicolumn{1}{c}{$\rm kT_{1}$} 
			&\multicolumn{1}{c}{$\rm \log\left(EM_{1}\right)$}  
			&\multicolumn{1}{c}{$\rm kT_{2}$} 
			&\multicolumn{1}{c}{$\rm \log\left(EM_{2}\right)$}
			&\multicolumn{1}{c}{$\rm \chi^{2}_{red}$}
			&\multicolumn{1}{c}{d.o.f.}
			&\multicolumn{1}{c}{$\rm T_{mean}$}\\  
			
			\multicolumn{1}{c}{} 
			&\multicolumn{1}{c}{[keV]}  
			&\multicolumn{1}{c}{[$\rm cm^{-3}$]}  
			&\multicolumn{1}{c}{[keV]} 
			&\multicolumn{1}{c}{[$\rm cm^{-3}$]}  
			&\multicolumn{1}{c}{}
			&\multicolumn{1}{c}{} 
			&\multicolumn{1}{c}{[keV]}\\
			\midrule[0.5mm]

			PM I08551+0132&0.11$\pm$0.03&50.75$\pm$0.37&$\cdot\cdot$&$\cdot\cdot$&0.4&\phantom{0}3&0.11$\pm$0.03\\
			PM I08590+0151&0.20$\pm$0.06&50.67$\pm$0.12&0.80$\pm$0.18&50.45$\pm$0.15&0.4&\phantom{0}9&0.42$\pm$0.07\\
			PM I09034$-$0023&0.30$\pm$0.02&51.23$\pm$0.06&1.25$\pm$0.20&51.05$\pm$0.08&1.0&19&0.67$\pm$0.07\\
			PM I09050+0226&0.23$\pm$0.03&50.67$\pm$0.07&0.99$\pm$0.16&50.40$\pm$0.07&0.8&\phantom{0}8&0.50$\pm$0.06\\
			PM I09050+0250&0.50$\pm$0.23&49.93$\pm$0.27&5.49$\pm$-1.10&50.39$\pm$0.22&1.2&\phantom{0}5&4.22$\pm$0.82\\
			PM I09161+0153&0.25$\pm$0.01&51.12$\pm$0.04&0.99$\pm$0.04&51.00$\pm$0.03&1.2&25&0.57$\pm$0.02\\
			PM I09201+0347&1.00$\pm$0.06&50.60$\pm$0.05&$\cdot\cdot$&$\cdot\cdot$&1.2&12&1.00$\pm$0.06\\
			PM I09205+0135&0.25$\pm$0.02&51.44$\pm$0.05&0.98$\pm$0.06&51.33$\pm$0.05&1.0&16&0.57$\pm$0.03\\
			PM I09238+0008&0.11$\pm$0.08&50.55$\pm$1.40&0.77$\pm$0.13&50.40$\pm$0.08&0.1&\phantom{0}6&0.38$\pm$0.07\\
			PM I09308+0227&0.31$\pm$0.02&50.73$\pm$0.06&1.22$\pm$0.20&50.44$\pm$0.11&0.5&10&0.62$\pm$0.07\\
			
			\bottomrule[0.5mm]
		\end{tabular}
		\label{tab:xspec_output}
	\end{center}
\end{table*}

Spectral analysis was performed with XSPEC\footnote{XSPEC NASA's HEASARC Software: \url{https://heasarc.gsfc.nasa.gov/xanadu/xspec/}} version 12.10. We carried out spectral fitting  only for the ten out of the $14$ detected sources that have more than $30$ net source counts. We used a two temperature thermal model (\rm{APEC\footnote{More information about APEC model used by XSPEC software can be found at \url{https://heasarc.gsfc.nasa.gov/xanadu/xspec/manual/node135.html}}}) except for one star, which is the faintest among those stars for which we have a reasonable spectrum and which can be described by a one-temperature APEC model.

Each APEC component has three parameters: the plasma temperature ($kT$), the global abundance ($Z$), and the emission measure ($EM$). The emission measure is the square of the number density of free electrons integrated over the volume of the emitting plasma, and it is obtained from the normalization factor of the XSPEC fit combined with the source distance. 
We fixed $Z$ at $0.3~{\rm Z_{\odot}}$, the typical global abundance for late-type stars, and we left $kT$ and $EM$ free to vary. 
We computed the mean coronal temperature ($T_{\rm mean}$) by weighting the temperatures of the individual {\sc apec} components by their EM, 
\begin{equation}
T_{\rm mean} = \frac{\sum\left( EM_{\rm n}\cdot T_{\rm n}\right)}{\sum\left(EM_{\rm n}\right)},
\end{equation} 
where $n=1, 2$ for the two components of the best fitting model. 
The parameters of the best-fitting model including the values of $T_{\rm mean}$ are listed in Table~\ref{tab:xspec_output} and the spectra are shown in Fig.~\ref{fig:eFEDS_spec}. 
One of the ten stars is the binary discussed in Sect.~\ref{subsect:sample_eFEDS}
that is unresolved with {\em eROSITA}, i.e. the spectra of two stars are summed up. Since the masses  of the two components are equal (see Table~\ref{table:stellar_par}) we can assume the X-ray spectra to be similar, and therefore we treat this spectrum in the same way as the others.

We computed the fluxes in the $0.2-5.0$\,keV band, $f_{\rm x}$, with the \textit{flux} routine provided by XSPEC.
For all stars that are too faint for spectral analysis we calculated a conversion factor ($CF_{\rm eFEDS}$) for transforming the count rate to flux. 
We defined $CF_{\rm eFEDS}$ as the ratio between flux and count rate of each source for which we have analyzed the spectrum. In particular, we used the fluxes computed with XSPEC and the count rates we found performing the source detection. Then we calculated the mean value,
\begin{equation}
\langle CF_{\rm eFEDS} \rangle = mean\left(\frac{f_{x}}{Ct.Rate}\right).
\end{equation} 
We excluded the two stars with $\rm d.o.f. \leq 5$ (see {Table~\ref{tab:xspec_output}) from the calculation of the mean, because the coronal temperatures we derive for them are not typical of M dwarf stars and this is likely the result of the poor statistics of the spectrum. From the eight stars with good quality spectra we obtained $\langle CF_{\rm eFEDS} \rangle = 7.81 \cdot 10^{-13} \pm 7.48 \cdot 10^{-14}\,\rm erg/cnt/cm^2$. 

We determined the X-ray fluxes of the four detected stars without spectral fit (i.e. those with less than $30$ net counts) and the two stars with poor spectral fit by combining their count rates with $\langle CF_{eFEDS} \rangle$.
The X-ray luminosities were determined by combining the fluxes with the distances from  Table~\ref{table:stellar_par}, and the X-ray to bolometric ratios, $\log(L_{\rm x}/L_{\rm bol})$ in Table~\ref{table:act_rot} were obtained using the $L_{\rm bol}$ values derived with the relations of \cite{Mann2015,Mann_2016}. The eFEDS X-ray luminosities are presented in  Table\,\ref{table:act_rot} together with {\em ROSAT} and {\em TESS} parameters derived in the following sections. 

 \begin{table*}
    \begin{center}
    \caption{Measurements of X-ray activity and rotation derived by us from {\em eROSITA}, {\em ROSAT} and {\em TESS} data.}
    \label{table:act_rot}
 \begin{tabular}{cccccccccc} 
 
            \midrule[0.5mm] 
                \multicolumn{1}{c}{LG11 Name} 
				&\multicolumn{1}{c}{$HR_{\rm 1}$}  
				&\multicolumn{1}{c}{$HR_{\rm 2}$}  
				&\multicolumn{1}{c}{${(\log{L_{\rm x}})}_{\rm eROSITA}$}
				&\multicolumn{1}{c}{${(\log{L_{\rm x}})}_{\rm ROSAT}$}
				&\multicolumn{1}{c}{$\log (L_{\rm x}/L_{bol})$}
				&\multicolumn{1}{c}{TIC number}
				&\multicolumn{1}{c}{$P_{\rm rot}$}
				&\multicolumn{1}{c}{flag\_p}
				&\multicolumn{1}{c}{$R_{\rm 0}$}\\

				\multicolumn{1}{c}{} 
				&\multicolumn{1}{c}{[cts/s]}  
				&\multicolumn{1}{c}{[cts/s]}  
				&\multicolumn{1}{c}{[erg/s]} 
				&\multicolumn{1}{c}{[erg/s]} 
				&\multicolumn{1}{c}{}
				&\multicolumn{1}{c}{}
				&\multicolumn{1}{c}{[d]}
				&\multicolumn{1}{c}{}
				&\multicolumn{1}{c}{}\\
            \midrule
            \multicolumn{10}{c}{LG11-{\it Gaia}/eFEDS sample}\\
            \midrule
            PM I08551$+$0132&-0.52&-1.00&27.10$\pm$0.10&$\cdot\cdot$&-5.47$\pm$0.04&265373654&$\cdot\cdot$&$\cdot\cdot$&$\cdot\cdot$\\
            PM I08570$+$0103&0.10&-1.00&27.52$\pm$0.15&$\cdot\cdot$&-4.87$\pm$0.04&265440550&$\cdot\cdot$&$\cdot\cdot$&$\cdot\cdot$\\
            PM I08590$+$0151&0.11&-1.00&28.29$\pm$0.05&$\cdot\cdot$&      -4.11$\pm$0.04&$\cdot\cdot$&$\cdot\cdot$&$\cdot\cdot$&$\cdot\cdot$\\
            $\cdot\cdot$&$\cdot\cdot$&$\cdot\cdot$&$\cdot\cdot$&$\cdot\cdot$&$\cdot\cdot$&$\cdot\cdot$&$\cdot\cdot$&$\cdot\cdot$&$\cdot\cdot$\\
            $\cdot\cdot$&$\cdot\cdot$&$\cdot\cdot$&$\cdot\cdot$&$\cdot\cdot$&$\cdot\cdot$&$\cdot\cdot$&$\cdot\cdot$&$\cdot\cdot$&$\cdot\cdot$\\
            \midrule
            \multicolumn{10}{c}{LG11-{\it Gaia}/eRASS1 sample}\\
            \midrule
         
      PM I00016$-$7613&0.35&-0.90&28.09$\pm$0.12&$\cdot\cdot$&-3.96$\pm$0.04&266878145&$\cdot\cdot$&$\cdot\cdot$&$\cdot\cdot$\\
      PM I00054$-$3721&0.63&-1.00&26.15$\pm$0.18&$\cdot\cdot$&-5.75$\pm$0.03&120461526&$\cdot\cdot$&$\cdot\cdot$&$\cdot\cdot$\\
      PM I00082$-$5705&0.04&-1.00&27.08$\pm$0.16&$\cdot\cdot$&-4.57$\pm$0.04&201287746&$\cdot\cdot$&$\cdot\cdot$&$\cdot\cdot$\\
      $\cdot\cdot$&$\cdot\cdot$&$\cdot\cdot$&$\cdot\cdot$&$\cdot\cdot$&$\cdot\cdot$&$\cdot\cdot$&$\cdot\cdot$&$\cdot\cdot$&$\cdot\cdot$\\
      $\cdot\cdot$&$\cdot\cdot$&$\cdot\cdot$&$\cdot\cdot$&$\cdot\cdot$&$\cdot\cdot$&$\cdot\cdot$&$\cdot\cdot$&$\cdot\cdot$&$\cdot\cdot$\\
            \bottomrule[0.5mm]
        \end{tabular}
     \end{center}
 \end{table*}

 We have verified with the eFEDS X-ray spectra that the $L_{\rm x}$ values we calculated for the $0.2-5.0$\,keV band differ from those for a softer energy band ($0.2-2.4$\,keV), on the level of $1$\,\% or less. This can be easily understood from Fig.~\ref{fig:eFEDS_spec} where the spectra drop steeply above $\sim 1$\,keV. Therefore, in the remainder of this paper we use for our {\em eROSITA} detections the broader standard {\em eROSITA} band, also where we compare {\em eROSITA} and {\em ROSAT} measurements.

\subsubsection{Light curve analysis}
\label{subsubsect:analysis_efeds_xrays_lcs}

Each of the four eFEDS fields was scanned by {\em eROSITA} in the direction of the longer side of the individual fields. Thus, {\em eROSITA} has visited a given object within the eFEDS fields several times with a time lapse between one and the next visit that depends on the position of the object. As a consequence of this observing mode, light curves of individual sources are defined by short ($<100$\,s) intervals of data taking (called one `visit') separated by longer data gaps during which the satellite scans through the rest of the field turns around and approaches again the source. Generally, the length of the data gaps is alternating between two values, 
except for sources that are located in the middle of the eFEDS fields along the scanning direction where all data gaps have approximately equal length.

We have performed the light curve extraction with the dedicated source products pipeline, {\tt srctool}, which is part of the eSASS software. We worked in the energy band between $0.2-5.0$\,keV and we used the same source and background regions that we have adopted for the spectral analysis. 

As explained above, in survey mode a regularly binned light curve is dominated by data gaps. We, therefore, used the {\tt REGULAR-} option to produce light curves with regularly spaced bins in which time intervals without data are automatically discarded. 
We performed tests with different bin sizes to identify the best value for each source in order to avoid bins with a very low number of counts - and correspondingly large uncertainty - because they start near the end of a visit. Since, as explained above, the visits are not regularly spaced a large binsize was required to reach this goal. The binning we determined from our tests is between $1-3$\,ks, chosen  individually for each source. For stars located in the scanning direction near the edge of the eFEDS fields this means that we are averaging over two successive visits. 

The light curves of all $14$ stars from our sample 
that are detected in the eFEDS fields are shown in Fig.~\ref{app:lcs},  
where the individual binsize is indicated for each star in the legend. The uncertainties of the count rate are automatically calculated by the eSASS pipeline and they depend on the uncertainties of the source and background counts, the fractional telescope collecting area and the fraction of the time bin which overlaps with the input good time intervals (GTIs) that have been calculated by the eSASS pipeline during the extraction of the events file.  
During its first and last visits the source is located at the edge of the field-of-view and the fractional telescope collecting area and the fractional temporal coverage become smaller, and consequently the error bars increase.

A systematic analysis of variability of all sources detected in the eFEDS field is presented by Boller et al. (A\&A subm.). That variability study refers to all objects in the official eFEDS source catalog which comprises our $14$ detected M dwarfs.
We have extracted the variability metrices for these $14$ stars from the catalog of Boller et al. (A\&A subm.) Specifically, that catalog provides the normalized excess variance (NEV) as defined by \cite{Boller16.0} and its uncertainty\footnote{Note that the eFEDS variability tests carried out by Boller et al. (A\&A subm.) on the full eFEDS sample were  performed on light curves with $100$\,s bin size.}. 
The ratio of these two quantities represents the probability that the source is variable in units of Gaussian $\sigma$. We find that five of the LG11-{\em Gaia}/eFEDS sample have a well-determined NEV and its uncertainty. In fact, in the eFEDS fields the net count statistics is low for most sources such that the NEV is unconstrained. Only for one of the M dwarfs identified as variable the NEV is $\sigma > 3$, and another one has $\sigma > 2.5$. These two stars are PM\,I09161+0153, the brightest star in our sample (in terms of X-ray count rate), for which the visual inspection of the light curve indicates a likely flare ongoing at the beginning of the observation, and PM\,I09201+0347 which shows a smoother and longer-lasting variability throughout the eFEDS light curve.

\subsection{TESS}
\label{subsect:analysis_efeds_TESS}

To retrieve {\em TESS} data we uploaded our target list of the $14$ M dwarfs from the LG11-{\em Gaia}/eFEDS catalog to the {\em Barbara A. Mikulski Archive for Space Telescopes} (MAST) interface and found {\em TESS} data for $13$ stars. 
For the match we have used the J2000 coordinates from LG11 with a match radius of $1^{\prime\prime}$. Since the pixel scale of {\em TESS} is so large ($21^{\prime\prime}$ per pixel) a P.M. correction or its omission does not influence the result.  
Note that many stars have two TIC numbers, and we determined the correct TIC counterpart by comparing the magnitudes of the multi-band photometry provided in the TIC with the values listed for the LG11 star in Simbad and ESASky. 

All but one of the $13$ stars have short (2-minute) cadence light curves available. The remaining star was observed in full frame image (FFI) mode only. The observation of the eFEDS fields was performed by \textit{TESS} in its Sectors 7 and 8 during January and February 2019, and we downloaded the data from the  MAST Portal.

\subsubsection{Analysis of two-minute cadence light curves}
\label{subsubsect:analysis_efeds_TESS_lcs}

For our analysis that consisted of three steps we used the Pre-search Data Conditioning Simple Aperture Photometry (PDCSAP) light curves. \textit{TESS} assigns a quality flag to all measurements including data that are of poor quality but also data that might be of lower quality or could cause problems for transit detection after applying detrending software \citep{Thompson2016}. Hence, removing all flagged data points by default could impede the detection of real astrophysical signals or the interpretation of systematics. Therefore, we removed all flagged data points except of `Impulsive outlier' (which could be real stellar flares) and `Cosmic ray in collateral data' (bits 10 and 11) in step 1 of the analysis. In a second step we normalized the light curves by dividing all data points by the median flux.

The third analysis step is the search for rotation periods. To this end we used three different methods, the generalized Lomb-Scargle periodogram \citep[\begin{scriptsize}GLS\end{scriptsize};][]{2009A&A...496..577Z}, the autocorrelation function (ACF), and fitting the light curves with a sine function. We have first used this approach on data from the K2 mission, and all details on our period search can be found in \cite{Stelzer2016} and \citet{2020A&A...637A..22R}. For the analysis with \begin{scriptsize}GLS\end{scriptsize} we had to bin the data by a factor of 3 because the implementation we use\footnote{Fortran Version v2.3.01 released: 2011-09-13 by Mathias Zechmeister.} can only deal with up to 10000 data points. 

\subsubsection{Analysis of the full-frame images}
\label{subsubsect:analysis_efeds_TESS_ffi}

Since one of our targets with available {\em TESS} data does not have a 2-minute cadence light curve we decided to extract the long (29-minute) cadence light curves of all $13$ targets from the FFIs. To create a light curve we performed aperture photometry. Instead of using the whole FFIs we made use of the ``postcards'', an intermediate data product from the FFI analysis tool \begin{scriptsize}ELEANOR\end{scriptsize} \citep{2019PASP..131i4502F}. Postcards are 148$\times$104 pixel background subtracted cutout regions of the FFIs that are time-stacked, including all cadences for which observations are available. We converted the postcard cubes into individual fits images. Sectors 7 and 8 include 1093 and 968 cadences, respectively. Photometry was performed following the procedures described by us in e.g. \citet{2016MNRAS.460.2834R}. In short, for the aperture photometry with ten different aperture radii we used a user script based on the standard \begin{scriptsize}IRAF\end{scriptsize} routine \textit{phot}. Our script allows us to obtain simultaneous photometry of all stars in an image. For this purpose a list of the pixel coordinates of all detectable stars was created using \begin{scriptsize}SOURCE EXTRACTOR\end{scriptsize} \citep[\begin{scriptsize}SEXTRACTOR\end{scriptsize};][]{1996A&AS..117..393B}. As the positions of the stars on the CCDs do not change during a sector a single file with pixel coordinates was used for all cadences. Finally, we derived differential magnitudes using an optimised artificial comparison star \citep{2005AN....326..134B}. Since \textit{TESS} has an image scale of $\sim$21\,arcsec per pixel we found the optimal aperture radius for all targets to be 2 pixels. The resulting light curves of sector 8 show strong systematic effects at the beginning of the observations and after the observation gap caused by the data downlink at Earth perigee. We removed all affected data points, leading to final light curves with 1093 and $\sim$650 data points for sectors 7 and 8, respectively.

The long cadence light curves do show a smaller scatter than the short cadence ones.
Figure~\ref{fig:TESS_example_lc} shows as an example the comparison of the $2$-min cadence PDCSAP light curve and the
$30$-min light curve extracted from the FFIs for PM\,I09034-0023 (TIC\,893123) observed by {\em TESS} in Sector\,8. Although both light curves agree within their error bars, the scatter is much lower in the FFI light curve. For this particular star, the FFI light curve hints to a double hump shape in contrast to a single sinusoidal shape in the
PDCSAP light curve. 
Furthermore, the detrending of the strong systematic effects of Sector\,8 might introduce  artifacts that could cause a false determination of the rotation period. Therefore we applied the period search as explained in the previous section also to the long cadence light curves. \begin{figure}[t]
	\begin{center}
	\includegraphics[width=6.5cm,angle=270]{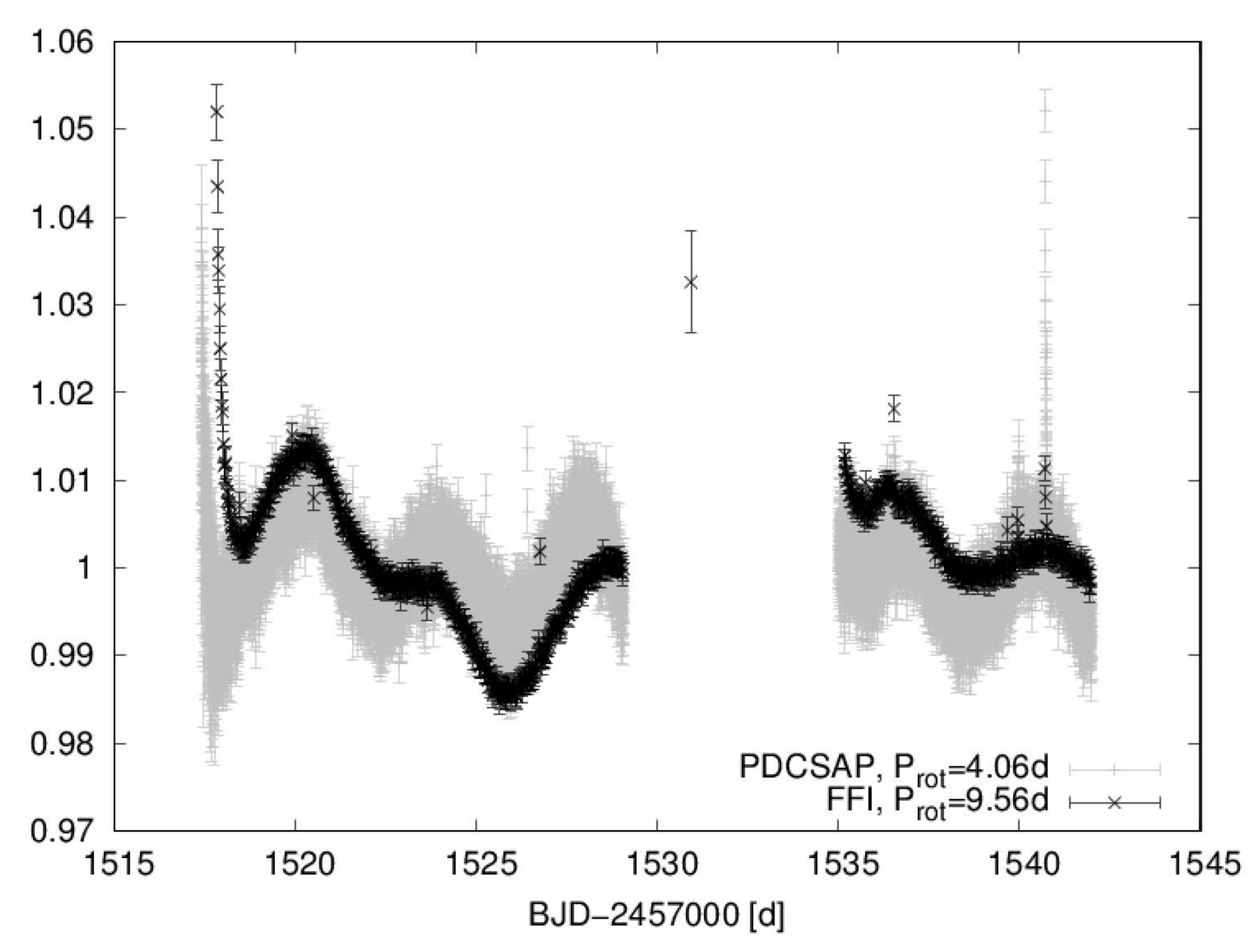}
	\caption{Comparison of the pipeline produced PDCSAP light curve of PM\,I09034-0023 (TIC\,893123, grey) and our own photometry (black) extracted from the full frame images. Within the error bars both light curves agree with each other while the scatter is much larger for the 2-min cadence light curve. The shape of the light curve is, however, different which results in different estimates of the rotation period.}
	\label{fig:TESS_example_lc}
		\end{center}
		\end{figure}

\subsubsection{Note on the binary star}\label{subsubsect:analysis_efeds_tess_binaries}

Our list of the $13$ LG11-{\em Gaia}/eFEDS stars 
observed with {\em TESS} includes the one close visual binary pair from Sect.~\ref{subsect:sample_eFEDS}. With the large pixel scale of \textit{TESS} these stars can not be resolved individually. 
Consequently, the automatic pixel masks of the short-cadence data which is centered on the star's position are slightly offset from each other, resulting in slightly different light curves. For the FFI we defined the pixel mask by ourselves, and we only created one light curve per binary pair.

\subsubsection{Outputs of the light curve analysis}\label{subsubsect:analysis_efeds_tess_output}

For each target we obtained six values for the rotation period, three for the long cadence light curves and three for the short cadence light curves. By eye-inspection of the phase-folded light curves we selected the best-fitting period. If several methods resulted in a similar value we computed the average. The standard deviation was used for the determination of the uncertainties. In addition, we used the formulas given in \citet{1985PASP...97..285G} to calculate an error for the rotation period. As final uncertainty we adopted the maximum of the standard deviation and the calculated error. We detected rotation periods for five targets from the LG11-{\em Gaia}/eFEDS sample.
By repeating the period search on the residuals of the light curve after subtracting the dominant period we found that one star shows an additional shorter period. 
Another one exhibits a double-hump shaped light curve which we discussed in Sect.~\ref{subsubsect:analysis_efeds_TESS_ffi}. 
We exclude these two stars with ambiguous period signal from our quantitative analysis, although we show them in the figures with separate symbols from the `reliable' sample. 
To summarize, only three of the $13$ eFEDS M dwarfs have a  reliable period detection, and these represent our LG11-{\em Gaia}/eFEDS/TESS sample (see Table~\ref{tab:sample_def}). All of them are `validated' with the \cite{Mann2015} relations. The most significant period is listed for all five stars in  Table~\ref{table:act_rot}, and a flag is provided for the ones that are not considered for the reasons described above.  
 
\section{Data analysis for eRASS1}\label{sect:analysis_erass1}

\subsection{eROSITA}\label{subsect:analysis_erass1_xrays}

In principle the same analysis steps described and carried out for the eFEDS data in Sect.~\ref{subsect:analysis_efeds_xrays} can be applied to eRASS data. However, we defer such an in-depth study to a future work. 

The X-ray parameters used in this work are directly extracted from the eRASS1 catalog. We list in Table~\ref{table:basic_xrayparams} the same parameters as for the X-ray detections found in the eFEDS fields for the $690$ LG11-{\em Gaia}/eRASS1 stars, i.e. next to the LG11 name (col.1) we provide the X-ray coordinates with their uncertainty (cols.~2$-$4), the offset between proper motion corrected optical position and the X-ray coordinates (col.4), the $0.2-5.0$\,keV count rate (col.5), and the $0.2-5.0$\,keV detection maximum likelihood (col.6). 

Since a systematic analysis of spectra and light curves such as that of Sect.~\ref{subsubsect:analysis_efeds_xrays_spectra} and~\ref{subsubsect:analysis_efeds_xrays_lcs} is beyond the scope of this work, to compute the X-ray fluxes for the eRASS1 detected M dwarfs we used the conversion factor $\langle CF_{\rm eFEDS} \rangle$ derived from the eFEDS data. X-ray luminosities and $L_{\rm x}/L_{\rm bol}$ ratios are then determined with the distances and $L_{\rm bol}$ values from Table~\ref{table:stellar_par}. 
The $L_{\rm x}$ and $L_{\rm x}/L_{\rm bol}$ values are listed in Table~\ref{table:act_rot}, together with the same parameters for the eFEDS detections.

\subsection{TESS}\label{subsect:analysis_erass1_TESS}

Analogous to the case of eFEDS (Sect.~\ref{subsect:analysis_efeds_TESS}), 
to define the sample of stars with {\em TESS} data we have loaded the list of $690$ X-ray detected stars from our LG11-{\em Gaia}/eRASS1 catalog into MAST, and we have  matched it with the target list of {\em TESS} using the J2000 coordinates in LG11 with a match radius of $1^{\prime\prime}$. We found that $488$ of the LG\,11-{\em Gaia} M dwarfs detected in eRASS1 have been observed with {\em TESS}. We recall the all-sky nature of eRASS1 which implies that the M dwarfs from this sample are distributed over all {\em TESS} sectors. Among the $488$ stars $127$ were observed in multiple sectors. The subset of the LG11-{\em Gaia}/eRASS1 sample observed with {\em TESS} includes three out of four co-moving binary systems from Sect.~\ref{subsect:sample_eRASS1}. 

For the stars from the LG11-{\em Gaia}/eRASS1 sample 
we have examined only the $2$-minute cadence {\em TESS} data in the way described in Sect.~\ref{subsubsect:analysis_efeds_TESS_lcs}, and the adopted values for the rotation periods were determined with the procedure described in Sect.~\ref{subsubsect:analysis_efeds_tess_output}.

We could find rotation periods for $222$ stars but we consider $39$ of these periods not reliable because the period is longer than half the duration of the observation. The periods of these stars are flagged in  Table~\ref{table:act_rot}.
Through by-eye inspection, we found that three more stars show a second period that was not identified as the dominant period with our period search methods, and an additional three stars have a light curve that looks double humped like the one case in the eFEDS sample discussed in Sect.~\ref{subsubsect:analysis_efeds_TESS_ffi}. 
We removed these six stars from the sample that we define as LG11-{\em Gaia}/eRASS1/TESS for our further analysis.
This sample comprises $177$ stars. We note that for completeness the removed six stars with ambiguous $P_{\rm rot}$ are shown in the figures distinguished with the plotting symbol from the stars from the  LG11-{\em Gaia}/eRASS1/TESS sample. 
The period with the highest significance is given in  Table\,~\ref{table:act_rot} together with a flag that identifies the stars that have been removed from the rotation sample as described above.  Taking into account the calibration range of \cite{Mann2015} we obtained a `validated' LG11-{\em Gaia}/eRASS1/TESS of $138$ stars. 

\section{Results \& discussion}\label{sect:results}

\subsection{The eROSITA M dwarf population}\label{subsect:results_mdwarfpopulation}

We have studied the X-ray emission of the M dwarfs from the LG11 catalog with matches in {\em Gaia}\,DR2 in two different {\em eROSITA} surveys, the eFEDS observation which covers $142$\,sq.deg in the Southern hemisphere and the first full all-sky survey, eRASS1. 
These two {\em eROSITA} samples 
together provide the X-ray luminosities of $\sim 704$ M dwarfs, exceeding our previously compiled sample \citep{Magaudda2020} in size by more than a factor of two, and historical samples from RASS by a factor of $7$ \citep[NEXXUS;][]{Schmitt2004} and from {\em Einstein} by a factor of $24$ \citep{Fleming88.0}.

From the list of $8229$ main-sequence M dwarfs provided by LG11 that have a  {\em Gaia}\,DR2 counterpart $30$ are located in the  eFEDS fields, almost exactly matching the all-sky space density average ($30/142 \approx 8229/41253$\,stars per sq.deg). We have detected $14$ out of these $30$ M dwarfs, i.e. nearly $50$\,\% of the sample. 
The average space density of X-ray detected M dwarfs in the eFEDS sample is, thus, $14/142 \approx 0.10$\,stars per sq.deg. 

Our analysis of the eRASS1 catalog has provided the to date largest sample of X-ray emitting M dwarfs, namely $690$ stars, which is a detection rate of $8.3$\,\% of our input sample LG11-{\em Gaia}, and the eRASS1 space density of X-ray detected M dwarfs is $690/20626 \approx 0.033$\,stars per sq.deg (considering half of the sky comprised in our version of the eRASS1 catalog). This is a factor three lower than for eFEDS which can be explained by the (on average) shorter exposure time during the all-sky survey.

The typical eFEDS exposure time is $\sim 1$\,ksec per sky position. 
The exposure time, and therefore the flux limit, during eRASS depends strongly on the sky position and, therefore, is not a universal value for a sample distributed over the sky. We can, however, give a rough value of $f_{\rm lim,eRASS1} \sim 3 \cdot 10^{-14}\,{\rm erg/cm^2/s}$. At this value the distribution of fluxes detected for our M dwarfs drops steeply, and only $\sim 2$\,\% of the detections have a flux lower than this. 
 Defining the flux limit in the same way as for eRASS1 (i.e. as that value $f_{\rm x}$ which is exceeded by $98$\,\% of the stars in the sample) we find $f_{\rm lim,eFEDS} \sim 2 \cdot 10^{-14}\,{\rm erg/cm^2/s}$, slightly deeper than that of eRASS1.
 We, thus, conclude that the eFEDS and eRASS1 detection statistics are qualitatively consistent with each other. A more detailed comparison is, however, prohibited by the low number statistics in the eFEDS fields.
 
\begin{figure*}[t]
 	\parbox{18cm}{
 	\parbox{9cm}{\includegraphics[width=0.5\textwidth]{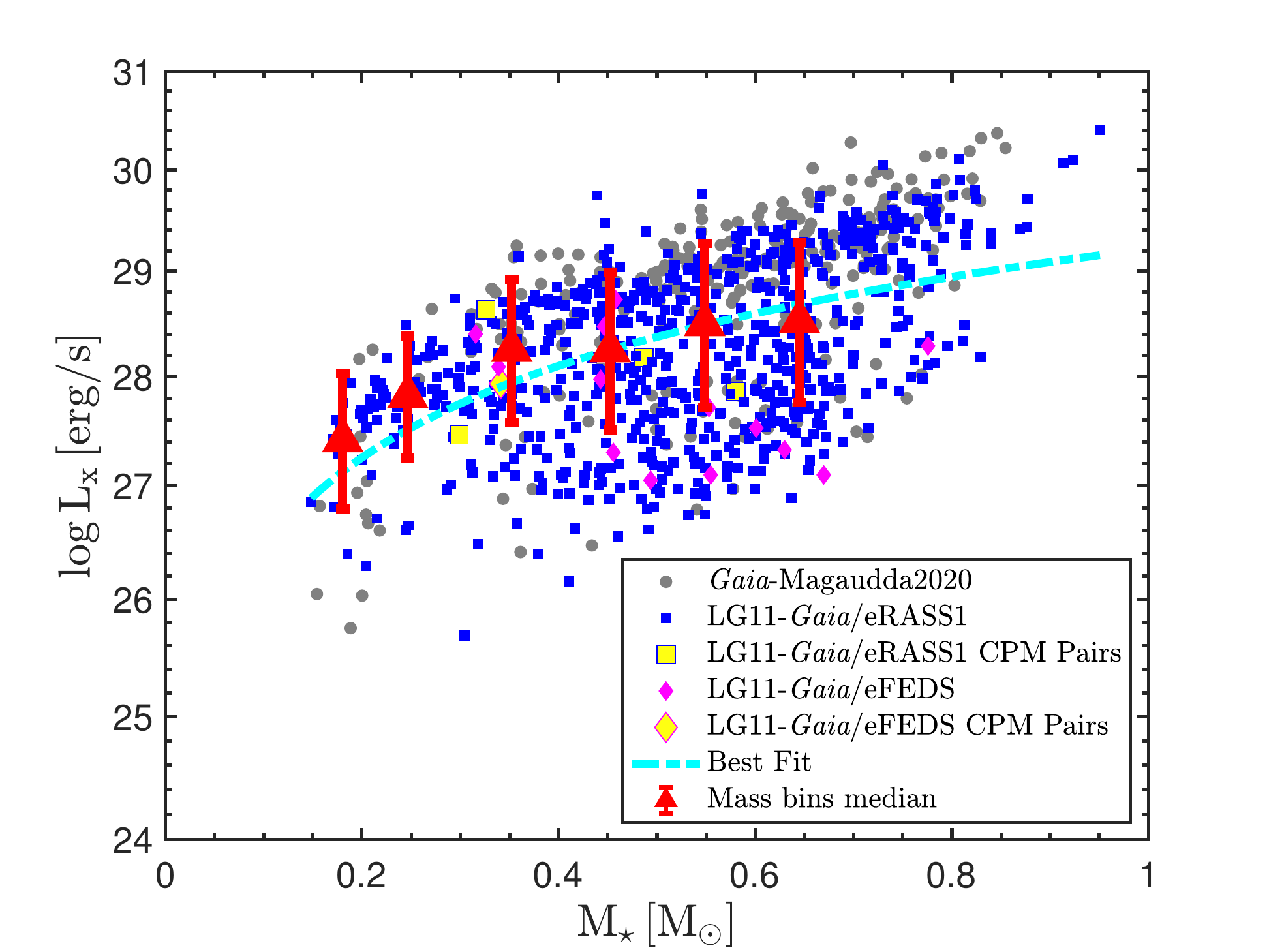}}
 	\parbox{9cm}{\includegraphics[width=0.5\textwidth]{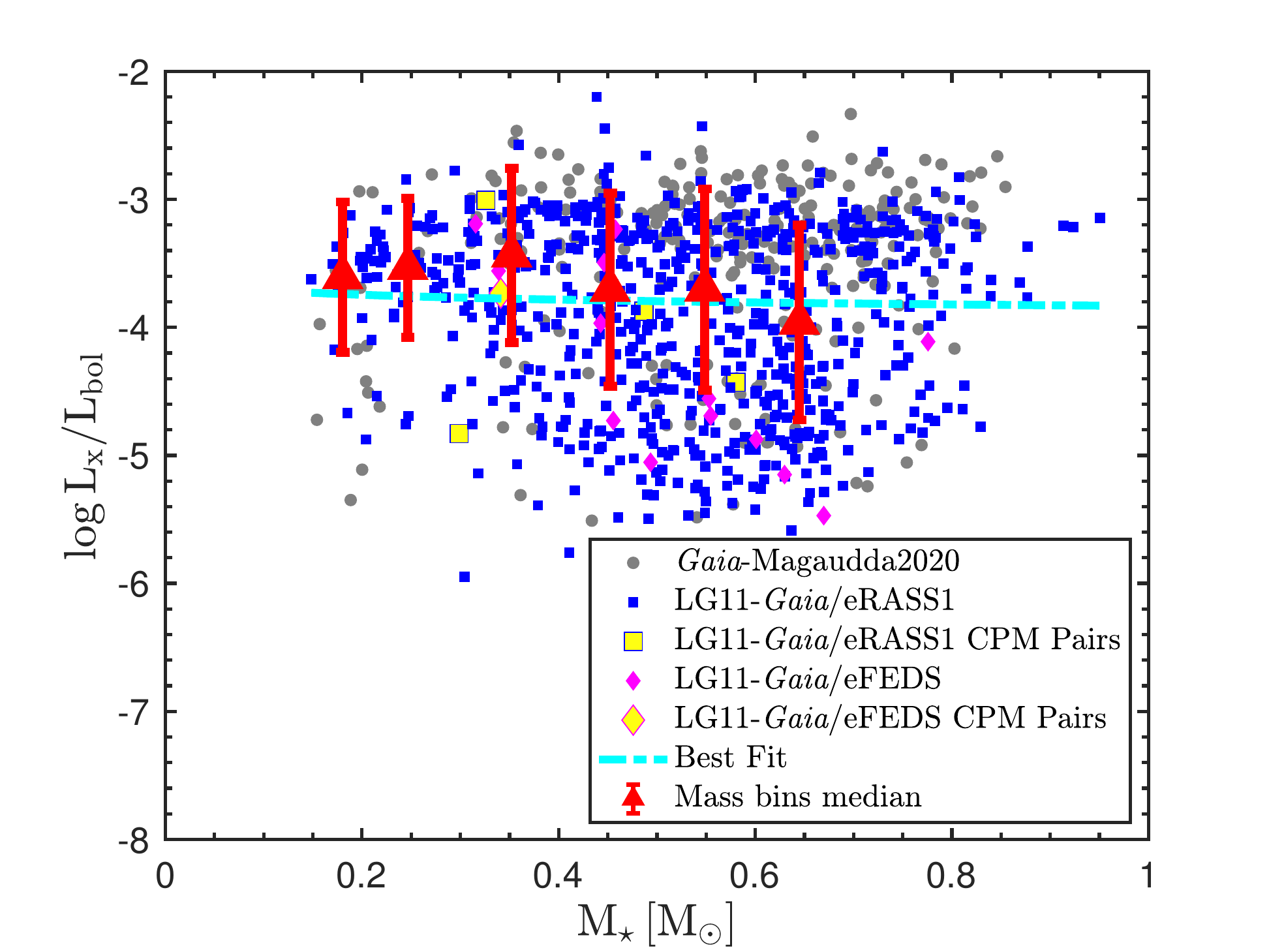}}	
 		}
 	\parbox{18cm}{	
 	\parbox{9cm}{\includegraphics[width=0.5\textwidth]{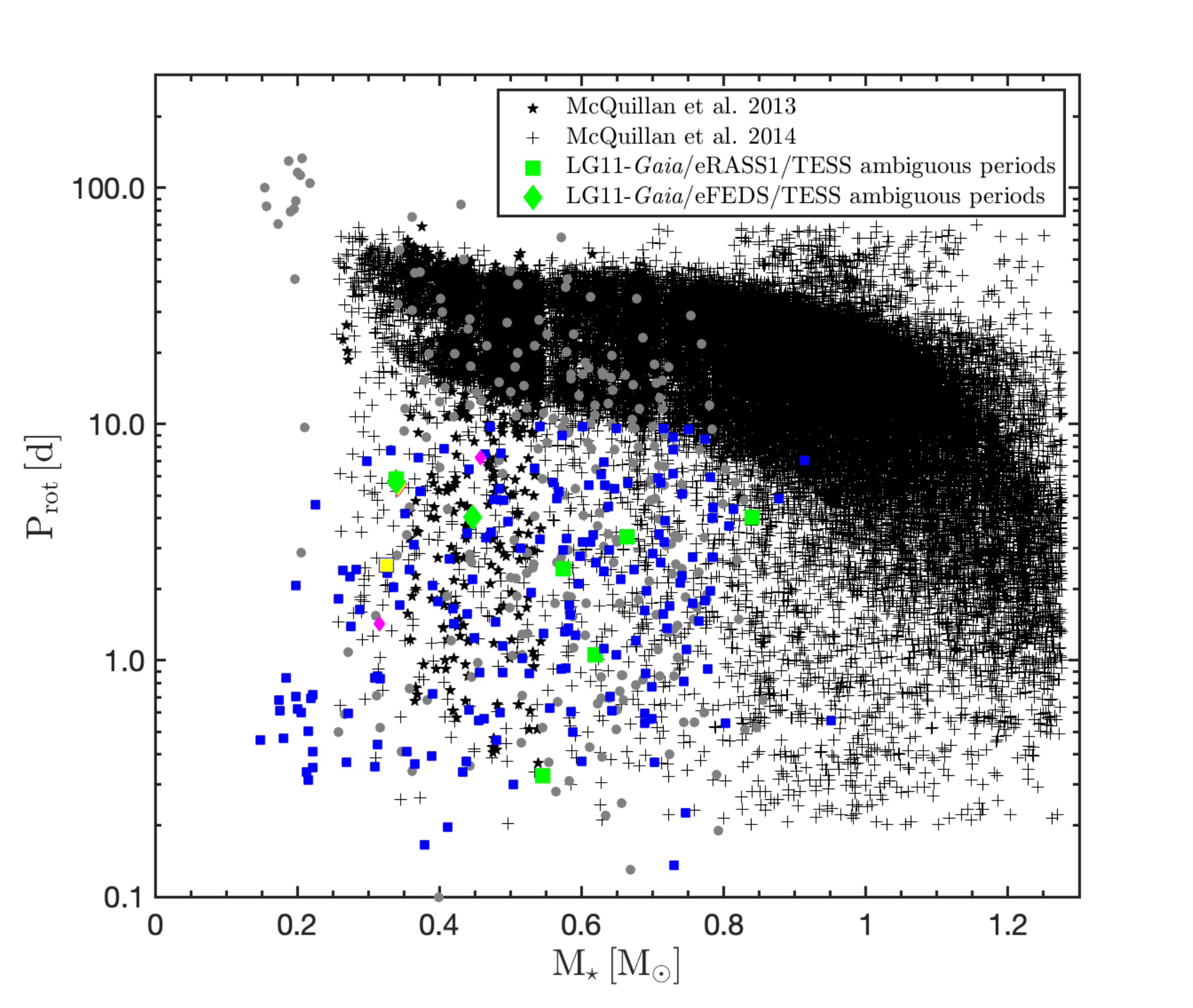}}
 	\parbox{9cm}{\caption{X-ray activity and rotation versus mass: blue and pink are the eRASS1 and eFEDS M dwarf samples, respectively, and grey is the revised sample from \citet{Magaudda2020}; see legend inside the panels for other literature data and highlighted specific subsamples. In the top panels the median and the standard deviation of the data are presented in red for bins of $0.1\,{\rm M_\odot}$ width. {\bf Top Left:} X-ray luminosity vs mass  and best fit (cyan) {\bf Top Right:} X-ray over bolometric luminosity vs mass and best fit (cyan). {\bf Bottom Left:}  Rotation period vs mass; for the stars with  double-humped {\em TESS} light curves the shorter of the two periods is shown.\label{fig:Lx_Mass}}}}
 \end{figure*}

\subsection{Mass-dependence of  activity and rotation}\label{subsect:results_mass_xrays}

Both activity and rotation are known to depend on stellar mass. While a detailed investigation of the rotation / mass relation has come into reach with the {\em Kepler} mission, statistical samples for X-ray / mass studies within a spectral subclass have not been available so far.
In Fig.~\ref{fig:Lx_Mass} we show X-ray activity diagnostics and rotation periods
versus stellar mass, i.e. $L_{\rm x}-M_{\star}$ (left panel), $L_{\rm x}/L_{\rm bol}-M_{\star}$ (right panel) and $P_{\rm rot}-M_{\star}$ (bottom panel) for M dwarfs. For the X-ray / mass relations we considered the LG11-{\em Gaia}/eFEDS and LG11-{\em Gaia}/eRASS1 samples, while for the 
$P_{\rm rot}-M_{\star}$ relation we show the X-ray detected M dwarfs with rotation period, that is the  
LG11-{\em Gaia}/eFEDS/TESS and LG11-{\em Gaia}/eRASS1/TESS samples and the stars with two periods (green symbols) that are not considered in our quantitative analysis. 
We include in all panels data from \citet{Magaudda2020} (in grey) which comprises new {\em XMM-Newton}, {\em Chandra}, and {\em K2} mission observations and a collection of results from the literature that we have updated in \cite{Magaudda2020}. 
The X-ray luminosities from \citet{Magaudda2020} were extracted in the {\em ROSAT} energy band ($0.1-2.4$\,keV). As explained in Sect.~\ref{subsubsect:analysis_efeds_xrays_spectra}, our use of a different energy band for the {\em eROSITA} data ($0.2-5.0$\,keV) does not bias the results.
To be consistent with the LG11-{\em Gaia}/{\em eROSITA} samples, we retrieved {\em Gaia} distances\footnote{In \cite{Magaudda2020} we had extracted distances from {\em Gaia-DR2} parallaxes \citep{Gaia2016,Gaia2018} and validated them using  \cite{Lindegren2018} and our own quality criteria.} 
from BJ18, and for Fig.~\ref{fig:Lx_Mass} and the  subsequent analysis in Sect.~\ref{subsect:results_rotact} we have cleaned the sample from \cite{Magaudda2020} removing all stars that lack {\em Gaia} photometry and/or distance and the $9$ upper limits presented in the original sample. 
For the calculation of stellar parameters we followed the recipes described in Sect.~\ref{sect:sample}. 
Moreover, we identified and removed $13$ stars with $M_{\rm G} < 5$\,mag. These updates are  motivated by a comparison of masses and spectral types analogous to the one carried out for the LG11 catalog in Sect.~\ref{sect:sample}. With these restrictions the sample from \citet[][henceforth referred to as {\it Gaia}-Magaudda2020]{Magaudda2020}  counts $259$ stars with $0.15 \leq M_{\star}/M_{\odot} \leq 0.85$. 

In Fig.~\ref{fig:Lx_Mass} we consider the full mass range obtained from the $M_{\rm Ks}$ values of the stars but strictly speaking masses above 0.7\,$M_\odot$ are not validated as those stars have $M_{\rm Ks}$ values outside the calibrated range of the relation from \cite{Mann2015}. We decided on showing the full mass range as there is no obvious qualitative change of the $L_{\rm x} - M_{\star}$ relation at the above-mentioned boundary, and we take this as a  justification for the extrapolation of the underlying $M_{\star} - M_{\rm Ks}$ relation.

Fig.~\ref{fig:Lx_Mass} shows that for a given stellar mass we observe a $2-3$ orders of magnitude spread of the X-ray activity level, except for the lowest ($M_{\star} \lesssim 0.3\,M_{\odot}$) and highest masses ($M_{\star} \gtrsim 0.7\,M_{\odot}$) where the spread is smaller. 
At the low-mass end clearly only the upper part of the $L_{\rm x}$ values are detectable in our flux-limited {\em eROSITA} observations,
while the high-mass end corresponds to the transition to spectral type K which is not 
fully sampled in the LG11 catalog. 
We calculated the median and the standard deviation in mass bins of $\Delta M_{\star} = 0.1\,{\rm M_\odot}$ for the combined M dwarf sample from the literature and from this {\em eROSITA} study. 
The resulting values are overlaid on the data in red for the `validated' mass range. In the intermediate mass range which is best sampled by our catalog the standard deviation is $\sim 0.74$ for $L_{\rm x}$ and $\sim 0.75$ for $L_{\rm x}/L_{\rm bol}$, in  logarithmic scale. 
We have fitted the two X-ray / mass relations for the combined 
literature and {\em eROSITA} sample with a linear function in log-log space, finding a slope of  $+2.81 \pm 0.25$. The best fit relations are given in Table~\ref{table:ActMass_FitRes}, and we display the fits in Fig.~\ref{fig:Lx_Mass}. 
Interestingly, a  historical study of M dwarfs by \cite{Fleming88.0} on a very limited sample ($\sim30$ field M stars with $0.15\leq M_{\star}/M_{\odot}\leq 0.6$) came up with 
a similar slope to ours for the $L_{\rm x} - M_\star$ relation within the uncertainty. The numbers for the average X-ray activity level across all masses in the M dwarf regime (cols.~2 and~4 in Table\,\ref{table:ActMass_FitRes}) must be considered an upper limit given that even in a volume of $10$\,pc around the Sun $\sim 40$\,\% of M dwarfs are still undetected in X-rays \citep{Stelzer2013}. We defer a more detailed discussion and comparison to  literature studies to Sect.~\ref{subsect:results_rotact}.

In the bottom panel of Fig.~\ref{fig:Lx_Mass} we inspect the $P_{\rm rot}-M_{\star}$ relation 
of our samples in comparison to data from \cite{McQuillan2013}, that cover the mass range of $0.3-0.55\,M_{\odot}$ with selection criteria based on $T_{\rm eff}$ and $\log g$ values from the {\em Kepler} input catalogue, and \cite{Mcquillan2014} which is the extension of that study to all stars with $T_{\rm eff} < 6500$\,K. 
The rotation periods of the sample 
from \cite{Magaudda2020} were extracted from light curves of the {\em K2} mission, the MEarth project and ground-based observations, and they cover a broad range of values from 0.1\,d to $\sim100$\,d. As explained by \cite{Mcquillan2014} and \cite{Stelzer2016}, the $P_{\rm rot}-M_{\star}$ relation shows a bimodal period distribution for lower masses, and an upper envelope of the period distribution that increases for decreasing masses.
With the data compiled by \cite{Magaudda2020} (grey in Fig.~\ref{fig:Lx_Mass}) we confirm the upwards trend for the longest periods and lowest-mass stars. 
The periods in our {\em eROSITA} and {\em TESS} samples are biased because {\em TESS} stares at a given field for only about a month, and we consider periods longer than about half the duration of a sector light curve not reliable. 
The {\em eROSITA/TESS} sample of LG11-{\em Gaia} stars is, therefore, located in the range of fast rotators which is a sparsely populated region in unbiased surveys for stellar rotation periods. Interestingly, {\em TESS} covers this  regime entirely, i.e. up to the transition (at $P_{\rm rot} \sim 10$\,d) where the bulk of the M dwarfs are situated.  
With eRASS1 we have added to 
the $P_{\rm rot} - M_\star$ relation some very low-mass stars with fast rotation, showing that the lowest mass stars span the largest range of periods,  and that the vast majority of rotation rates in between the extremes have not yet been covered by X-ray observations for this mass range.

\begin{figure*}[t]
	\begin{multicols}{2}
	\includegraphics[width=0.5\textwidth]{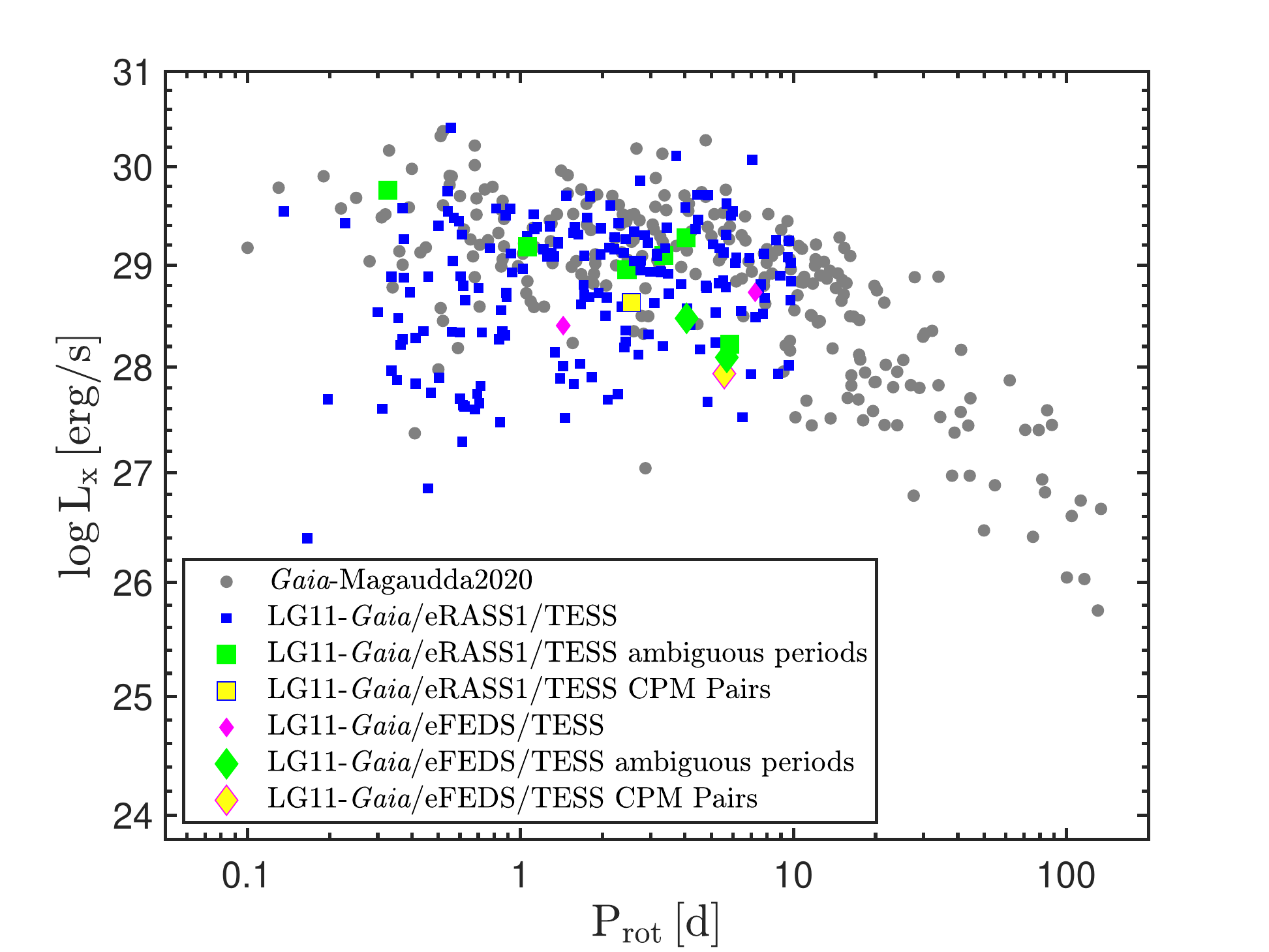}\par	\includegraphics[width=0.5\textwidth]{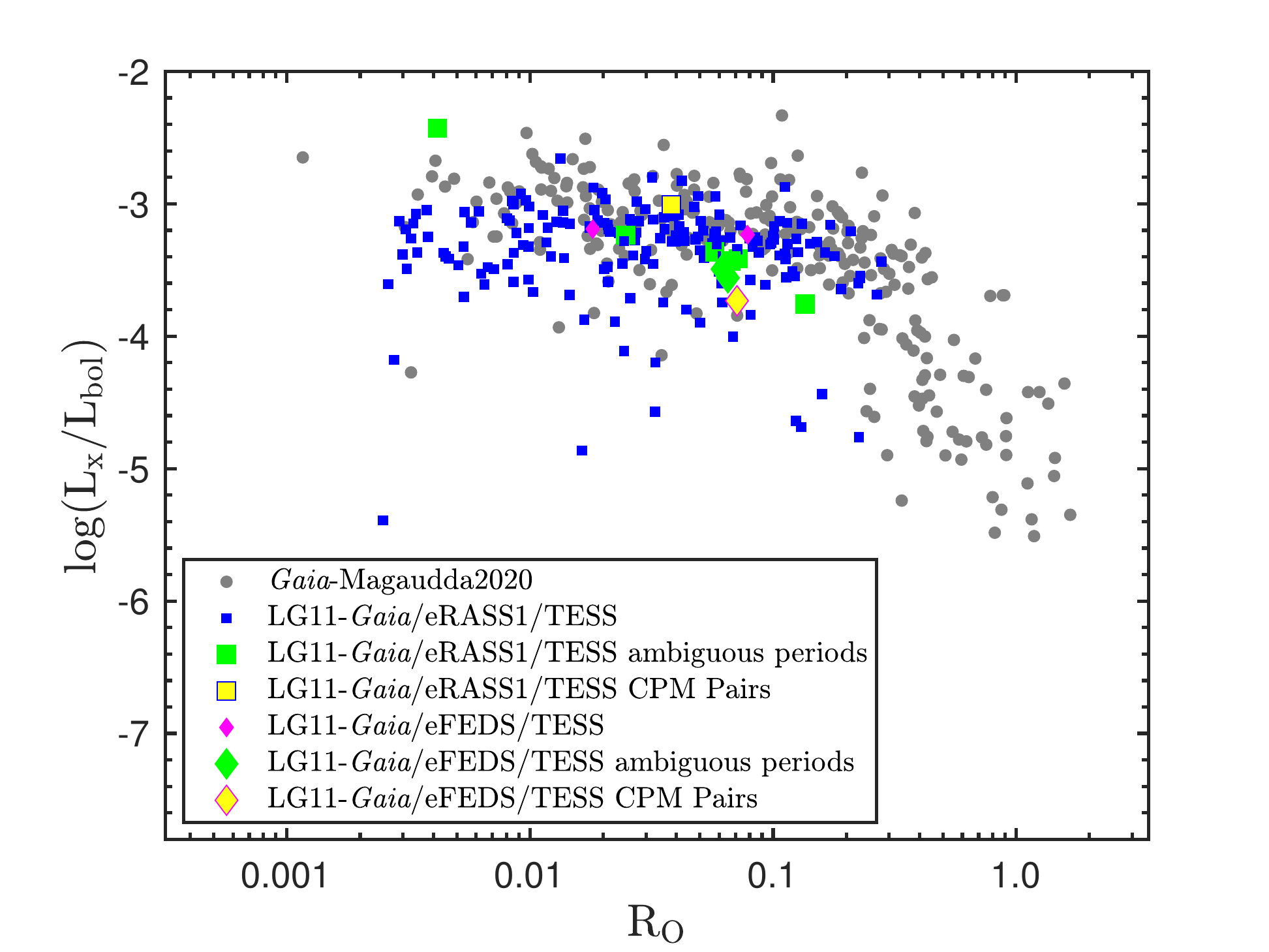}\par 
	\end{multicols}
	\caption{ X-ray activity-rotation relations \textbf{Left}: X-ray luminosity versus rotation period. 
	\textbf{Right}: X-ray luminosity as fraction of  bolometric luminosity 	vs. Rossby number. See legend, caption of Fig.~\ref{fig:Lx_Mass} and text in Sect.~\ref{subsect:results_rotact} for the subsamples at display.}
	\label{LxProt}
\end{figure*}

\subsection{Rotation-activity  relation}\label{subsect:results_rotact}

\begin{table*}
    \caption{Results obtained from the fitting of $L_{\rm x}-M_{\star}$ and $L_{\rm}/L_{\rm bol}-M_{\star}$ for the full sample (LG11-{\em Gaia}-eFEDS, LG11-{\em Gaia}-eRASS1 and {\em Gaia}-Magaudda20) and for the saturated sub-samples, i.e. stars with $P_{\rm rot}\leq P_{\rm sat}$; see Sect.~\ref{subsect:results_mass_xrays}\&\ref{subsect:results_rotact} for details. We show the slopes for the $L_{\rm x}-M_{\star}$ and $L_{\rm x}/L_{\rm bol}-M_{\star}$ relations (cols.~2 \& 4), and in cols.~3 \& 5 the $L_{\rm x}-$, $L_{\rm x}/L_{\rm bol}-$level referred to the mass distribution center ($M_{\star} = 0.5\,M_{\odot}$).}
    \begin{center}
        \begin{tabular}{ccccc}
            \midrule[0.5mm]                    
            \multicolumn{1}{c}{Sample} 
            &\multicolumn{1}{c}{\textbf{$\beta_{\rm L_{x}-M_{\star}}$}}  
            &\multicolumn{1}{c}{\textbf{$\log \left(L_{\rm x}\right)~(M_{\star}=0.5\,M_{\odot})$}}
            &\multicolumn{1}{c}{\textbf{$\beta_{\rm L_{x}/L_{bol}-M_{\star}}$}}  
            &\multicolumn{1}{c}{\textbf{$\log \left(\frac{L_{\rm x}}{L_{bol}}\right)~(M_{\star}=0.5\,M_{\odot})$}}\\
                    
            \multicolumn{1}{c}{}  
            &\multicolumn{1}{c}{} 
            &\multicolumn{1}{c}{\textbf{[erg/s]}}
            &\multicolumn{1}{c}{}
            &\multicolumn{1}{c}{}\\
            \midrule[0.5mm]
             Full&$+2.81\pm$0.25&28.37$\pm$0.07&$-0.12\pm$0.10&$-3.79\pm$0.03\\
             Saturated&$+3.14\pm$0.26&28.89$\pm$0.08&$+0.41\pm$0.26&$-3.24\pm$0.08\\
             \bottomrule[0.5mm]
        \end{tabular}
         \label{table:ActMass_FitRes}
    \end{center}
\end{table*}

In the previous section we have show that with the new {\em eROSITA/TESS} sample we can study the regime up to $P_{\rm rot} \sim 10$\,d, which corresponds to the `saturated' regime in the rotation-activity relation. Here, we present the X-ray activity-rotation relation we constructed with the results obtained from {\em eROSITA} and {\em TESS} observations of the `validated' LG11-{\em Gaia} sample, combined with the stars of the sample from \cite{Magaudda2020} revised as described in Sect.~\ref{subsect:results_mass_xrays}
which we also restrict to the stars with $M_{\rm Ks}$ in the calibration range of \cite{Mann2015}.
This `validated' {\em Gaia}-Magaudda20 sample counts $197$ M dwarfs. The plots of X-ray activity versus rotation diagnostics are  shown in Fig.~\ref{LxProt} with the same color code as in Fig.~\ref{fig:Lx_Mass}. 

As described in Sects.~\ref{subsubsect:analysis_efeds_tess_output} and~\ref{subsect:analysis_erass1_TESS} the LG11-{\em Gaia}/eFEDS and LG11-{\em Gaia}/eRASS1 stars that have both {\em eROSITA} X-ray detection and {\em TESS} observations are in total $501$ out of $704$. The sub-sample with reliable {\em TESS} rotation period consist of $3$ and $138$ stars, for eFEDS and eRASS1  respectively. The new {\em eROSITA}/{\em TESS} data, thus, nearly double the sample that was previously available for studies of the X-ray activity-rotation relation.

Previous studies \citep{Pizzolato2003,Wright2011,Wright2016,Wright2018,Magaudda2020} revealed two different regimes of the rotation-activity relation, a saturated regime for fast-rotating stars with $P_{\rm rot} \leq 10$\,d and an unsaturated regime for slowly rotating stars with $P_{\rm rot} > 10$\,d. These two regimes are clearly present in Fig.~\ref{LxProt} but the new {\em eROSITA/TESS} samples cover only the saturated part. The relation was always thought to be better defined in the
$L_{\rm x}/L_{\rm bol} - R_{\rm O}$ space, i.e. the fractional X-ray luminosity vs the Rossby Number, $R_{\rm 0}$. The latter one is defined as the ratio between the rotation period and the convective turnover time ($\tau_{\rm conv}$), which is not a directly observable parameter. 
Here - as well as in \cite{Magaudda2020} - we adopted the empirical calibration of $\tau_{\rm conv}$ with
$V-K_{s}$ magnitude provided by \cite{Wright2018}.
The Rossby numbers obtained from the {\em TESS} rotation period values are provided in Table\,~\ref{table:act_rot}.
In the right panel of Fig.~\ref{LxProt} we can see that, in fact, switching from the $L_{\rm x}-P_{\rm rot}$ space to the $L_{\rm x}/L_{\rm bol}-R_{\rm O}$ space, the rotation-activity relation changes its structure.  
For the {\em eROSITA/TESS} samples the most relevant difference  between the two diagrams is the decrease of the vertical spread in the saturated regime when the X-ray luminosity is normalized by the bolometric luminosity. 

We can use the unprecedented statistics in the saturated regime to examine the mass dependence of the X-ray emission of fast rotators within the M spectral class. These represent the younger population of M dwarfs (age $<1$\,Gyr according to \citeauthor{Magaudda2020}) for partially convective stars, and up to $\sim 4$\,Gyr for stars with $M_\star < 0.4\,{\rm M_\odot}$). To this end we define a subsample of M dwarfs - combining our new data and the revised `validated' {\em Gaia}-Magaudda20 sample - which we limit to stars with detected rotation period that fulfill the criterion $P_{\rm rot} < P_{\rm sat} = 8.5$\,d. Here $P_{\rm sat}$ is the period at the transition from the saturated to the non-saturated regime. Our adopted value for $P_{\rm sat}$ is the one from \cite{Magaudda2020} in their full M dwarf sample. It can be seen from Fig.~\ref{LxProt} that with this choice we avoid to include non-saturated stars. 

The distributions of X-ray activity versus stellar mass for this fast rotator sample are shown in Fig.~\ref{fig:Lx_Mass_Psat8}. As in Fig.~\ref{fig:Lx_Mass} we also display the median and the standard deviation of the data in bins of $0.1\,{\rm M_\odot }$ width. The linear fit obtained for these stars (blue) yields a slope of $3.14\pm0.26$ 
for $L_{\rm x}$ and $0.41 \pm 0.26$ for $L_{\rm x}/L_{\rm bol}$ (see Table~\ref{table:ActMass_FitRes}). 
For comparison we have inserted in Fig.~\ref{fig:Lx_Mass_Psat8} the fits obtained in Sect.~\ref{subsect:results_mdwarfpopulation} for our full sample and the result of \cite{Preibisch2005} who performed the same type of linear fit but for the T\,Tauri stars in the Orion Nebular Cluster (ONC) with $M_{\star}<2\,M_{\odot}$.

\begin{figure*}[t]
 	\parbox{18cm}{
 	\parbox{9cm}{\includegraphics[width=0.5\textwidth]{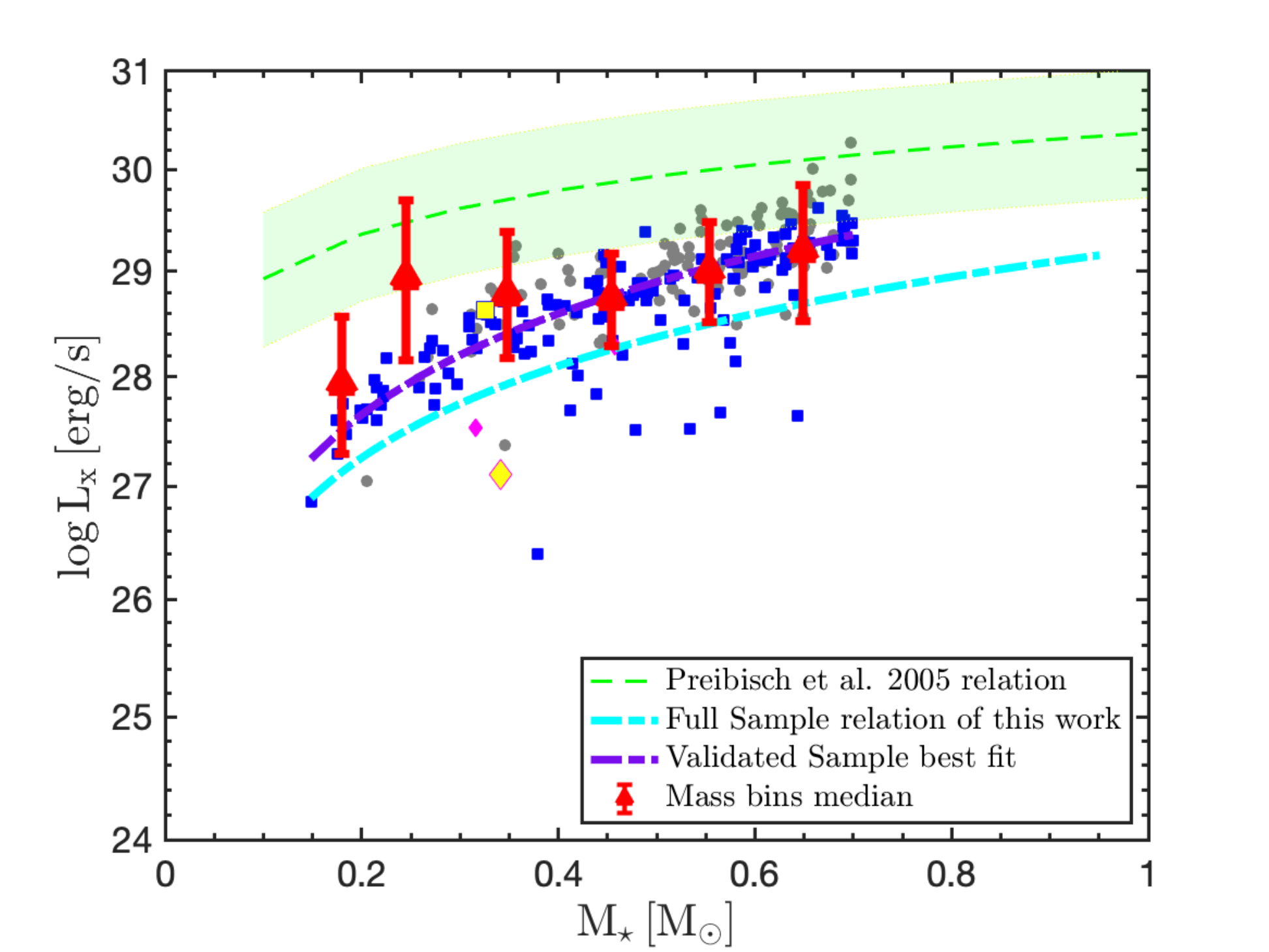}}
 	\parbox{9cm}{\includegraphics[width=0.5\textwidth]{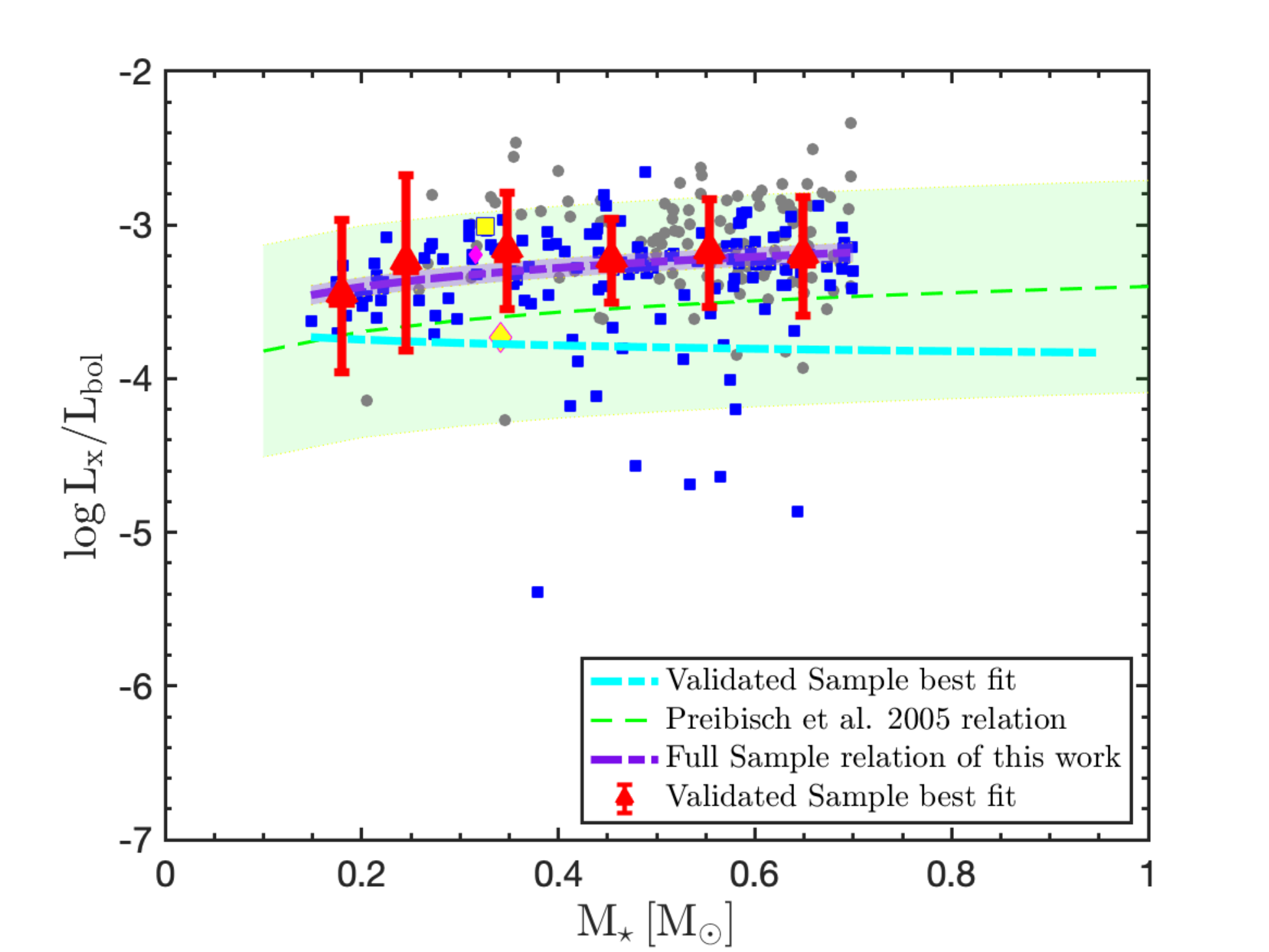}}
 		}
 	\caption{X-ray activity vs mass for the `validated' samples (see legend in Figs.~\ref{fig:Lx_Mass} and~\ref{LxProt}) restricted to the saturated stars, i.e. stars with $P_{\rm rot}\leq 8.5$\,d, best fit (violet)  and median plus standard deviation of the data (red) in bins of $0.1\,{\rm M_\odot}$. Also shown is the fit to the young stars in the Orion Nebular cluster provided by \cite{Preibisch+2005} together with  the standard deviation of this fit (green) and the best fit we found for the full M dwarf sample presented in the top panels of  Fig.~\ref{fig:Lx_Mass} (cyan). {\bf Left:} X-ray luminosity vs mass; {\bf Right:} X-ray over bolometric luminosity vs mass.}
	\label{fig:Lx_Mass_Psat8}
 \end{figure*}

First, we can observe that the exclusion of non-saturated stars barely changes the slope of the $L_{\rm x}-M_\star$ relation but it converts the marginally negative slope in $L_{\rm x}/L_{\rm bol}-M_\star$ space to a marginally positive one. This is due to the fact that, despite the dedicated studies of slowly rotating fully convective stars by \cite{Wright2016,Wright2018}, the non-saturated regime is dominated by the more massive M dwarfs which have shorter spin-down  timescales. In the center of our mass distribution
($\approx 0.5\,M_{\odot}$) the X-ray / mass relation for the saturated subsample is shifted by $\sim 0.5$\,dex to higher activity levels as compared to the full sample.
Since the full sample is likely still incomplete and it includes the saturated subsample this poses only a lower limit to the change of the X-ray emission level between $<1$\,Gyr- and several Gyr-old M dwarfs.  

The slope of $L_{\rm x} - M_\star$ we derived for our field M dwarfs
is significantly higher than the one for the ONC from \cite{Preibisch+2005} ($\beta_{\rm ONC} = 1.44 \pm 0.10$).
The X-ray luminosities of the Orion sample are  shifted upwards with respect to the field dwarfs because of the young age 
of the ONC. The decrease of the $L_{\rm x} - M_\star$ relation between the ONC and 
our saturated subsample encodes the evolution between $1$\,Myr and $\lesssim 1$\,Gyr, which is $\sim 2.0$\,dex in logarithmic space for the low-mass end ($\sim 0.15\,{\rm M_\odot}$) and $\sim 0.8$\,dex at the high-mass end ($0.7\,{\rm M_\odot}$).  
In terms of normalized X-ray luminosity both our M dwarf distributions are within the uncertainty of the $L_{\rm x}/L_{\rm bol} - M_\star$ relation of the ONC. Remarkably, our saturated sample - which spans the same range of periods as the ONC  \citep[e.g.][]{Choi96.0, RodriguezLedesma09.0} - is located in the upper half of the ONC distribution, i.e.  fast-rotating field M dwarfs have $L_{\rm x}/L_{\rm bol}$ levels at least as high as pre-main-sequence stars. As noted by \cite{Preibisch+2005} this apparently reduced activity level for the pre-main-sequence stars is due to the fact that the ONC sample includes accreting stars which have lower X-ray luminosities than non-accretors. 

\subsection{The coronal temperature - luminosity relation}\label{subsect:results_tx_lx}

In Fig.~\ref{fig:Tx_logLx} we show the eFEDS M dwarfs that have X-ray temperature and luminosity determined from the spectral analysis (see Table~\ref{tab:xspec_output}) in a scatter plot. 
As we explained in \ref{subsect:analysis_efeds_xrays}, we exclude the two stars with $\rm d.o.f. \leq 5$ because of the poor statistics of their spectra.
For comparison we show also the results from \cite{Johnstone15.0} for a sample of GKM stars collected by these authors from the literature, and the recent {\em XMM-Newton} measurement for the planet host star GJ\,357 from \cite{ModirroustaGalian20.0}. This latter one is a  representative of the faintest and coolest M dwarf coronae studied so far.

\begin{figure}[t]
	\begin{center}
	\includegraphics[width=9.0cm]{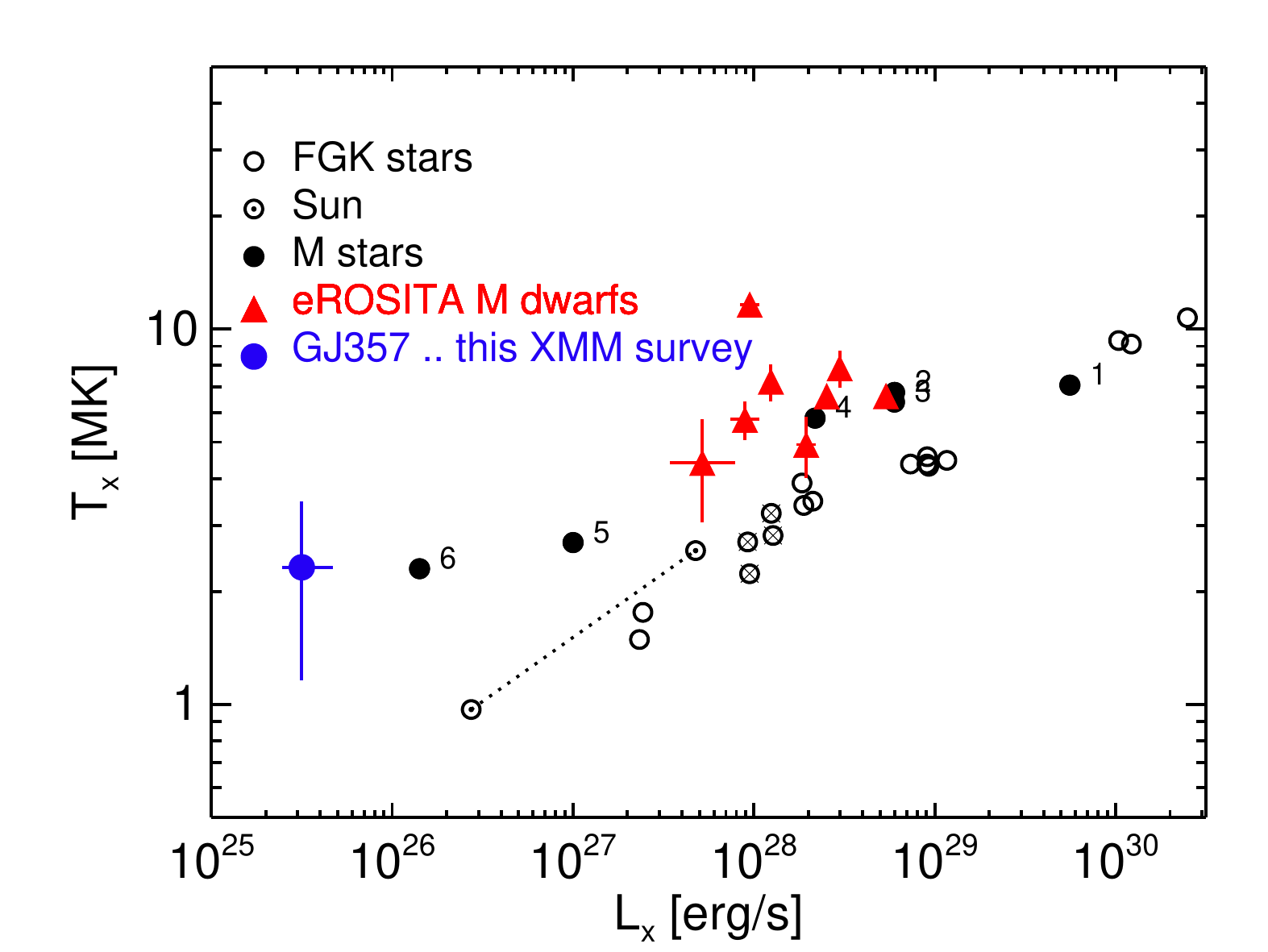}
	\caption{Coronal temperature vs X-ray luminosity for M dwarfs with {\em eROSITA}  spectra (red circles), compared to FGKM stars from \cite{Johnstone15.0} (black circles) for which according to our updated analysis there are six M dwarfs (filled  circles and individually labelled: 1 - AU\,Mic, 2 - EV\,Lac, 3 - AD\,Leo, 4 - YZ\,CMi, 5 - Prox\, Cen,
and 6 - SCR\,1845).
Stars labeled low-mass ($M_\star \leq 0.65\,{\rm M_\odot}$) in \cite{Johnstone15.0} but which have SpT K are marked with a cross symbol. At the faint and cool end is the planet hosting M dwarf GJ\,357 from \cite{ModirroustaGalian20.0} (blue circle).}
	\label{fig:Tx_logLx}
	\end{center}
\end{figure}

The stars from the \cite{Johnstone15.0} sample are among the most well-known dwarf stars in the solar neighborhood, and for some of them the stellar parameters have been determined very precisely in dedicated studies. However, for the sake of homogeneity we selected the M dwarfs from their sample with the same procedure, explained in Sect.~\ref{sect:sample}, that we applied to the full LG11 catalog. 
Specifically, we computed the SpTs  from $G_{\rm BP} - G_{\rm RP}$ (see Sect.~\ref{sect:sample} and footnote~\ref{note1}).
This provided six M dwarfs, the same ones for which SIMBAD gives an M spectral type. 
We note in passing that all of them except for SCR J1845-6357 (hence-forth SCR1845) are within the validation range of the \cite{Mann2015} relations. 
According to the literature SCR J1845-6357 is a late-M dwarf \citep[SpT M8.5;][]{Robrade10.0}, and its $M_{\rm Ks}$ value is slightly larger than the upper boundary of the range calibrated by \cite{Mann2015}. The six stars we have selected have $M_\star \lesssim  0.7\,{\rm M_\odot}$ according to the \cite{Mann2015} relation. These stars are highlighted as filled, black circles in Fig.~\ref{fig:Tx_logLx}. 
Their X-ray properties are adopted from \cite{Johnstone15.0}, except for the X-ray luminosity of
Prox Cen where we use the value from \cite{Ribas16.0},  $\log{L_{\rm x}}\,{\rm [erg/s]} = 27$. 

The six M dwarfs from \cite{Johnstone15.0} alone delineate a rather well-defined correlation between $T_{\rm x}$ and $L_{\rm x}$, and GJ\,357 is roughly consistent with an extension of this relation at the faint and cool end. 
We note that \cite{Johnstone15.0} have distinguished stars in two mass bins, above and below  $M_\star = 0.65\,{\rm M_\odot}$. They determined the stellar masses from $B-V$ colors using the evolutionary models of \cite{An07.0}. This way they include four  more stars in their low-mass group with respect to the six we have selected. In Fig.~\ref{fig:Tx_logLx} these are found among the open circles where they are marked with a cross. These stars have SpT early-K and it can be seen that they are displaced downwards with respect to the M dwarfs. 

The {\em eROSITA} sample comprises a narrow range of X-ray luminosities limited by the sensitivity of the eFEDS  observation. However, contrary to the literature sample the stars from eFEDS display a significant spread in terms of $T_{\rm x}$. As explained in Sect.~\ref{subsubsect:analysis_efeds_xrays_spectra}, the X-ray luminosities of the eFEDS stars are insensitive to variations in the energy band within the typical range of {\em ROSAT} and {\em eROSITA} data.
The mean upwards shift of the eFEDS sample, therefore, is unlikely to be the result of an observational bias. Moreover, scatter is seen within the eFEDS sample itself which has been analyzed in a homogeneous way.  
We caution, however, that the star with the highest $T_{\rm x}$ value is the only one from the eFEDS sample that was analyzed with a $1$-T spectral model,  and it presents evidence for variability in its light curve (see App.~\ref{app:lcs}).

A larger data base of homogeneous  coronal temperature measurements for M dwarfs is needed to explore the $T_{\rm x} - L_{\rm x}$ relation and, in particular, its spread and a possible influence of flares.  
The eRASS represents a such valuable data base. While a detailed spectral analysis of hundreds of M dwarfs detected in eRASS will be presented in a later work, here we use hardness ratios (HR) as a proxy for coronal temperature.

\subsection{eROSITA hardness ratios}\label{subsect:results_hr}

 The analysis of hardness ratios does not involve the stellar parameters, thus, based on the argument put forth in Sect.~\ref{subsect:results_mdwarfpopulation} we consider 
 also the stars outside the validation range of the \cite{Mann2015} relations.

We used the three energy bands provided by the eRASS1 catalog,  the soft ($0.2-0.6$\,keV), medium ($0.6-2.3$\,keV) and hard ($2.3-5.0$\,keV) band,  to define hardness ratios as follows: 
 
\begin{eqnarray}
    HR_{\rm 1} = \frac{Rate_{m}-Rate_{s}}{Rate_{m}+Rate_{s}}\\
    HR_{\rm 2} = \frac{Rate_{h}-Rate_{m}}{Rate_{h}+Rate_{m}}
\end{eqnarray} 
where $HR_{\rm 1}$ is the count rate ratio between the medium and soft energy bands,
while $HR_{\rm 2}$ is calculated between the hard and the medium energy bands.
The two hardness ratios are given for the {\em eROSITA} detected stars in Table\,~\ref{table:act_rot}. 

In Fig.~\ref{fig:HR_eFEDSeRASS1} we show the scatter plot of $HR_{\rm 2}$ vs $HR_{\rm 1}$ for both the LG11-{\em Gaia}/eRASS1 
and the LG11-{\em Gaia}/eFEDS sample. 
It is evident that a large fraction of the stars ($\sim 61$\,\%) have no counts in the hard energy band ($HR_{\rm 2}=-1$), and most of them are clustered at $0.2 \leq HR_{\rm 1} \leq 0.6$ ($\sim$ 
$65\,\%$ of the total sample).
This range of $HR_{\rm 1}$ includes all but two stars for which we have analyzed the eFEDS spectra. 
We thus can conclude that moderately positive values of $HR_{\rm 1}$ are associated to soft plasma of $\lesssim 0.5$\,keV. In fact, according to Foster et al. (A\&A subm.) the above-mentioned range of $HR_{\rm 1}$ values corresponds to plasma of $0.2...0.35$\,keV. We caution, however, that their analysis is based on a $1$-T model while our sample shows that two  temperatures are required to adequately describe the {\em eROSITA} spectra of M dwarfs. Therefore, the calibration between $HR_{\rm 1}$ and $kT$ from Foster et al. (A\&A subm.) may not be applicable to our sample. 

Concerning $HR_{\rm 2}$, the curious objects are the ones that are not at the soft limit, i.e. those with $HR_{\rm 2} > -1$. 
These can be broadly distinguished in two groups: one being clustered near the $HR_{\rm 2}$ soft bound and intermediate $HR_{\rm 1}$ values ($\sim 25$\,\% of the whole sample) and the other scattered through the parameter space ($\sim 9$\,\%). 
The former ones are likely represented by M~dwarfs with a slightly hotter corona than the ones at $HR_{\rm 2} = -1$, while for the latter ones a plausible hypothesis for their harder spectra is flaring activity. The analysis of all individual eRASS1 light curves in a future study will allow us to further examine this interpretation.

 \begin{figure}[t]
	\begin{center}
		\includegraphics[width=9.5cm]{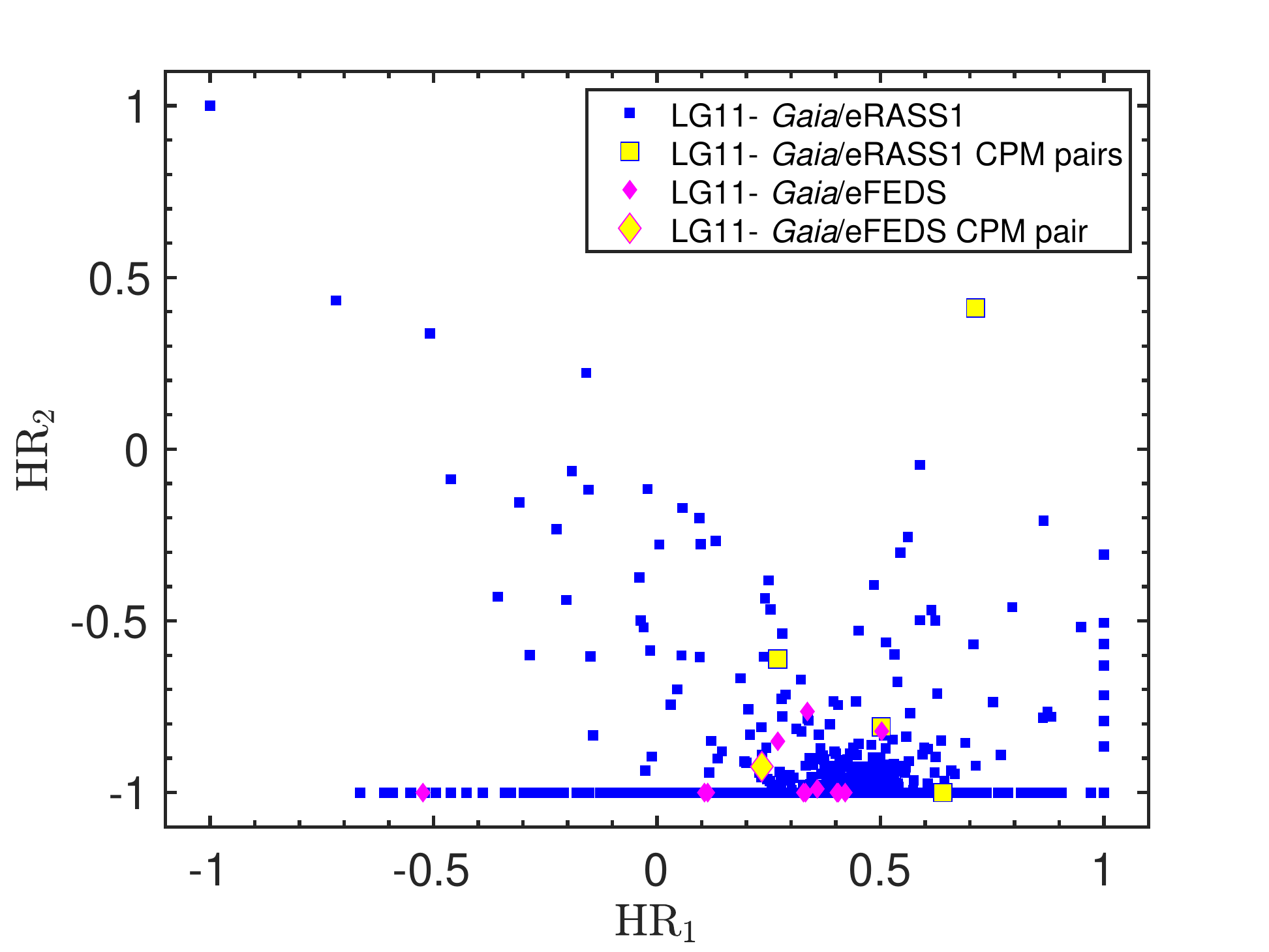}
		\caption{{\em eROSITA} hardness ratios of LG11-{\em Gaia} M dwarfs detected in eFEDS (pink filled diamonds) and in eRASS1 (blue filled squares); see text in Sect.~\ref{subsect:results_hr} for the definition of the hardness ratios. In yellow we show the CPM pairs in eFEDS and in eRASS1 samples.}
	\label{fig:HR_eFEDSeRASS1}
	\end{center}
\end{figure}

\subsection{X-ray variability}\label{subsect:results_var}

We have analyzed the {\em eROSITA} light curves of $14$ M dwarfs detected during eFEDS. While $12$ of them did not show any significant variation, two showed a likely flare: an event ongoing at the beginning of the observation for PM\,I09161+0153 and a smoother longer-lasting variability throughout the detection for PM\,I09201+0347. These  results are also confirmed by the eFEDS variability study performed by Boller et al. (A\&A subm.). We refer to Sect.~\ref{subsubsect:analysis_efeds_xrays_lcs} and App.~\ref{app:lcs} for more details.

An analysis of short-term variability in the much larger eRASS1 sample is beyond the scope of this work. However, we present here a comparison between {\em ROSAT}, eFEDS and eRASS1 X-ray luminosities, based on the LG11-{\em Gaia}/eFEDS and LG11-{\em Gaia}/eRASS1 samples. 
This study probes variability on  time-scales of years. 
While the {\em ROSAT} all-sky survey (RASS) predates the mean eRASS1 epoch by about $29$ years - the middle of the surveys were on 1st June 1991 and 10th March 2020, respectively - eFEDS and eRASS1 differ by a few months. eFEDS was carried out between Nov 4 and Nov 6 2019, and eRASS1 started on 13 Dec 2019 and lasted for 6 months.

\subsubsection{Comparison between eFEDS and eRASS1}\label{subsubsect:results_longtermvar_erosita}

We found that seven LG11-{\em Gaia} M dwarfs are detected with {\em eROSITA} during both eFEDS and eRASS1. We used the $L_{\rm x}$ values calculated in Sects.~\ref{subsect:analysis_efeds_xrays}  and~\ref{subsect:analysis_erass1_xrays} for eFEDS and eRASS1 stars, respectively. 

In Fig.~\ref{fig:eFEDS_eRASS1_comp} we show the comparison between the observed X-ray luminosities during the two different {\em eROSITA} epochs. 
We can see that two stars have exhibited more than a factor two higher $L_{\rm x}$ during eRASS1 as compared to eFEDS.
From the eFEDS light curves in Fig.~\ref{fig:eFEDS_lcs} we know that PM\,I09201+0347 was detected during a flare. Therefore,  
the fact that the star was on average brighter during eRASS1 indicates that likely another, brighter flare occurred during the all-sky survey.
A flare during eRASS1 is also a probable explanation for the change in X-ray luminosity of the other star, PM\,I08551+0132. In a future work we will examine in detail the X-ray variability of M dwarfs during eRASS.

\begin{figure}[t]
	\begin{center}
		\includegraphics[width=0.5\textwidth]{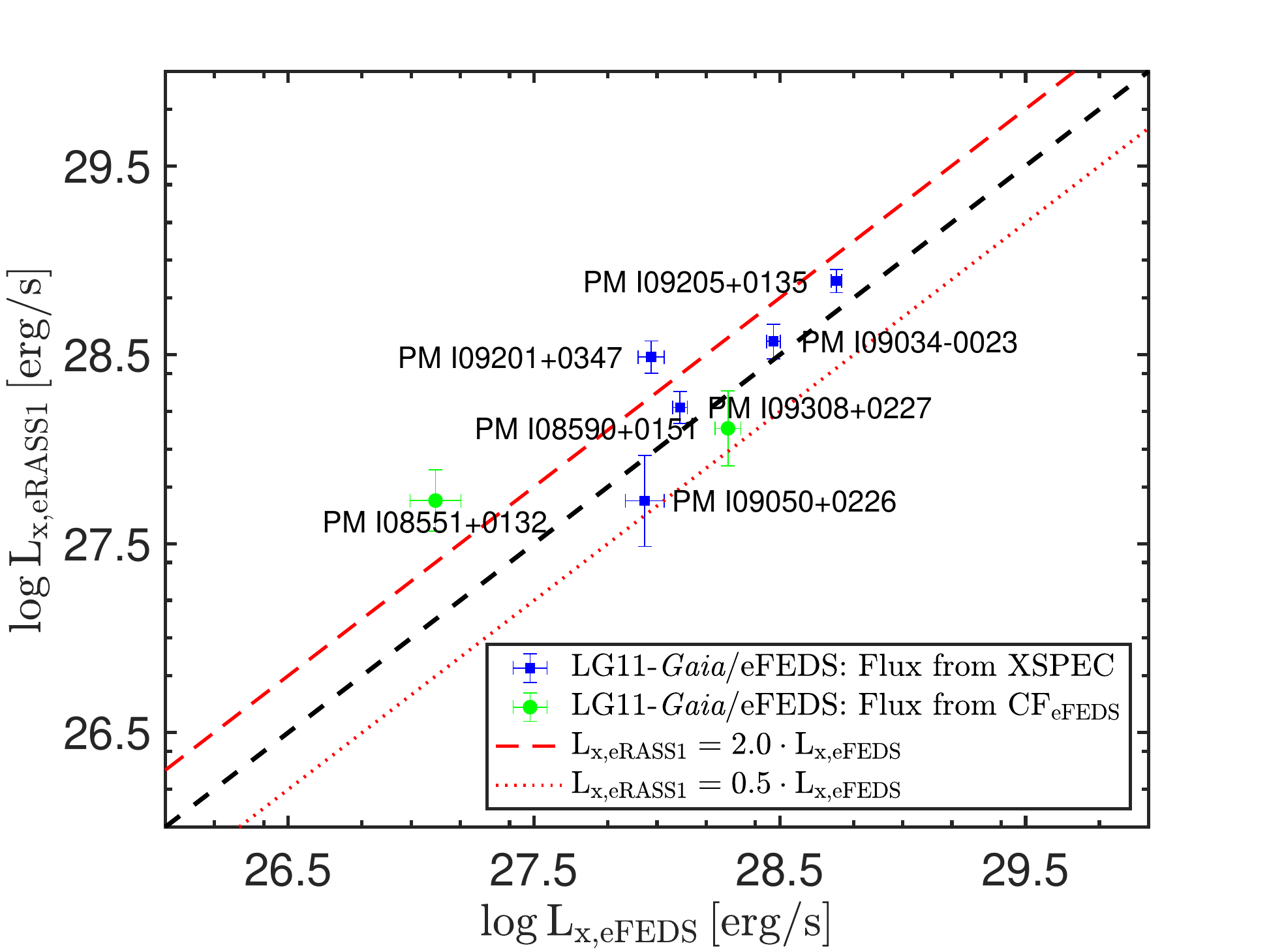}
		\caption{Comparison between X-ray luminosities observed during eFEDS ($\log{L_{\rm x, eFEDS}}$) and eRASS1 ($\log{L_{\rm x, eRASS1}}$). The two stars above the red dashed line have $L_{\rm x,eRASS1}$ more than twice as high as $L_{\rm x,eFEDS}$.}
	\label{fig:eFEDS_eRASS1_comp}
	\end{center}
\end{figure}

\subsubsection{Comparison between {\em eROSITA} and {\em ROSAT}}\label{subsubsect:results_longtermvar_rosat}

\begin{figure}[t]
 	\parbox{18cm}{
 	\parbox{9cm}{\includegraphics[width=0.5\textwidth]{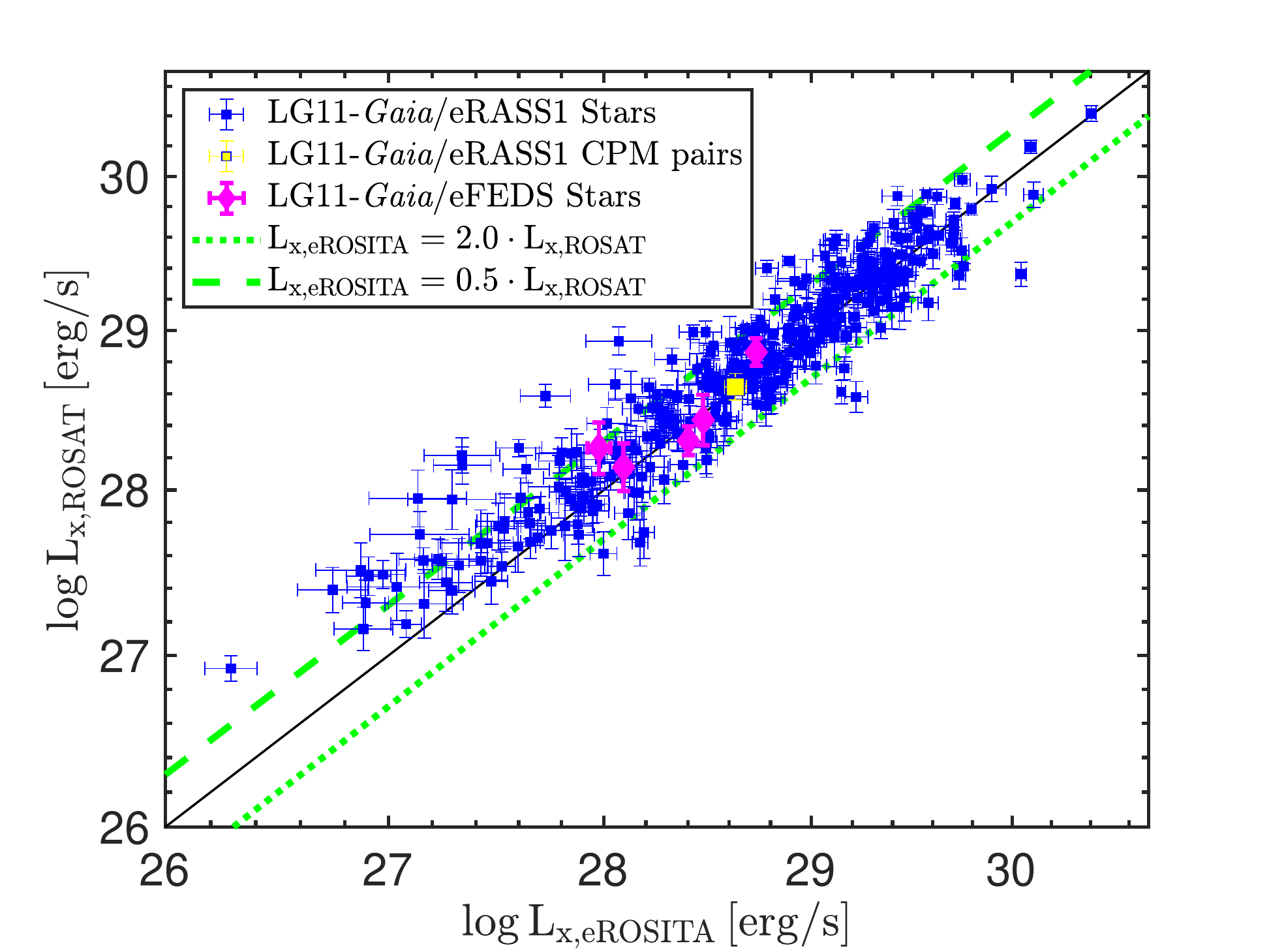}}
 	}
 	\parbox{18cm}{
 	\parbox{9cm}{\includegraphics[width=0.5\textwidth]{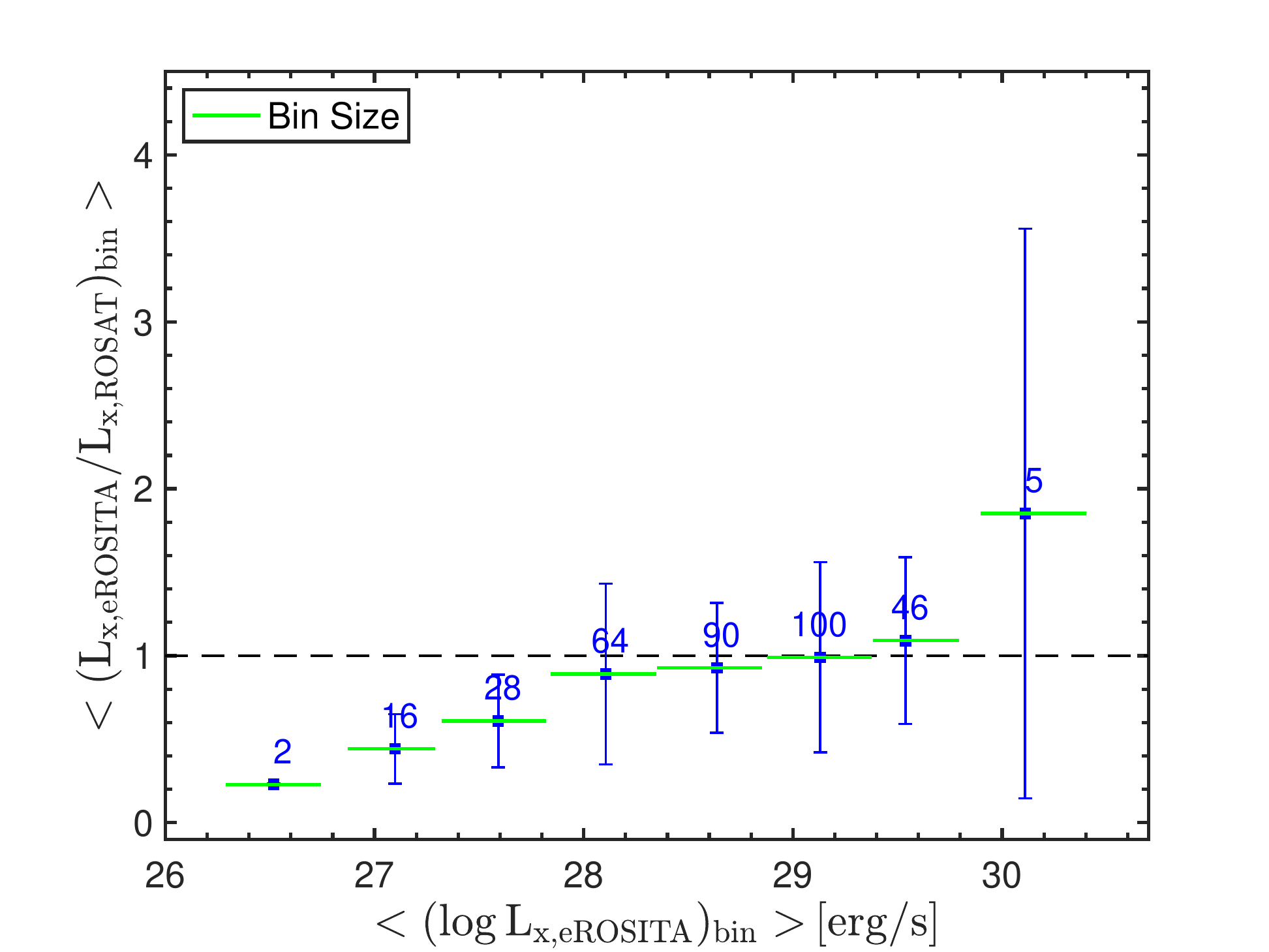}}
 		}
 	\caption{Comparison between the X-ray luminosities from {\em eROSITA} (eRASS1 and eFEDS) and those from {\em ROSAT} (RASS) for the LG11-{\em Gaia} sample.   
 	{\bf Top panel} - scatter plot; CPM pairs are highlighted in yellow. {\bf Bottom panel} - Ratio between the {\em eROSITA} and {\em ROSAT} luminosities in logarithmic bins of width $0.5$ in X-ray luminosity, quantifying the systematic trend. Labels on top of the data points represent the number of stars in each bin, the vertical bars are the standard deviations and the horizontal bars denote the bin width.}
		\label{eROSITA_Rosat}
 \end{figure}

For the comparison on longer (decades) time-scales we cross matched the LG11-{\em Gaia}/eFEDS and LG11-{\em Gaia}/eRASS1 samples with the {\em ROSAT} catalogs  after having corrected their {\em Gaia}-DR2 coordinates with the proper motions to the mean {\em RASS} epoch (June 1991).
Following our conservative approach we use a search radius of $\rm 30\,^{\prime\prime}$, somewhat smaller than the values applied in other works on RASS data \cite[e.g.][]{Neuhaeuser95.0}. 
We found five stars from the LG11-{\em Gaia}/eFEDS sample and $346$ from the LG11-{\em Gaia}/eRASS1 sample in the {\em Second ROSAT All-Sky Survey Point Source Catalog} \cite[2RXS;][]{Boller16.0}.
We converted the count rates listed in the {\em ROSAT} catalogs into flux by adopting the {\em ROSAT} conversion factor from \citet{Magaudda2020} ($CF = 5.77 \cdot 10^{-12}\,{\rm erg/cm^2/cnt}$). From the fluxes and distances we then computed the X-ray luminosities observed with {\em ROSAT} ($\log L_{\rm x,ROSAT}$), which refer to the $0.1-2.4$\,keV energy band.
These values are presented in Table\,~\ref{table:act_rot}.

In Fig.~\ref{eROSITA_Rosat} we show the comparison between $L_{\rm x,eROSITA}$ and $L_{\rm x,ROSAT}$. We note again that this comparison is not affected by the different energy bands used for data from the two instruments. 
We find that $12$ stars have $L_{\rm x,eROSITA} > 2 \cdot L_{\rm x,ROSAT}$ and $55$ stars have $L_{\rm x,eROSITA} < 0.5 \cdot L_{\rm x,ROSAT}$. All these variations are significant, i.e. the error bars do not reach to the 1:1 line. It strikes that the X-ray luminosities in Fig.~\ref{eROSITA_Rosat} are distributed symmetrically around the 1:1 line for $\log{L_{\rm x}}\,{\rm [erg/s]} \geq 28$, but the distribution shows an upward curvature for lower luminosities. {\em eROSITA} has higher sensitivity than the RASS\footnote{The flux limit of eRASS1 is a factor $2-10$ fainter than that of the 2\,RXS catalog; see values given by \cite{Boller16.0} and \cite{Predehl2021}.}.
Therefore, the stars at the {\em eROSITA} detection limit are expected to be undetected during RASS, unless they were in a higher activity state (e.g. a flare) during the {\em ROSAT} survey. This is a likely explanation for the observation that at the faint end the {\em ROSAT} luminosities are higher than the ones measured by {\em eROSITA}.
 
In order to quantify the systematic deviation between the brightness seen in {\em eROSITA} and RASS for low-luminosity sources, we grouped $\log(L_{\rm x,eROSITA})$ with a bin size of 0.5 and calculated for each bin the mean of the ratio between the luminosities observed with {\em eROSITA}  
 and with {\em ROSAT}, $ {\langle L_{\rm x,eROSITA}/L_{\rm x,ROSAT} \rangle}_{\rm bin}$. In the bottom panel of Fig.~\ref{eROSITA_Rosat} we show the result of this statistical test, indicating the bin width (horizontal green lines), the number of stars in each bin (label on top of each data point) and the standard deviation of the mean of  
 $L_{\rm x,eROSITA}/L_{\rm x,ROSAT}$ in each bin as error bars. 
 This ratio gets increasingly smaller for lower $L_{\rm x}$, and it is significantly lower than one at the faint end. This systematic trend supports our above interpretation as an effect of sensitivity limits combined with intrinsic source variability. 

\section{Conclusions \& outlook}\label{sect:conclusions}

We have presented a comprehensive study of the X-ray activity of M dwarfs and its relation with stellar rotation.
The careful match of our input target list, the proper motion catalog by \cite{L_pine_2011}, with {\em Gaia}\,DR2 data provides more reliable stellar parameters for these stars than the values used in previous studies on the same argument. 
Among the $\sim 8300$ nearby M dwarfs from this LG11-{\em Gaia} catalog $\sim 8$\,\% have an X-ray detection in the first {\em eROSITA} All-Sky survey, and this X-ray sample is skewed to nearby stars (with a peak in their distance distribution at $\sim 20$\,pc). 
A subset of only  about $1/4$ of the {\em eROSITA}-detected M dwarfs have detectable rotation periods in {\em TESS} light curves.
We can, therefore, state that {\em eROSITA} X-ray measurements are a  much more  sensitive diagnostic for magnetic activity than star spot amplitudes measured with {\em TESS}. 

From an {\em eROSITA} survey of the CalPV phase (the so-called eFEDS fields) we  derived the first  coronal luminosities and temperatures for M dwarfs obtained from {\em eROSITA} X-ray spectra. The resulting rate-to-flux conversion factor was the basis for the X-ray luminosities we  determined for the faint majority of our sample stars,  and the same $CF$ can be used in future {\em eROSITA} studies of faint M dwarfs.

We have examined the mass-dependence of M dwarf X-ray activity on an unprecedentedly large sample, and we have quantified its slope ($L_{\rm x} \sim M_\star^\beta$,  $\beta=2.81 \pm 0.25$), which is considerably steeper than the slope
measured for the pre-main-sequence stars from the ONC study of \cite{Preibisch+2005} ($\beta_{\rm ONC} = 1.4$), and offset towards lower luminosities by a factor that  depends on the mass, i.e. $2.0$ logarithmic dex for $M_\star = 0.15\,{\rm M_\odot}$ and $0.8$\,dex for $M_\star = 0.65\,{\rm M_\odot}$.  
For a given mass our `validated' M dwarfs display a spread in $L_{\rm x}$ of $0.6$  decades. The true scatter is likely significantly larger because of the incompleteness of eRASS1 and eFEDS related to the flux limit discussed above.

The most obvious candidate for explaining the broad range of X-ray activity levels for a given M dwarf mass is a distribution of rotation periods (and ensuing dynamo efficiency). Our X-ray selected sample in the `validated' M dwarf mass range   presents periods between $0.10$ and $8.35$\,d, where the upper boundary is mostly produced by the duration of the {\em TESS} campaigns ($\sim 27$\,d). The possible influence of the  flux-limit of {\em eROSITA} on the $P_{\rm rot}$ distribution of the sample can be examined in a direct study of the activity-rotation relation. We found that in the saturated regime, which by coincidence reaches roughly up to our period boundary of $\sim 10$\,d, the {\em eROSITA} data comprises some stars with very faint X-ray emission ($\log{L_{\rm x}/L_{\rm bol}} \approx -4...-5$). These are downward outliers in the saturated part of the rotation-activity relation, and their origin needs further investigation. 
On the basis of their $\log{(L_{\rm x}/L_{\rm bol})}$  values we might conjecture that eRASS can reach into the bulk of the stars in the non-saturated regime of slow rotators, but that facilities capable of providing longer periods must be used to detect their corresponding rotation signal.

With respect to {\em ROSAT} even the first all-sky survey of {\em eROSITA} reaches down to lower X-ray activity levels, as we have shown in our direct comparison of RASS and eRASS1 luminosities for the $\sim 350$ M dwarfs that are detected in both surveys, 
and which make about half of the eRASS1 sample. The majority of the other half, not detected in RASS, are near the eRASS1 detection limit (with ML\_CTS\_0$\lesssim 20$).
Adding in data from the other seven {\em eROSITA} surveys will in the near future provide access to X-ray detections of more and fainter M dwarfs. 
 
Similarly, significant quantitative  progress can be expected in our understanding of the coronal temperature distribution on M dwarfs, for which so far no statistical samples have been available. Our results from the spectral analysis of about a dozen X-ray bright M dwarfs in eFEDS combined with the {\em eROSITA} hardness ratios for the full (but mostly faint) X-ray detected  sample has shown that at least $2/3$ of the {\em eROSITA}-detected M dwarfs have typical coronal temperatures of $\sim 0.5$\,keV (corresponding to $\sim 6$\,MK). 
Our analysis of {\em eROSITA}/eFEDS  spectra also shows that the relation between $L_{\rm x}$ and $T_{\rm x}$ of M dwarfs is poorly known so far. But the huge number of relatively bright X-ray emitters in eRASS will put new constraints.

Finally, {\em eROSITA} has opened a new window for variability studies of coronal X-ray emission. Our comparison with RASS data has confirmed the evidence from previous much smaller samples 
of M dwarfs \cite[e.g.][]{Marino00.0} that large changes of the X-ray luminosity are rare. Specifically, only $17$\,\% of our combined {\em eROSITA/RASS} sample displays variability by more than a factor of two between the two surveys.  
Since the most obvious candidates for X-ray variability in M dwarfs are flares, this low variability amplitude arises  clearly from the averaging over survey exposures (typically $6-8$ intervals of $\sim 40$\,s duration each and separated by $\sim 4$\,h in case of eRASS).
About one-third  of the $14$ M dwarfs detected in the eFEDS have significantly variable {\em eROSITA} light curves. 
Detailed studies of these short-term light curves are required to extract better estimates for the brightness scale of the variations. Moreover, the huge number of such light curves available with {\em eROSITA} should enable for the first time to put limits on the flare frequency of M dwarfs in the X-ray domain. 

\begin{acknowledgements} 
    EM is supported by the \textsl{Bundesministerium f\"{u}r Wirtschaft und Energie} through the \textsl{Deutsches Zentrum f\"{u}r Luft- und Raumfahrt e.V. (DLR)} under grant number FKZ 50 OR 1808.
        
     AK is supported by the Deutsche Forschungsgemeinschaft (DFG) project number 413113723.

    This work is based on data from eROSITA, the soft X-ray instrument aboard SRG, a joint Russian-German science mission supported by the Russian Space Agency (Roskosmos), in the interests of the Russian Academy of Sciences represented by its Space Research Institute (IKI), and the Deutsches Zentrum für Luft- und Raumfahrt (DLR). The SRG spacecraft was built by Lavochkin Association (NPOL) and its subcontractors, and is operated by NPOL with support from the Max Planck Institute for Extraterrestrial Physics (MPE).

    The development and construction of the eROSITA X-ray instrument was led by MPE, with contributions from the Dr. Karl Remeis Observatory Bamberg \& ECAP (FAU Erlangen-Nuernberg), the University of Hamburg Observatory, the Leibniz Institute for Astrophysics Potsdam (AIP), and the Institute for Astronomy and Astrophysics of the University of T\"{u}bingen, with the support of DLR and the Max Planck Society. The Argelander Institute for Astronomy of the University of Bonn and the Ludwig Maximilians Universit\"{a}t Munich also participated in the science preparation for eROSITA.
    
    The eROSITA data shown here were processed using the eSASS/NRTA software system developed by the German eROSITA consortium.

    This paper includes data collected with the TESS mission, obtained from the MAST data archive at the Space Telescope Science Institute (STScI). Funding for the TESS mission is provided by the NASA Explorer Program. STScI is operated by the Association of Universities for Research in Astronomy, Inc., under NASA contract NAS 5–26555.
    
    This work has made use of data from the European Space Agency (ESA) mission {\it Gaia} (\url{https://www.cosmos.esa.int/gaia}), processed by the {\it Gaia} Data Processing and Analysis Consortium (DPAC, \url{https://www.cosmos.esa.int/web/gaia/dpac/consortium}). Funding for the DPAC has been provided by national institutions, in particular the institutions participating in the {\it Gaia} Multilateral Agreement.
    
    This publication makes use of data products from the Two Micron All Sky Survey, which is a joint project of the University of Massachusetts and the Infrared Processing and Analysis Center/California Institute of Technology, funded by the National Aeronautics and Space Administration and the National Science Foundation and of data products from the Wide-field Infrared Survey Explorer, which is a joint project of the University of California, Los Angeles, and the Jet Propulsion Laboratory/California Institute of Technology, funded by the National Aeronautics and Space Administration.
\end{acknowledgements}

\bibliographystyle{aa} 
\bibliography{bibliography,literatur}

\begin{thebibliography}{56}
\expandafter\ifx\csname natexlab\endcsname\relax\def\natexlab#1{#1}\fi

\bibitem[{{An} {et~al.}(2007){An}, {Terndrup}, {Pinsonneault}, {Paulson},
  {Hanson}, \& {Stauffer}}]{An07.0}
{An}, D., {Terndrup}, D.~M., {Pinsonneault}, M.~H., {et~al.} 2007, \apj, 655,
  233

\bibitem[{{Bailer-Jones} {et~al.}(2018){Bailer-Jones}, {Rybizki}, {Fouesneau},
  {Mantelet}, \& {Andrae}}]{BailerJones2018}
{Bailer-Jones}, C.~A.~L., {Rybizki}, J., {Fouesneau}, M., {Mantelet}, G., \&
  {Andrae}, R. 2018, \aj, 156, 58

\bibitem[{{Bertin} \& {Arnouts}(1996)}]{1996A&AS..117..393B}
{Bertin}, E. \& {Arnouts}, S. 1996, \aaps, 117, 393

\bibitem[{{Boller} {et~al.}(2016){Boller}, {Freyberg}, {Tr{\"u}mper}, {Haberl},
  {Voges}, \& {Nandra}}]{Boller16.0}
{Boller}, T., {Freyberg}, M.~J., {Tr{\"u}mper}, J., {et~al.} 2016, \aap, 588,
  A103

\bibitem[{{Broeg} {et~al.}(2005){Broeg}, {Fern{\'a}ndez}, \&
  {Neuh{\"a}user}}]{2005AN....326..134B}
{Broeg}, C., {Fern{\'a}ndez}, M., \& {Neuh{\"a}user}, R. 2005, Astronomische
  Nachrichten, 326, 134

\bibitem[{{Choi} \& {Herbst}(1996)}]{Choi96.0}
{Choi}, P.~I. \& {Herbst}, W. 1996, \aj, 111, 283

\bibitem[{{Feinstein} {et~al.}(2019){Feinstein}, {Montet}, {Foreman-Mackey},
  {Bedell}, {Saunders}, {Bean}, {Christiansen}, {Hedges}, {Luger}, {Scolnic},
  \& {Cardoso}}]{2019PASP..131i4502F}
{Feinstein}, A.~D., {Montet}, B.~T., {Foreman-Mackey}, D., {et~al.} 2019,
  \pasp, 131, 094502

\bibitem[{{Fleming} \& {Stone}(2003)}]{Fleming2003}
{Fleming}, T. \& {Stone}, J.~M. 2003, \apj, 585, 908

\bibitem[{{Fleming}(1998)}]{Fleming98.0}
{Fleming}, T.~A. 1998, \apj, 504, 461

\bibitem[{{Fleming} {et~al.}(1988){Fleming}, {Liebert}, {Gioia}, \&
  {Maccacaro}}]{Fleming88.0}
{Fleming}, T.~A., {Liebert}, J., {Gioia}, I.~M., \& {Maccacaro}, T. 1988, \apj,
  331, 958

\bibitem[{{Gaia Collaboration} {et~al.}(2018{\natexlab{a}}){Gaia
  Collaboration}, {Babusiaux}, {van Leeuwen}, {Barstow}, {Jordi}, {Vallenari},
  {Bossini}, {Bressan}, {Cantat-Gaudin}, {van Leeuwen}, {Brown}, {Prusti}, {de
  Bruijne}, {Bailer-Jones}, {Biermann}, {Evans}, {Eyer}, {Jansen}, {Klioner},
  {Lammers}, {Lindegren}, {Luri}, {Mignard}, {Panem}, {Pourbaix}, {Randich},
  {Sartoretti}, {Siddiqui}, {Soubiran}, {Walton}, {Arenou}, {Bastian},
  {Cropper}, {Drimmel}, {Katz}, {Lattanzi}, {Bakker}, {Cacciari},
  {Casta{\~n}eda}, {Chaoul}, {Cheek}, {De Angeli}, {Fabricius}, {Guerra},
  {Holl}, {Masana}, {Messineo}, {Mowlavi}, {Nienartowicz}, {Panuzzo},
  {Portell}, {Riello}, {Seabroke}, {Tanga}, {Th{\'e}venin}, {Gracia-Abril},
  {Comoretto}, {Garcia-Reinaldos}, {Teyssier}, {Altmann}, {Andrae}, {Audard},
  {Bellas-Velidis}, {Benson}, {Berthier}, {Blomme}, {Burgess}, {Busso},
  {Carry}, {Cellino}, {Clementini}, {Clotet}, {Creevey}, {Davidson}, {De
  Ridder}, {Delchambre}, {Dell'Oro}, {Ducourant},
  {Fern{\'a}ndez-Hern{\'a}ndez}, {Fouesneau}, {Fr{\'e}mat}, {Galluccio},
  {Garc{\'\i}a-Torres}, {Gonz{\'a}lez-N{\'u}{\~n}ez}, {Gonz{\'a}lez-Vidal},
  {Gosset}, {Guy}, {Halbwachs}, {Hambly}, {Harrison}, {Hern{\'a}ndez},
  {Hestroffer}, {Hodgkin}, {Hutton}, {Jasniewicz}, {Jean-Antoine-Piccolo},
  {Jordan}, {Korn}, {Krone-Martins}, {Lanzafame}, {Lebzelter}, {L{\"o}ffler},
  {Manteiga}, {Marrese}, {Mart{\'\i}n-Fleitas}, {Moitinho}, {Mora}, {Muinonen},
  {Osinde}, {Pancino}, {Pauwels}, {Petit}, {Recio-Blanco}, {Richards},
  {Rimoldini}, {Robin}, {Sarro}, {Siopis}, {Smith}, {Sozzetti}, {S{\"u}veges},
  {Torra}, {van Reeven}, {Abbas}, {Abreu Aramburu}, {Accart}, {Aerts},
  {Altavilla}, {{\'A}lvarez}, {Alvarez}, {Alves}, {Anderson}, {Andrei},
  {Anglada Varela}, {Antiche}, {Antoja}, {Arcay}, {Astraatmadja}, {Bach},
  {Baker}, {Balaguer-N{\'u}{\~n}ez}, {Balm}, {Barache}, {Barata}, {Barbato},
  {Barblan}, {Barklem}, {Barrado}, {Barros}, {Bartholom{\'e} Mu{\~n}oz},
  {Bassilana}, {Becciani}, {Bellazzini}, {Berihuete}, {Bertone}, {Bianchi},
  {Bienaym{\'e}}, {Blanco-Cuaresma}, {Boch}, {Boeche}, {Bombrun}, {Borrachero},
  {Bouquillon}, {Bourda}, {Bragaglia}, {Bramante}, {Breddels}, {Brouillet},
  {Br{\"u}semeister}, {Brugaletta}, {Bucciarelli}, {Burlacu}, {Busonero},
  {Butkevich}, {Buzzi}, {Caffau}, {Cancelliere}, {Cannizzaro}, {Carballo},
  {Carlucci}, {Carrasco}, {Casamiquela}, {Castellani}, {Castro-Ginard},
  {Charlot}, {Chemin}, {Chiavassa}, {Cocozza}, {Costigan}, {Cowell}, {Crifo},
  {Crosta}, {Crowley}, {Cuypers}, {Dafonte}, {Damerdji}, {Dapergolas}, {David},
  {David}, {de Laverny}, {De Luise}, {De March}, {de Martino}, {de Souza}, {de
  Torres}, {Debosscher}, {del Pozo}, {Delbo}, {Delgado}, {Delgado}, {Diakite},
  {Diener}, {Distefano}, {Dolding}, {Drazinos}, {Dur{\'a}n}, {Edvardsson},
  {Enke}, {Eriksson}, {Esquej}, {Eynard Bontemps}, {Fabre}, {Fabrizio},
  {Faigler}, {Falc{\~a}o}, {Farr{\`a}s Casas}, {Federici}, {Fedorets},
  {Fernique}, {Figueras}, {Filippi}, {Findeisen}, {Fonti}, {Fraile}, {Fraser},
  {Fr{\'e}zouls}, {Gai}, {Galleti}, {Garabato}, {Garc{\'\i}a-Sedano},
  {Garofalo}, {Garralda}, {Gavel}, {Gavras}, {Gerssen}, {Geyer}, {Giacobbe},
  {Gilmore}, {Girona}, {Giuffrida}, {Glass}, {Gomes}, {Granvik}, {Gueguen},
  {Guerrier}, {Guiraud}, {Guti{\'e}}, {Haigron}, {Hatzidimitriou}, {Hauser},
  {Haywood}, {Heiter}, {Helmi}, {Heu}, {Hilger}, {Hobbs}, {Hofmann}, {Holland},
  {Huckle}, {Hypki}, {Icardi}, {Jan{\ss}en}, {Jevardat de Fombelle}, {Jonker},
  {Juh{\'a}sz}, {Julbe}, {Karampelas}, {Kewley}, {Klar}, {Kochoska}, {Kohley},
  {Kolenberg}, {Kontizas}, {Kontizas}, {Koposov}, {Kordopatis},
  {Kostrzewa-Rutkowska}, {Koubsky}, {Lambert}, {Lanza}, {Lasne}, {Lavigne}, {Le
  Fustec}, {Le Poncin-Lafitte}, {Lebreton}, {Leccia}, {Leclerc},
  {Lecoeur-Taibi}, {Lenhardt}, {Leroux}, {Liao}, {Licata}, {Lindstr{\o}m},
  {Lister}, {Livanou}, {Lobel}, {L{\'o}pez}, {Managau}, {Mann}, {Mantelet},
  {Marchal}, {Marchant}, {Marconi}, {Marinoni}, {Marschalk{\'o}}, {Marshall},
  {Martino}, {Marton}, {Mary}, {Massari}, {Matijevi{\v{c}}}, {Mazeh},
  {McMillan}, {Messina}, {Michalik}, {Millar}, {Molina}, {Molinaro},
  {Moln{\'a}r}, {Montegriffo}, {Mor}, {Morbidelli}, {Morel}, {Morris},
  {Mulone}, {Muraveva}, {Musella}, {Nelemans}, {Nicastro}, {Noval},
  {O'Mullane}, {Ord{\'e}novic}, {Ord{\'o}{\~n}ez-Blanco}, {Osborne}, {Pagani},
  {Pagano}, {Pailler}, {Palacin}, {Palaversa}, {Panahi}, {Pawlak},
  {Piersimoni}, {Pineau}, {Plachy}, {Plum}, {Poggio}, {Poujoulet},
  {Pr{\v{s}}a}, {Pulone}, {Racero}, {Ragaini}, {Rambaux}, {Ramos-Lerate},
  {Regibo}, {Reyl{\'e}}, {Riclet}, {Ripepi}, {Riva}, {Rivard}, {Rixon},
  {Roegiers}, {Roelens}, {Romero-G{\'o}mez}, {Rowell}, {Royer}, {Ruiz-Dern},
  {Sadowski}, {Sagrist{\`a} Sell{\'e}s}, {Sahlmann}, {Salgado}, {Salguero},
  {Sanna}, {Santana-Ros}, {Sarasso}, {Savietto}, {Schultheis}, {Sciacca},
  {Segol}, {Segovia}, {S{\'e}gransan}, {Shih}, {Siltala}, {Silva}, {Smart},
  {Smith}, {Solano}, {Solitro}, {Sordo}, {Soria Nieto}, {Souchay}, {Spagna},
  {Spoto}, {Stampa}, {Steele}, {Steidelm{\"u}ller}, {Stephenson}, {Stoev},
  {Suess}, {Surdej}, {Szabados}, {Szegedi-Elek}, {Tapiador}, {Taris}, {Tauran},
  {Taylor}, {Teixeira}, {Terrett}, {Teyssandier}, {Thuillot}, {Titarenko},
  {Torra Clotet}, {Turon}, {Ulla}, {Utrilla}, {Uzzi}, {Vaillant}, {Valentini},
  {Valette}, {van Elteren}, {Van Hemelryck}, {Vaschetto}, {Vecchiato},
  {Veljanoski}, {Viala}, {Vicente}, {Vogt}, {von Essen}, {Voss}, {Votruba},
  {Voutsinas}, {Walmsley}, {Weiler}, {Wertz}, {Wevers}, {Wyrzykowski},
  {Yoldas}, {{\v{Z}}erjal}, {Ziaeepour}, {Zorec}, {Zschocke}, {Zucker},
  {Zurbach}, \& {Zwitter}}]{Gaia2018}
{Gaia Collaboration}, {Babusiaux}, C., {van Leeuwen}, F., {et~al.}
  2018{\natexlab{a}}, \aap, 616, A10

\bibitem[{{Gaia Collaboration} {et~al.}(2018{\natexlab{b}}){Gaia
  Collaboration}, {Brown}, {Vallenari}, {Prusti}, {de Bruijne}, {Babusiaux},
  {Bailer-Jones}, {Biermann}, {Evans}, {Eyer}, \& et~al.}]{2018A&A...616A...1G}
{Gaia Collaboration}, {Brown}, A.~G.~A., {Vallenari}, A., {et~al.}
  2018{\natexlab{b}}, \aap, 616, A1

\bibitem[{{Gaia Collaboration} {et~al.}(2016){Gaia Collaboration}, {Prusti},
  {de Bruijne}, {Brown}, {Vallenari}, {Babusiaux}, {Bailer-Jones}, {Bastian},
  {Biermann}, {Evans}, {Eyer}, {Jansen}, {Jordi}, {Klioner}, {Lammers},
  {Lindegren}, {Luri}, {Mignard}, {Milligan}, {Panem}, {Poinsignon},
  {Pourbaix}, {Randich}, {Sarri}, {Sartoretti}, {Siddiqui}, {Soubiran},
  {Valette}, {van Leeuwen}, {Walton}, {Aerts}, {Arenou}, {Cropper}, {Drimmel},
  {H{\o}g}, {Katz}, {Lattanzi}, {O'Mullane}, {Grebel}, {Holland}, {Huc},
  {Passot}, {Bramante}, {Cacciari}, {Casta{\~n}eda}, {Chaoul}, {Cheek}, {De
  Angeli}, {Fabricius}, {Guerra}, {Hern{\'a}ndez}, {Jean-Antoine-Piccolo},
  {Masana}, {Messineo}, {Mowlavi}, {Nienartowicz}, {Ord{\'o}{\~n}ez-Blanco},
  {Panuzzo}, {Portell}, {Richards}, {Riello}, {Seabroke}, {Tanga},
  {Th{\'e}venin}, {Torra}, {Els}, {Gracia-Abril}, {Comoretto},
  {Garcia-Reinaldos}, {Lock}, {Mercier}, {Altmann}, {Andrae}, {Astraatmadja},
  {Bellas-Velidis}, {Benson}, {Berthier}, {Blomme}, {Busso}, {Carry},
  {Cellino}, {Clementini}, {Cowell}, {Creevey}, {Cuypers}, {Davidson}, {De
  Ridder}, {de Torres}, {Delchambre}, {Dell'Oro}, {Ducourant}, {Fr{\'e}mat},
  {Garc{\'\i}a-Torres}, {Gosset}, {Halbwachs}, {Hambly}, {Harrison}, {Hauser},
  {Hestroffer}, {Hodgkin}, {Huckle}, {Hutton}, {Jasniewicz}, {Jordan},
  {Kontizas}, {Korn}, {Lanzafame}, {Manteiga}, {Moitinho}, {Muinonen},
  {Osinde}, {Pancino}, {Pauwels}, {Petit}, {Recio-Blanco}, {Robin}, {Sarro},
  {Siopis}, {Smith}, {Smith}, {Sozzetti}, {Thuillot}, {van Reeven}, {Viala},
  {Abbas}, {Abreu Aramburu}, {Accart}, {Aguado}, {Allan}, {Allasia},
  {Altavilla}, {{\'A}lvarez}, {Alves}, {Anderson}, {Andrei}, {Anglada Varela},
  {Antiche}, {Antoja}, {Ant{\'o}n}, {Arcay}, {Atzei}, {Ayache}, {Bach},
  {Baker}, {Balaguer-N{\'u}{\~n}ez}, {Barache}, {Barata}, {Barbier}, {Barblan},
  {Baroni}, {Barrado y Navascu{\'e}s}, {Barros}, {Barstow}, {Becciani},
  {Bellazzini}, {Bellei}, {Bello Garc{\'\i}a}, {Belokurov}, {Bendjoya},
  {Berihuete}, {Bianchi}, {Bienaym{\'e}}, {Billebaud}, {Blagorodnova},
  {Blanco-Cuaresma}, {Boch}, {Bombrun}, {Borrachero}, {Bouquillon}, {Bourda},
  {Bouy}, {Bragaglia}, {Breddels}, {Brouillet}, {Br{\"u}semeister},
  {Bucciarelli}, {Budnik}, {Burgess}, {Burgon}, {Burlacu}, {Busonero}, {Buzzi},
  {Caffau}, {Cambras}, {Campbell}, {Cancelliere}, {Cantat-Gaudin}, {Carlucci},
  {Carrasco}, {Castellani}, {Charlot}, {Charnas}, {Charvet}, {Chassat},
  {Chiavassa}, {Clotet}, {Cocozza}, {Collins}, {Collins}, {Costigan}, {Crifo},
  {Cross}, {Crosta}, {Crowley}, {Dafonte}, {Damerdji}, {Dapergolas}, {David},
  {David}, {De Cat}, {de Felice}, {de Laverny}, {De Luise}, {De March}, {de
  Martino}, {de Souza}, {Debosscher}, {del Pozo}, {Delbo}, {Delgado},
  {Delgado}, {di Marco}, {Di Matteo}, {Diakite}, {Distefano}, {Dolding}, {Dos
  Anjos}, {Drazinos}, {Dur{\'a}n}, {Dzigan}, {Ecale}, {Edvardsson}, {Enke},
  {Erdmann}, {Escolar}, {Espina}, {Evans}, {Eynard Bontemps}, {Fabre},
  {Fabrizio}, {Faigler}, {Falc{\~a}o}, {Farr{\`a}s Casas}, {Faye}, {Federici},
  {Fedorets}, {Fern{\'a}ndez-Hern{\'a}ndez}, {Fernique}, {Fienga}, {Figueras},
  {Filippi}, {Findeisen}, {Fonti}, {Fouesneau}, {Fraile}, {Fraser}, {Fuchs},
  {Furnell}, {Gai}, {Galleti}, {Galluccio}, {Garabato}, {Garc{\'\i}a-Sedano},
  {Gar{\'e}}, {Garofalo}, {Garralda}, {Gavras}, {Gerssen}, {Geyer}, {Gilmore},
  {Girona}, {Giuffrida}, {Gomes}, {Gonz{\'a}lez-Marcos},
  {Gonz{\'a}lez-N{\'u}{\~n}ez}, {Gonz{\'a}lez-Vidal}, {Granvik}, {Guerrier},
  {Guillout}, {Guiraud}, {G{\'u}rpide}, {Guti{\'e}rrez-S{\'a}nchez}, {Guy},
  {Haigron}, {Hatzidimitriou}, {Haywood}, {Heiter}, {Helmi}, {Hobbs},
  {Hofmann}, {Holl}, {Holland}, {Hunt}, {Hypki}, {Icardi}, {Irwin}, {Jevardat
  de Fombelle}, {Jofr{\'e}}, {Jonker}, {Jorissen}, {Julbe}, {Karampelas},
  {Kochoska}, {Kohley}, {Kolenberg}, {Kontizas}, {Koposov}, {Kordopatis},
  {Koubsky}, {Kowalczyk}, {Krone-Martins}, {Kudryashova}, {Kull}, {Bachchan},
  {Lacoste-Seris}, {Lanza}, {Lavigne}, {Le Poncin-Lafitte}, {Lebreton},
  {Lebzelter}, {Leccia}, {Leclerc}, {Lecoeur-Taibi}, {Lemaitre}, {Lenhardt},
  {Leroux}, {Liao}, {Licata}, {Lindstr{\o}m}, {Lister}, {Livanou}, {Lobel},
  {L{\"o}ffler}, {L{\'o}pez}, {Lopez-Lozano}, {Lorenz}, {Loureiro},
  {MacDonald}, {Magalh{\~a}es Fernandes}, {Managau}, {Mann}, {Mantelet},
  {Marchal}, {Marchant}, {Marconi}, {Marie}, {Marinoni}, {Marrese},
  {Marschalk{\'o}}, {Marshall}, {Mart{\'\i}n-Fleitas}, {Martino}, {Mary},
  {Matijevi{\v{c}}}, {Mazeh}, {McMillan}, {Messina}, {Mestre}, {Michalik},
  {Millar}, {Miranda}, {Molina}, {Molinaro}, {Molinaro}, {Moln{\'a}r},
  {Moniez}, {Montegriffo}, {Monteiro}, {Mor}, {Mora}, {Morbidelli}, {Morel},
  {Morgenthaler}, {Morley}, {Morris}, {Mulone}, {Muraveva}, {Musella},
  {Narbonne}, {Nelemans}, {Nicastro}, {Noval}, {Ord{\'e}novic},
  {Ordieres-Mer{\'e}}, {Osborne}, {Pagani}, {Pagano}, {Pailler}, {Palacin},
  {Palaversa}, {Parsons}, {Paulsen}, {Pecoraro}, {Pedrosa}, {Pentik{\"a}inen},
  {Pereira}, {Pichon}, {Piersimoni}, {Pineau}, {Plachy}, {Plum}, {Poujoulet},
  {Pr{\v{s}}a}, {Pulone}, {Ragaini}, {Rago}, {Rambaux}, {Ramos-Lerate},
  {Ranalli}, {Rauw}, {Read}, {Regibo}, {Renk}, {Reyl{\'e}}, {Ribeiro},
  {Rimoldini}, {Ripepi}, {Riva}, {Rixon}, {Roelens}, {Romero-G{\'o}mez},
  {Rowell}, {Royer}, {Rudolph}, {Ruiz-Dern}, {Sadowski}, {Sagrist{\`a}
  Sell{\'e}s}, {Sahlmann}, {Salgado}, {Salguero}, {Sarasso}, {Savietto},
  {Schnorhk}, {Schultheis}, {Sciacca}, {Segol}, {Segovia}, {Segransan},
  {Serpell}, {Shih}, {Smareglia}, {Smart}, {Smith}, {Solano}, {Solitro},
  {Sordo}, {Soria Nieto}, {Souchay}, {Spagna}, {Spoto}, {Stampa}, {Steele},
  {Steidelm{\"u}ller}, {Stephenson}, {Stoev}, {Suess}, {S{\"u}veges}, {Surdej},
  {Szabados}, {Szegedi-Elek}, {Tapiador}, {Taris}, {Tauran}, {Taylor},
  {Teixeira}, {Terrett}, {Tingley}, {Trager}, {Turon}, {Ulla}, {Utrilla},
  {Valentini}, {van Elteren}, {Van Hemelryck}, {van Leeuwen}, {Varadi},
  {Vecchiato}, {Veljanoski}, {Via}, {Vicente}, {Vogt}, {Voss}, {Votruba},
  {Voutsinas}, {Walmsley}, {Weiler}, {Weingrill}, {Werner}, {Wevers},
  {Whitehead}, {Wyrzykowski}, {Yoldas}, {{\v{Z}}erjal}, {Zucker}, {Zurbach},
  {Zwitter}, {Alecu}, {Allen}, {Allende Prieto}, {Amorim},
  {Anglada-Escud{\'e}}, {Arsenijevic}, {Azaz}, {Balm}, {Beck}, {Bernstein},
  {Bigot}, {Bijaoui}, {Blasco}, {Bonfigli}, {Bono}, {Boudreault}, {Bressan},
  {Brown}, {Brunet}, {Bunclark}, {Buonanno}, {Butkevich}, {Carret}, {Carrion},
  {Chemin}, {Ch{\'e}reau}, {Corcione}, {Darmigny}, {de Boer}, {de Teodoro}, {de
  Zeeuw}, {Delle Luche}, {Domingues}, {Dubath}, {Fodor}, {Fr{\'e}zouls},
  {Fries}, {Fustes}, {Fyfe}, {Gallardo}, {Gallegos}, {Gardiol}, {Gebran},
  {Gomboc}, {G{\'o}mez}, {Grux}, {Gueguen}, {Heyrovsky}, {Hoar}, {Iannicola},
  {Isasi Parache}, {Janotto}, {Joliet}, {Jonckheere}, {Keil}, {Kim},
  {Klagyivik}, {Klar}, {Knude}, {Kochukhov}, {Kolka}, {Kos}, {Kutka}, {Lainey},
  {LeBouquin}, {Liu}, {Loreggia}, {Makarov}, {Marseille}, {Martayan},
  {Martinez-Rubi}, {Massart}, {Meynadier}, {Mignot}, {Munari}, {Nguyen},
  {Nordlander}, {Ocvirk}, {O'Flaherty}, {Olias Sanz}, {Ortiz}, {Osorio},
  {Oszkiewicz}, {Ouzounis}, {Palmer}, {Park}, {Pasquato}, {Peltzer}, {Peralta},
  {P{\'e}turaud}, {Pieniluoma}, {Pigozzi}, {Poels}, {Prat}, {Prod'homme},
  {Raison}, {Rebordao}, {Risquez}, {Rocca-Volmerange}, {Rosen}, {Ruiz-Fuertes},
  {Russo}, {Sembay}, {Serraller Vizcaino}, {Short}, {Siebert}, {Silva},
  {Sinachopoulos}, {Slezak}, {Soffel}, {Sosnowska}, {Strai{\v{z}}ys}, {ter
  Linden}, {Terrell}, {Theil}, {Tiede}, {Troisi}, {Tsalmantza}, {Tur},
  {Vaccari}, {Vachier}, {Valles}, {Van Hamme}, {Veltz}, {Virtanen}, {Wallut},
  {Wichmann}, {Wilkinson}, {Ziaeepour}, \& {Zschocke}}]{Gaia2016}
{Gaia Collaboration}, {Prusti}, T., {de Bruijne}, J.~H.~J., {et~al.} 2016,
  \aap, 595, A1

\bibitem[{{Gilliland} \& {Fisher}(1985)}]{1985PASP...97..285G}
{Gilliland}, R.~L. \& {Fisher}, R. 1985, \pasp, 97, 285

\bibitem[{{Johnstone} \& {G{\"u}del}(2015)}]{Johnstone15.0}
{Johnstone}, C.~P. \& {G{\"u}del}, M. 2015, \aap, 578, A129

\bibitem[{L{\'{e}}pine \& Gaidos(2011)}]{L_pine_2011}
L{\'{e}}pine, S. \& Gaidos, E. 2011, AJ, 142, 138

\bibitem[{Lindegren {et~al.}(2018)Lindegren, Hern{\'{a}}ndez, Bombrun, Klioner,
  Bastian, Ramos-Lerate, de~Torres, Steidelm{\"{u}}ller, Stephenson, Hobbs,
  Lammers, Biermann, Geyer, Hilger, Michalik, Stampa, McMillan,
  Casta{\~{n}}eda, Clotet, Comoretto, Davidson, Fabricius, Gracia, Hambly,
  Hutton, Mora, Portell, van Leeuwen, Abbas, Abreu, Altmann, Andrei, Anglada,
  Balaguer-N{\'{u}}{\~{n}}ez, Barache, Becciani, Bertone, Bianchi, Bouquillon,
  Bourda, Br{\"{u}}semeister, Bucciarelli, Busonero, Buzzi, Cancelliere,
  Carlucci, Charlot, Cheek, Crosta, Crowley, de~Bruijne, de~Felice, Drimmel,
  Esquej, Fienga, Fraile, Gai, Garralda, Gonz{\'{a}}lez-Vidal, Guerra, Hauser,
  Hofmann, Holl, Jordan, Lattanzi, Lenhardt, Liao, Licata, Lister,
  L{\"{o}}ffler, Marchant, Martin-Fleitas, Messineo, Mignard, Morbidelli,
  Poggio, Riva, Rowell, Salguero, Sarasso, Sciacca, Siddiqui, Smart, Spagna,
  Steele, Taris, Torra, van Elteren, van Reeven, \& Vecchiato}]{Lindegren2018}
Lindegren, L., Hern{\'{a}}ndez, J., Bombrun, A., {et~al.} 2018, A{\&}A, 616, A2

\bibitem[{{Magaudda} {et~al.}(2020){Magaudda}, {Stelzer}, {Covey}, {Raetz},
  {Matt}, \& {Scholz}}]{Magaudda2020}
{Magaudda}, E., {Stelzer}, B., {Covey}, K.~R., {et~al.} 2020, \aap, 638, A20

\bibitem[{{Mann} {et~al.}(2015){Mann}, {Feiden}, {Gaidos}, {Boyajian}, \& {von
  Braun}}]{Mann2015}
{Mann}, A.~W., {Feiden}, G.~A., {Gaidos}, E., {Boyajian}, T., \& {von Braun},
  K. 2015, \apj, 804, 64

\bibitem[{Mann {et~al.}(2016)Mann, Feiden, Gaidos, Boyajian, \& von
  Braun}]{Mann_2016}
Mann, A.~W., Feiden, G.~A., Gaidos, E., Boyajian, T., \& von Braun, K. 2016,
  \apj, 819, 87

\bibitem[{{Marino} {et~al.}(2000){Marino}, {Micela}, \& {Peres}}]{Marino00.0}
{Marino}, A., {Micela}, G., \& {Peres}, G. 2000, \aap, 353, 177

\bibitem[{Matt {et~al.}(2015)Matt, Brun, Baraffe, Bouvier, \&
  Chabrier}]{Matt_2015}
Matt, S.~P., Brun, A.~S., Baraffe, I., Bouvier, J., \& Chabrier, G. 2015, ApJ,
  799, L23

\bibitem[{{McQuillan} {et~al.}(2013){McQuillan}, {Mazeh}, \&
  {Aigrain}}]{McQuillan2013}
{McQuillan}, A., {Mazeh}, T., \& {Aigrain}, S. 2013, \apjl, 775, L11

\bibitem[{{McQuillan} {et~al.}(2014){McQuillan}, {Mazeh}, \&
  {Aigrain}}]{Mcquillan2014}
{McQuillan}, A., {Mazeh}, T., \& {Aigrain}, S. 2014, \apjs, 211, 24

\bibitem[{{Mer{\'\i}n} {et~al.}(2017){Mer{\'\i}n}, {Salgado}, {Giordano},
  {Baines}, {Sarmiento}, {Mart{\'\i}}, {Racero}, {Guti{\'e}rrez}, {Pollock},
  {Rosa}, {Castellanos}, {Gonz{\'a}lez}, {Le{\'o}n}, {Ortiz de Landaluce}, {de
  Teodoro}, {Nieto}, {Lennon}, {Arviset}, {de Marchi}, \&
  {O'Mullane}}]{Merin2017}
{Mer{\'\i}n}, B., {Salgado}, J., {Giordano}, F., {et~al.} 2017, in Astronomical
  Society of the Pacific Conference Series, Vol. 512, Astronomical Data
  Analysis Software and Systems XXV, ed. N.~P.~F. {Lorente}, K.~{Shortridge},
  \& R.~{Wayth}, 495

\bibitem[{{Modirrousta-Galian} {et~al.}(2020){Modirrousta-Galian}, {Stelzer},
  {Magaudda}, {Maldonado}, {G{\"u}del}, {Sanz-Forcada}, {Edwards}, \&
  {Micela}}]{ModirroustaGalian20.0}
{Modirrousta-Galian}, D., {Stelzer}, B., {Magaudda}, E., {et~al.} 2020, \aap,
  641, A113

\bibitem[{{Neuhaeuser} {et~al.}(1995){Neuhaeuser}, {Sterzik}, {Schmitt},
  {Wichmann}, \& {Krautter}}]{Neuhaeuser95.0}
{Neuhaeuser}, R., {Sterzik}, M.~F., {Schmitt}, J.~H.~M.~M., {Wichmann}, R., \&
  {Krautter}, J. 1995, \aap, 297, 391

\bibitem[{{Pallavicini} {et~al.}(1981){Pallavicini}, {Golub}, {Rosner},
  {Vaiana}, {Ayres}, \& {Linsky}}]{Pallavicini1981}
{Pallavicini}, R., {Golub}, L., {Rosner}, R., {et~al.} 1981, \apj, 248, 279

\bibitem[{{Parker}(1993)}]{Parker1993}
{Parker}, E.~N. 1993, \apj, 408, 707

\bibitem[{Pizzolato {et~al.}(2003)Pizzolato, Maggio, Micela, Sciortino, \&
  Ventura}]{Pizzolato2003}
Pizzolato, N., Maggio, A., Micela, G., Sciortino, S., \& Ventura, P. 2003,
  A{\&}A, 397, 147

\bibitem[{{Predehl} {et~al.}(2021){Predehl}, {Andritschke}, {Arefiev},
  {Babyshkin}, {Batanov}, {Becker}, {B{\"o}hringer}, {Bogomolov}, {Boller},
  {Borm}, {Bornemann}, {Br{\"a}uninger}, {Br{\"u}ggen}, {Brunner}, {Brusa},
  {Bulbul}, {Buntov}, {Burwitz}, {Burkert}, {Clerc}, {Churazov}, {Coutinho},
  {Dauser}, {Dennerl}, {Doroshenko}, {Eder}, {Emberger}, {Eraerds},
  {Finoguenov}, {Freyberg}, {Friedrich}, {Friedrich}, {F{\"u}rmetz},
  {Georgakakis}, {Gilfanov}, {Granato}, {Grossberger}, {Gueguen}, {Gureev},
  {Haberl}, {H{\"a}lker}, {Hartner}, {Hasinger}, {Huber}, {Ji}, {Kienlin},
  {Kink}, {Korotkov}, {Kreykenbohm}, {Lamer}, {Lomakin}, {Lapshov}, {Liu},
  {Maitra}, {Meidinger}, {Menz}, {Merloni}, {Mernik}, {Mican}, {Mohr},
  {M{\"u}ller}, {Nandra}, {Nazarov}, {Pacaud}, {Pavlinsky}, {Perinati},
  {Pfeffermann}, {Pietschner}, {Ramos-Ceja}, {Rau}, {Reiffers}, {Reiprich},
  {Robrade}, {Salvato}, {Sanders}, {Santangelo}, {Sasaki}, {Scheuerle},
  {Schmid}, {Schmitt}, {Schwope}, {Shirshakov}, {Steinmetz}, {Stewart},
  {Str{\"u}der}, {Sunyaev}, {Tenzer}, {Tiedemann}, {Tr{\"u}mper}, {Voron},
  {Weber}, {Wilms}, \& {Yaroshenko}}]{Predehl2021}
{Predehl}, P., {Andritschke}, R., {Arefiev}, V., {et~al.} 2021, \aap, 647, A1

\bibitem[{{Preibisch} \& {Feigelson}(2005)}]{Preibisch2005}
{Preibisch}, T. \& {Feigelson}, E.~D. 2005, \apjs, 160, 390

\bibitem[{{Preibisch} {et~al.}(2005){Preibisch}, {Kim}, {Favata}, {Feigelson},
  {Flaccomio}, {Getman}, {Micela}, {Sciortino}, {Stassun}, {Stelzer}, \&
  {Zinnecker}}]{Preibisch+2005}
{Preibisch}, T., {Kim}, Y.-C., {Favata}, F., {et~al.} 2005, \apjs, 160, 401

\bibitem[{{Raetz} {et~al.}(2016){Raetz}, {Schmidt}, {Czesla}, {Klocov{\'a}},
  {Holmes}, {Errmann}, {Kitze}, {Fern{\'a}ndez}, {Sota}, {Brice{\~n}o},
  {Hern{\'a}ndez}, {Downes}, {Dimitrov}, {Kjurkchieva}, {Radeva}, {Wu}, {Zhou},
  {Takahashi}, {Henych}, {Seeliger}, {Mugrauer}, {Adam}, {Marka}, {Schmidt},
  {Hohle}, {Ginski}, {Pribulla}, {Trepl}, {Moualla}, {Pawellek}, {Gelszinnis},
  {Buder}, {Masda}, {Maciejewski}, \& {Neuh{\"a}user}}]{2016MNRAS.460.2834R}
{Raetz}, S., {Schmidt}, T.~O.~B., {Czesla}, S., {et~al.} 2016, \mnras, 460,
  2834

\bibitem[{{Raetz} {et~al.}(2020){Raetz}, {Stelzer}, {Damasso}, \&
  {Scholz}}]{2020A&A...637A..22R}
{Raetz}, S., {Stelzer}, B., {Damasso}, M., \& {Scholz}, A. 2020, \aap, 637, A22

\bibitem[{{Reiners} {et~al.}(2012){Reiners}, {Joshi}, \&
  {Goldman}}]{Reiners2012}
{Reiners}, A., {Joshi}, N., \& {Goldman}, B. 2012, \aj, 143, 93

\bibitem[{{Reiners} {et~al.}(2014){Reiners}, {Sch{\"u}ssler}, \&
  {Passegger}}]{Reiners2014}
{Reiners}, A., {Sch{\"u}ssler}, M., \& {Passegger}, V.~M. 2014, \apj, 794, 144

\bibitem[{{Ribas} {et~al.}(2016){Ribas}, {Bolmont}, {Selsis}, {Reiners},
  {Leconte}, {Raymond}, {Engle}, {Guinan}, {Morin}, {Turbet}, {Forget}, \&
  {Anglada-Escud{\'e}}}]{Ribas16.0}
{Ribas}, I., {Bolmont}, E., {Selsis}, F., {et~al.} 2016, \aap, 596, A111

\bibitem[{{Ricker} {et~al.}(2014){Ricker}, {Winn}, {Vanderspek}, {Latham},
  {Bakos}, {Bean}, {Berta-Thompson}, {Brown}, {Buchhave}, {Butler}, {Butler},
  {Chaplin}, {Charbonneau}, {Christensen-Dalsgaard}, {Clampin}, {Deming},
  {Doty}, {De Lee}, {Dressing}, {Dunham}, {Endl}, {Fressin}, {Ge}, {Henning},
  {Holman}, {Howard}, {Ida}, {Jenkins}, {Jernigan}, {Johnson}, {Kaltenegger},
  {Kawai}, {Kjeldsen}, {Laughlin}, {Levine}, {Lin}, {Lissauer}, {MacQueen},
  {Marcy}, {McCullough}, {Morton}, {Narita}, {Paegert}, {Palle}, {Pepe},
  {Pepper}, {Quirrenbach}, {Rinehart}, {Sasselov}, {Sato}, {Seager},
  {Sozzetti}, {Stassun}, {Sullivan}, {Szentgyorgyi}, {Torres}, {Udry}, \&
  {Villasenor}}]{Ricker14.0}
{Ricker}, G.~R., {Winn}, J.~N., {Vanderspek}, R., {et~al.} 2014, in Society of
  Photo-Optical Instrumentation Engineers (SPIE) Conference Series, Vol. 9143,
  Space Telescopes and Instrumentation 2014: Optical, Infrared, and Millimeter
  Wave, ed. J.~{Oschmann}, Jacobus~M., M.~{Clampin}, G.~G. {Fazio}, \& H.~A.
  {MacEwen}, 914320

\bibitem[{{Robrade} {et~al.}(2010){Robrade}, {Poppenhaeger}, \&
  {Schmitt}}]{Robrade10.0}
{Robrade}, J., {Poppenhaeger}, K., \& {Schmitt}, J.~H.~M.~M. 2010, \aap, 513,
  A12

\bibitem[{{Rodr{\'\i}guez-Ledesma} {et~al.}(2009){Rodr{\'\i}guez-Ledesma},
  {Mundt}, \& {Eisl{\"o}ffel}}]{RodriguezLedesma09.0}
{Rodr{\'\i}guez-Ledesma}, M.~V., {Mundt}, R., \& {Eisl{\"o}ffel}, J. 2009,
  \aap, 502, 883

\bibitem[{{Rosner} {et~al.}(1985){Rosner}, {Golub}, \& {Vaiana}}]{Rosner1985}
{Rosner}, R., {Golub}, L., \& {Vaiana}, G.~S. 1985, \araa, 23, 413

\bibitem[{{Schmitt} \& {Liefke}(2004)}]{Schmitt2004}
{Schmitt}, J.~H.~M.~M. \& {Liefke}, C. 2004, \aap, 417, 651

\bibitem[{{Shu} {et~al.}(2019){Shu}, {Koposov}, {Evans}, {Belokurov},
  {McMahon}, {Auger}, \& {Lemon}}]{Shu2019}
{Shu}, Y., {Koposov}, S.~E., {Evans}, N.~W., {et~al.} 2019, \mnras, 489, 4741

\bibitem[{{Skrutskie} {et~al.}(2006){Skrutskie}, {Cutri}, {Stiening},
  {Weinberg}, {Schneider}, {Carpenter}, {Beichman}, {Capps}, {Chester},
  {Elias}, {Huchra}, {Liebert}, {Lonsdale}, {Monet}, {Price}, {Seitzer},
  {Jarrett}, {Kirkpatrick}, {Gizis}, {Howard}, {Evans}, {Fowler}, {Fullmer},
  {Hurt}, {Light}, {Kopan}, {Marsh}, {McCallon}, {Tam}, {Van Dyk}, \&
  {Wheelock}}]{2006AJ....131.1163S}
{Skrutskie}, M.~F., {Cutri}, R.~M., {Stiening}, R., {et~al.} 2006, \aj, 131,
  1163

\bibitem[{Stelzer {et~al.}(2016)Stelzer, Damasso, Scholz, \&
  Matt}]{Stelzer2016}
Stelzer, B., Damasso, M., Scholz, A., \& Matt, S.~P. 2016, MNRAS, 463, 1844

\bibitem[{Stelzer {et~al.}(2013)Stelzer, Marino, Micela, L{\'{o}}pez-Santiago,
  \& Liefke}]{Stelzer2013}
Stelzer, B., Marino, A., Micela, G., L{\'{o}}pez-Santiago, J., \& Liefke, C.
  2013, MNRAS, 431, 2063

\bibitem[{{Tarter} {et~al.}(2007){Tarter}, {Backus}, {Mancinelli}, {Aurnou},
  {Backman}, {Basri}, {Boss}, {Clarke}, {Deming}, {Doyle}, {Feigelson},
  {Freund}, {Grinspoon}, {Haberle}, {Hauck}, {Heath}, {Henry}, {Hollingsworth},
  {Joshi}, {Kilston}, {Liu}, {Meikle}, {Reid}, {Rothschild}, {Scalo}, {Segura},
  {Tang}, {Tiedje}, {Turnbull}, {Walkowicz}, {Weber}, \& {Young}}]{Tarter2007}
{Tarter}, J.~C., {Backus}, P.~R., {Mancinelli}, R.~L., {et~al.} 2007,
  Astrobiology, 7, 30

\bibitem[{{Taylor}(2005)}]{2005ASPC..347...29T}
{Taylor}, M.~B. 2005, in Astronomical Society of the Pacific Conference Series,
  Vol. 347, Astronomical Data Analysis Software and Systems XIV, ed.
  P.~{Shopbell}, M.~{Britton}, \& R.~{Ebert}, 29

\bibitem[{{Thompson} {et~al.}(2016){Thompson}, {Fraquelli}, {Van Cleve}, \&
  {Caldwell}}]{Thompson2016}
{Thompson}, S.~E., {Fraquelli}, D., {Van Cleve}, J.~E., \& {Caldwell}, D.~A.
  2016, {Kepler Archive Manual}, Kepler Science Document KDMC-10008-006

\bibitem[{{Tu} {et~al.}(2015){Tu}, {Johnstone}, {G{\"u}del}, \&
  {Lammer}}]{Tu2015}
{Tu}, L., {Johnstone}, C.~P., {G{\"u}del}, M., \& {Lammer}, H. 2015, \aap, 577,
  L3

\bibitem[{{Wenger} {et~al.}(2000){Wenger}, {Ochsenbein}, {Egret}, {Dubois},
  {Bonnarel}, {Borde}, {Genova}, {Jasniewicz}, {Lalo{\"e}}, {Lesteven}, \&
  {Monier}}]{2000A&AS..143....9W}
{Wenger}, M., {Ochsenbein}, F., {Egret}, D., {et~al.} 2000, \aaps, 143, 9

\bibitem[{Wright \& Drake(2016)}]{Wright2016}
Wright, N.~J. \& Drake, J.~J. 2016, Nature, 535, 526

\bibitem[{Wright {et~al.}(2011)Wright, Drake, Mamajek, \& Henry}]{Wright2011}
Wright, N.~J., Drake, J.~J., Mamajek, E.~E., \& Henry, G.~W. 2011, ApJ, 743

\bibitem[{Wright {et~al.}(2018)Wright, Newton, Williams, Drake, \&
  Yadav}]{Wright2018}
Wright, N.~J., Newton, E.~R., Williams, P.~K., Drake, J.~J., \& Yadav, R.~K.
  2018, MNRAS, 479, 2351

\bibitem[{{Zechmeister} \& {K{\"u}rster}(2009)}]{2009A&A...496..577Z}
{Zechmeister}, M. \& {K{\"u}rster}, M. 2009, \aap, 496, 577

\end{thebibliography}

\begin{appendix}

\section{Best fit spectral model}\label{app:spec}
Fig.~\ref{fig:eFEDS_spec} displays the {\em eROSITA} spectra from the eFEDS observation for the ten brightest M dwarfs in the field, as discussed in Sect.~\ref{subsubsect:analysis_efeds_xrays_spectra}. The best fitting model from Table~\ref{tab:xspec_output} is overlaid, and the bottom panels show the residuals between data and model.

       \renewcommand{\thetable}{A.\arabic{figure}} 
       \begin{figure*}[htbp]
 	\begin{multicols}{3}
 		\subfigure{\label{fig:a}\includegraphics[width=0.33\textwidth]{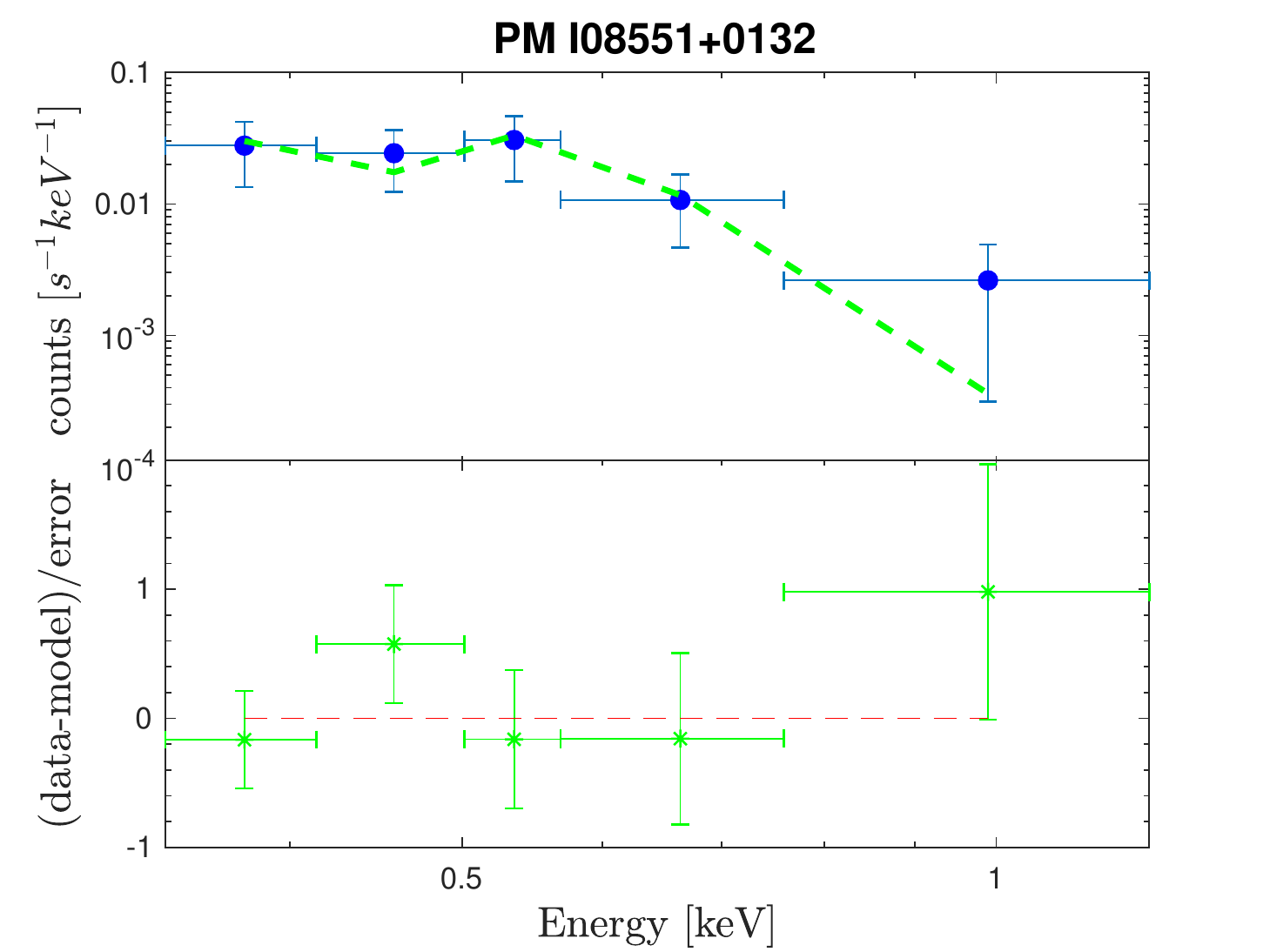}}
 		\subfigure{\label{fig:b}\includegraphics[width=0.33\textwidth]{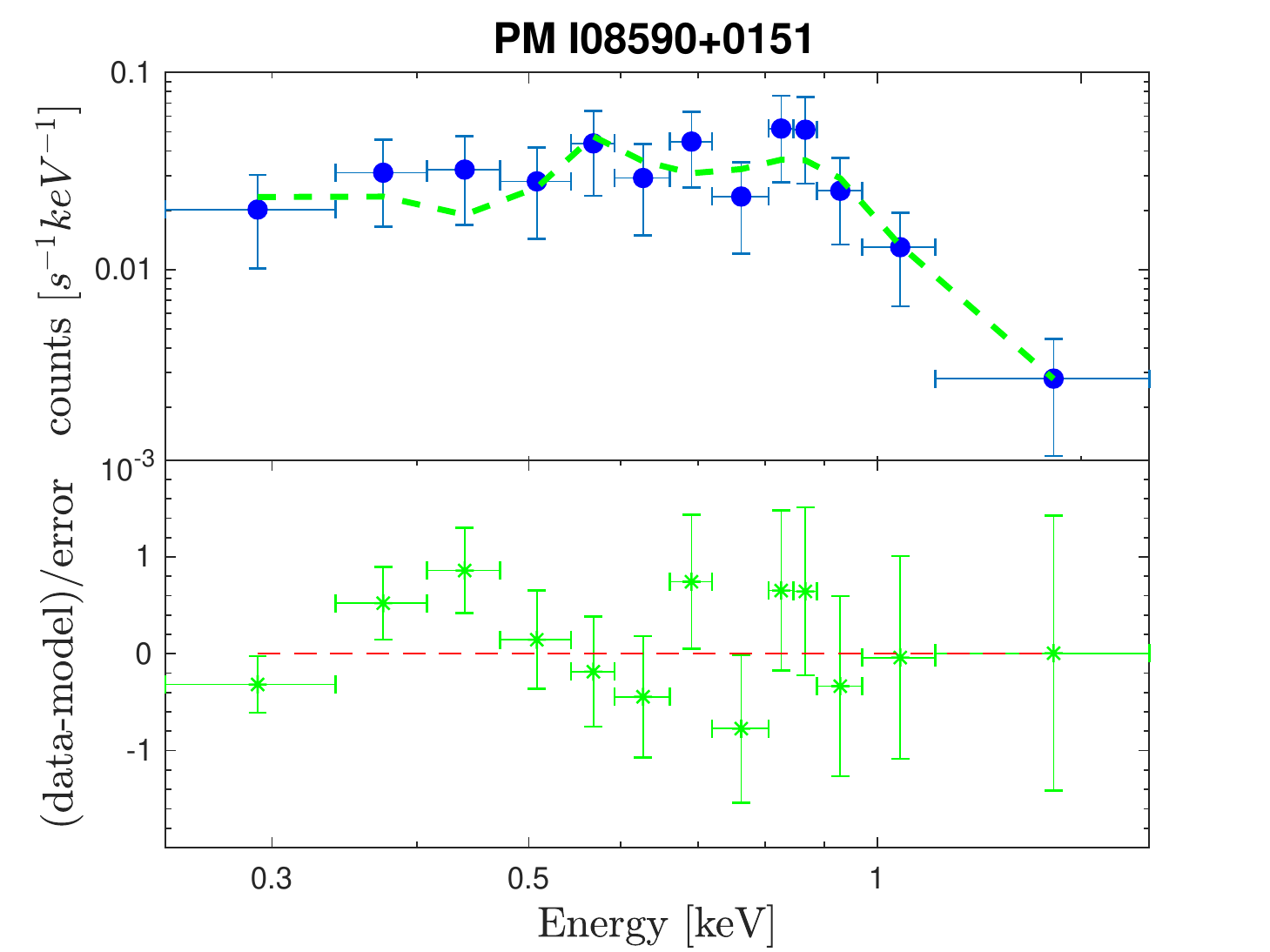}}
 		\subfigure{\label{fig:b}\includegraphics[width=0.33\textwidth]{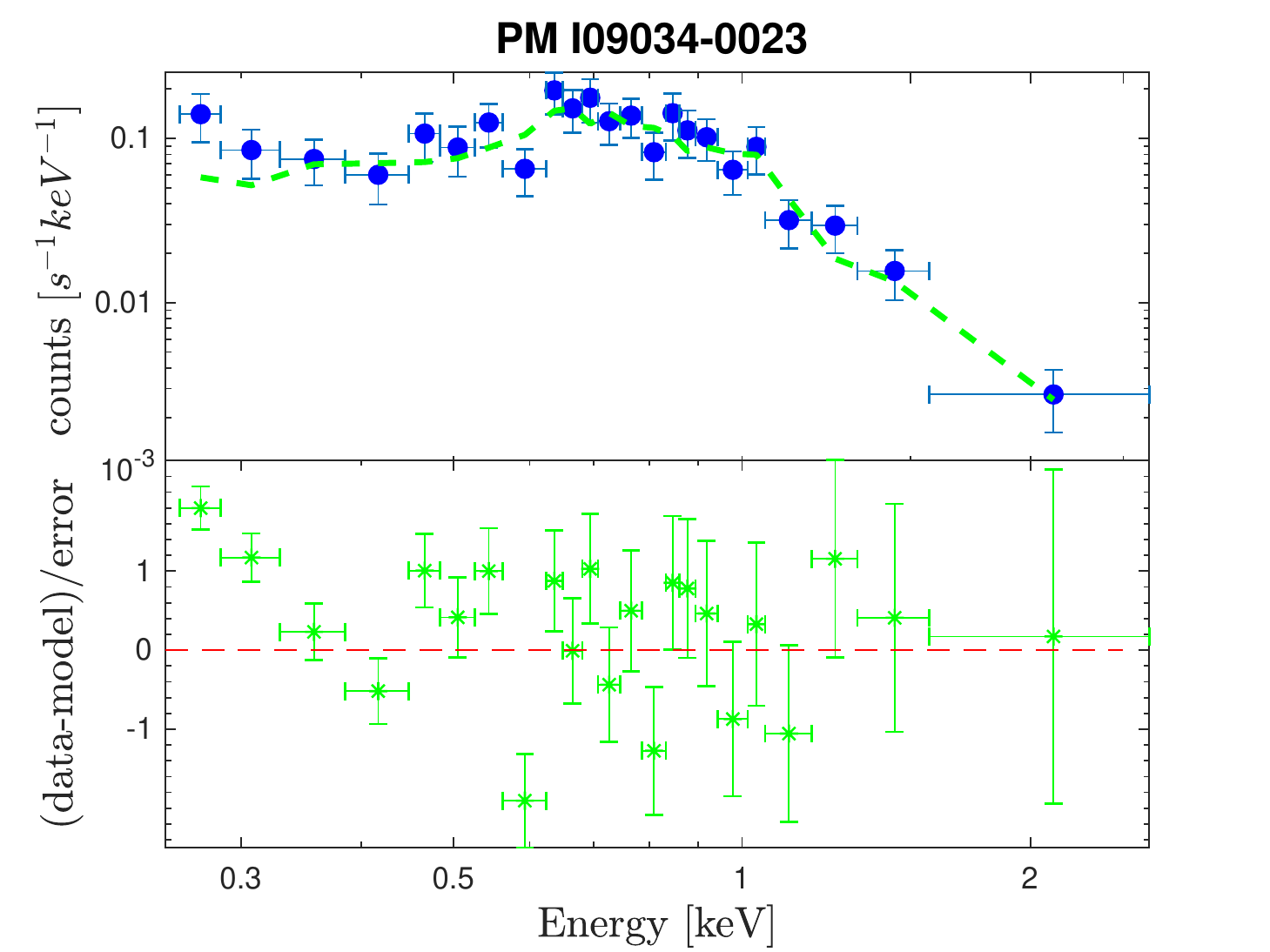}}
 		\subfigure{\label{fig:b}\includegraphics[width=0.33\textwidth]{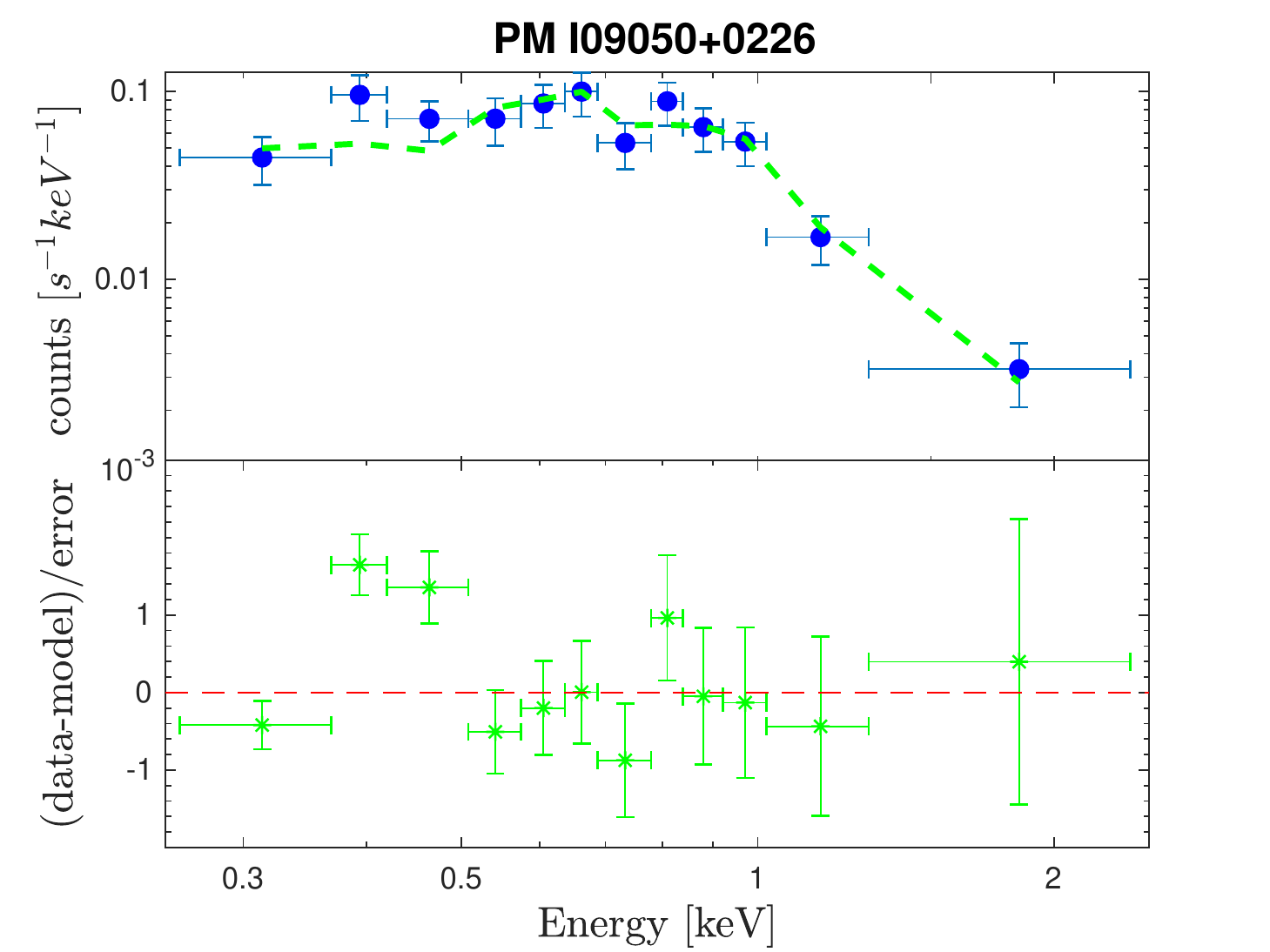}}
        \subfigure{\label{fig:b}\includegraphics[width=0.33\textwidth]{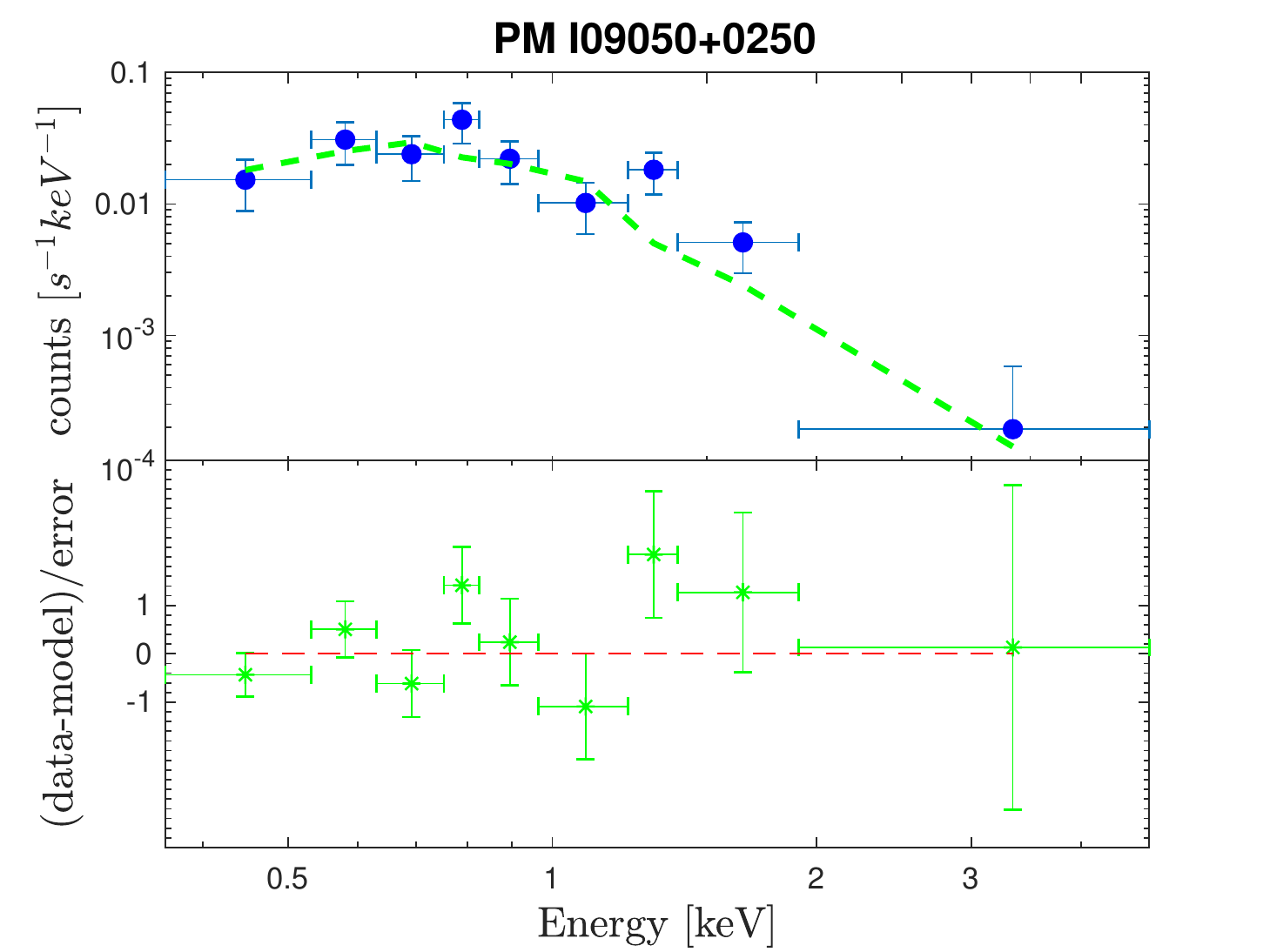}}
 		\subfigure{\label{fig:b}\includegraphics[width=0.33\textwidth]{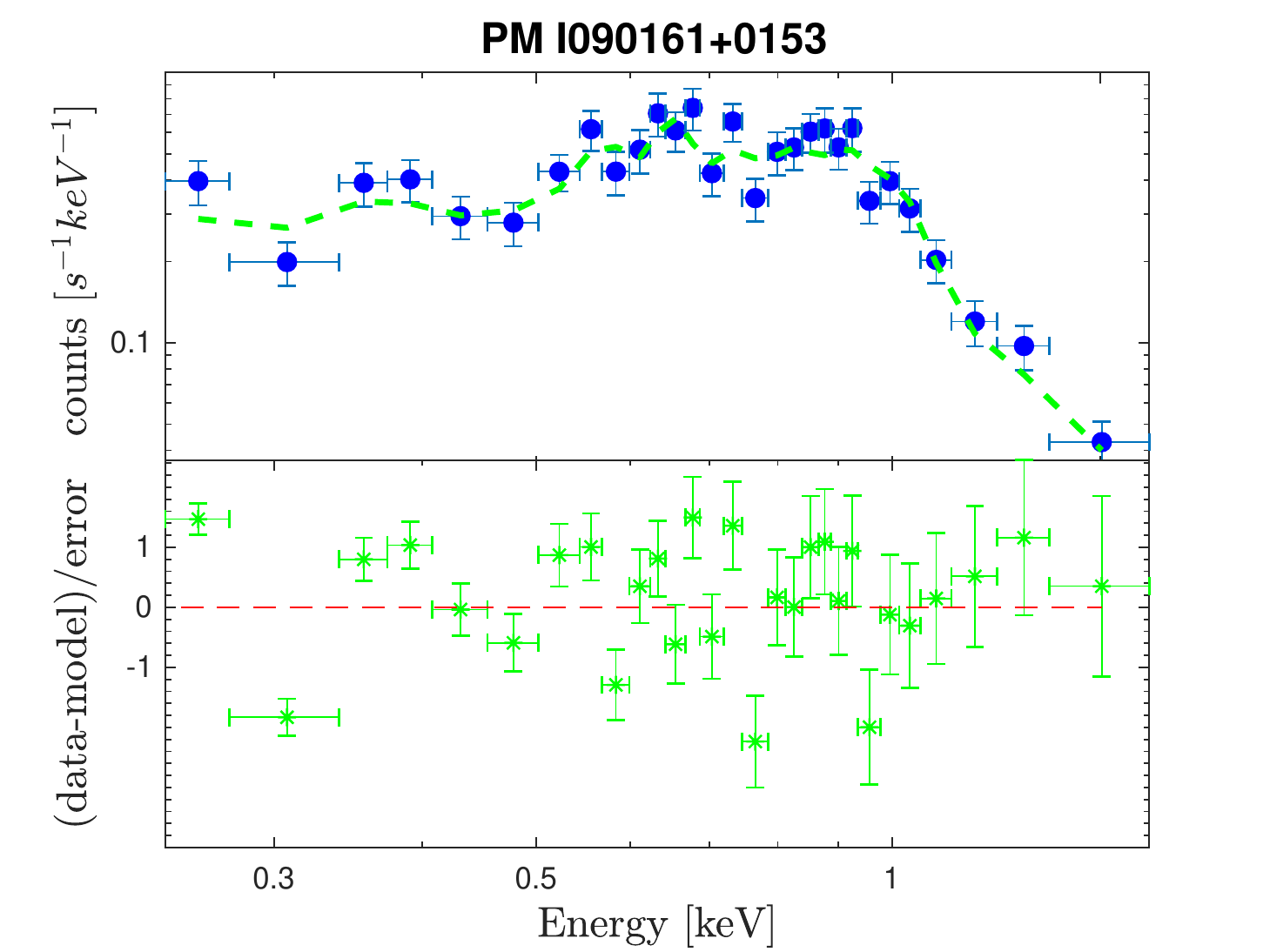}}
 		\subfigure{\label{fig:b}\includegraphics[width=0.33\textwidth]{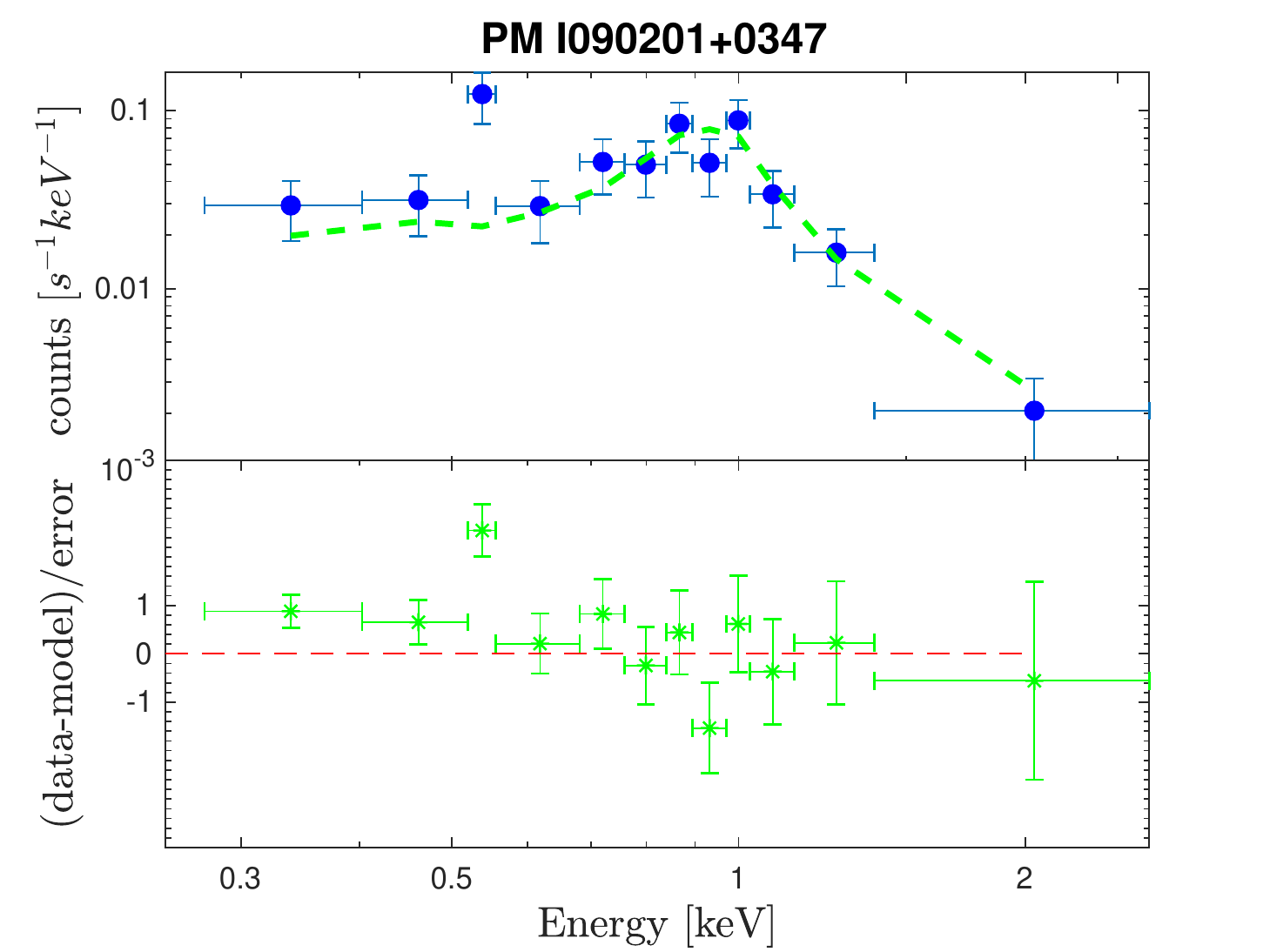}}
 		\subfigure{\label{fig:b}\includegraphics[width=0.33\textwidth]{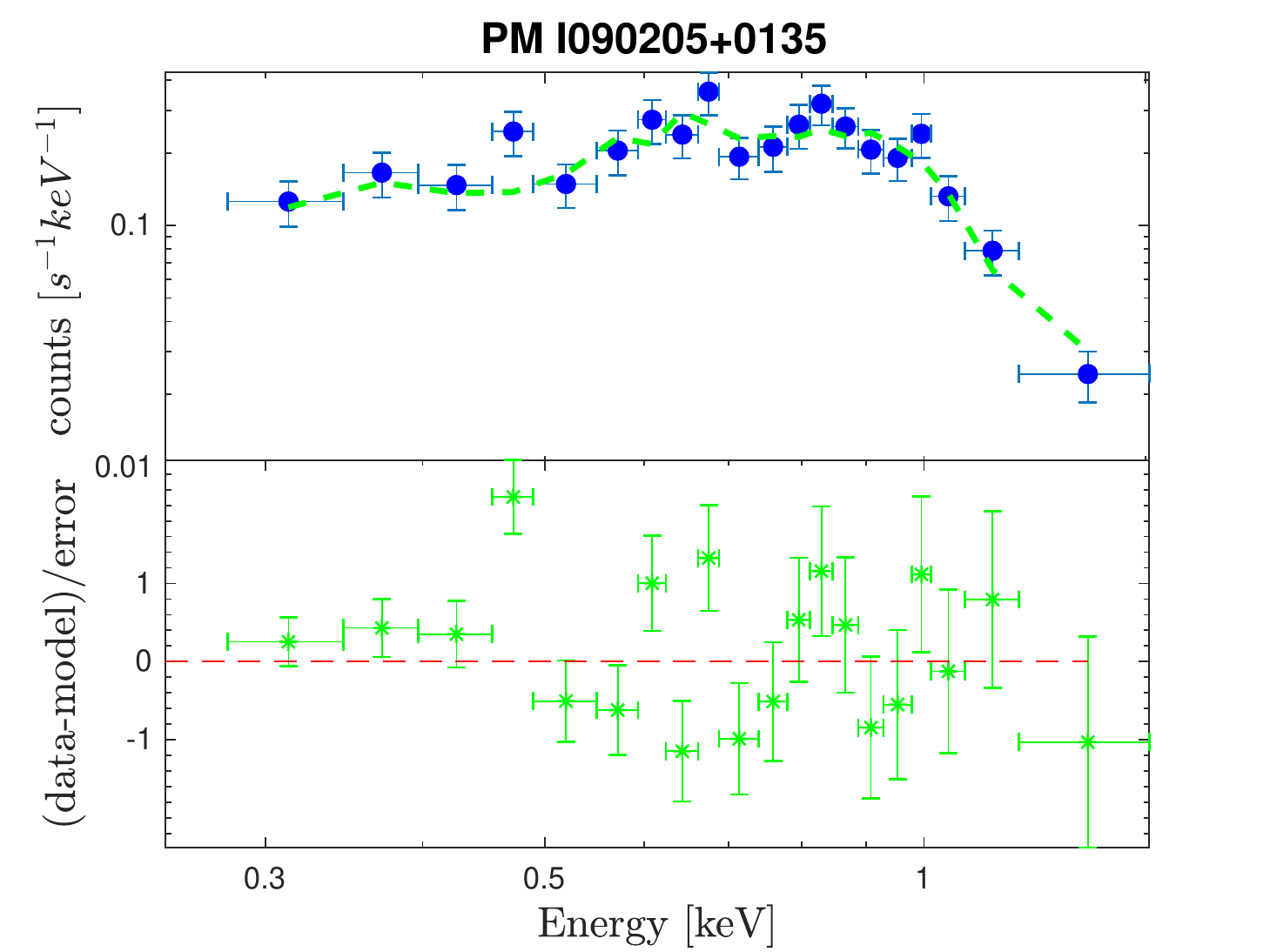}}
 		\subfigure{\label{fig:b}\includegraphics[width=0.33\textwidth]{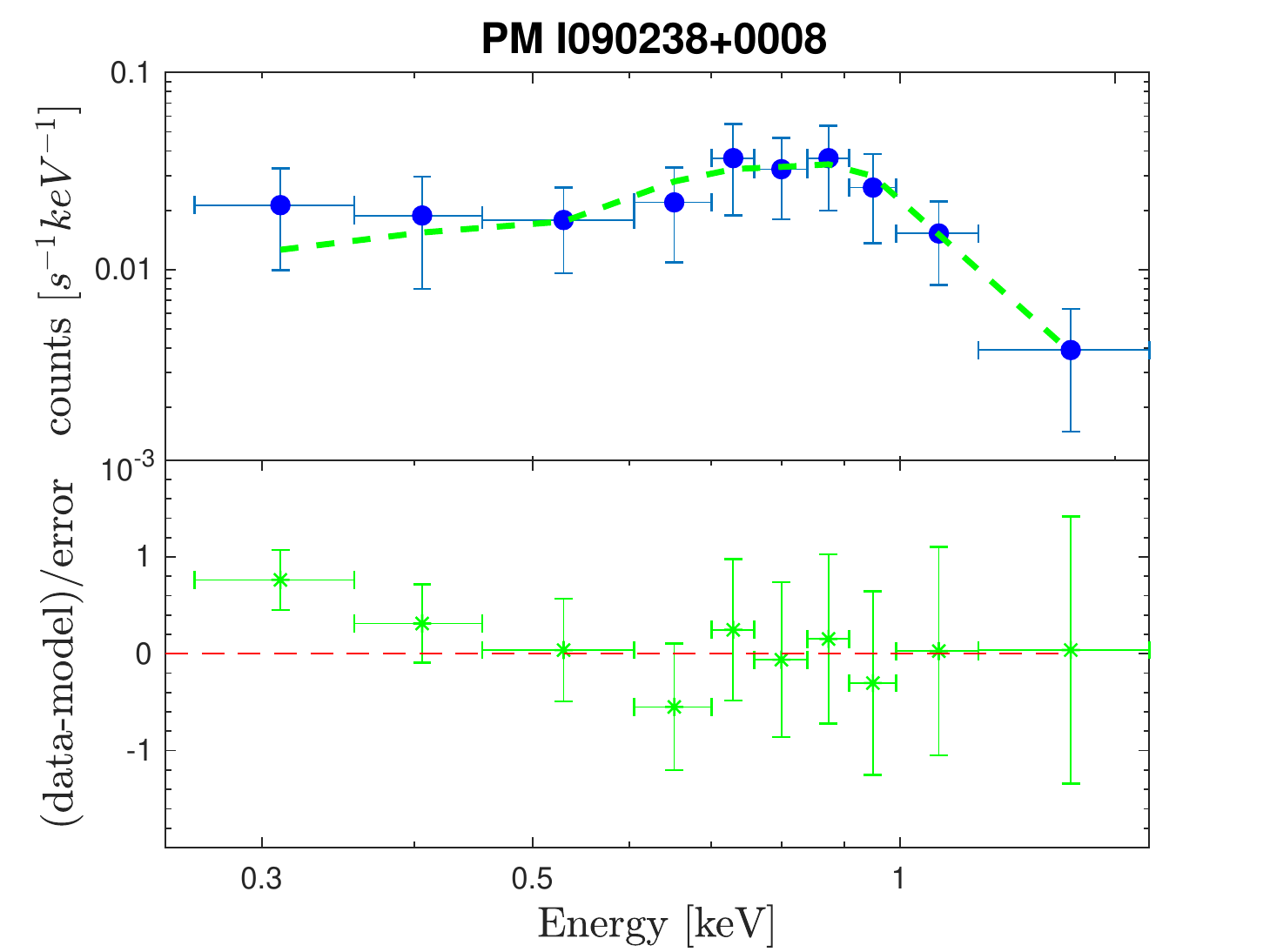}}
 		\subfigure{\label{fig:b}\includegraphics[width=0.33\textwidth]{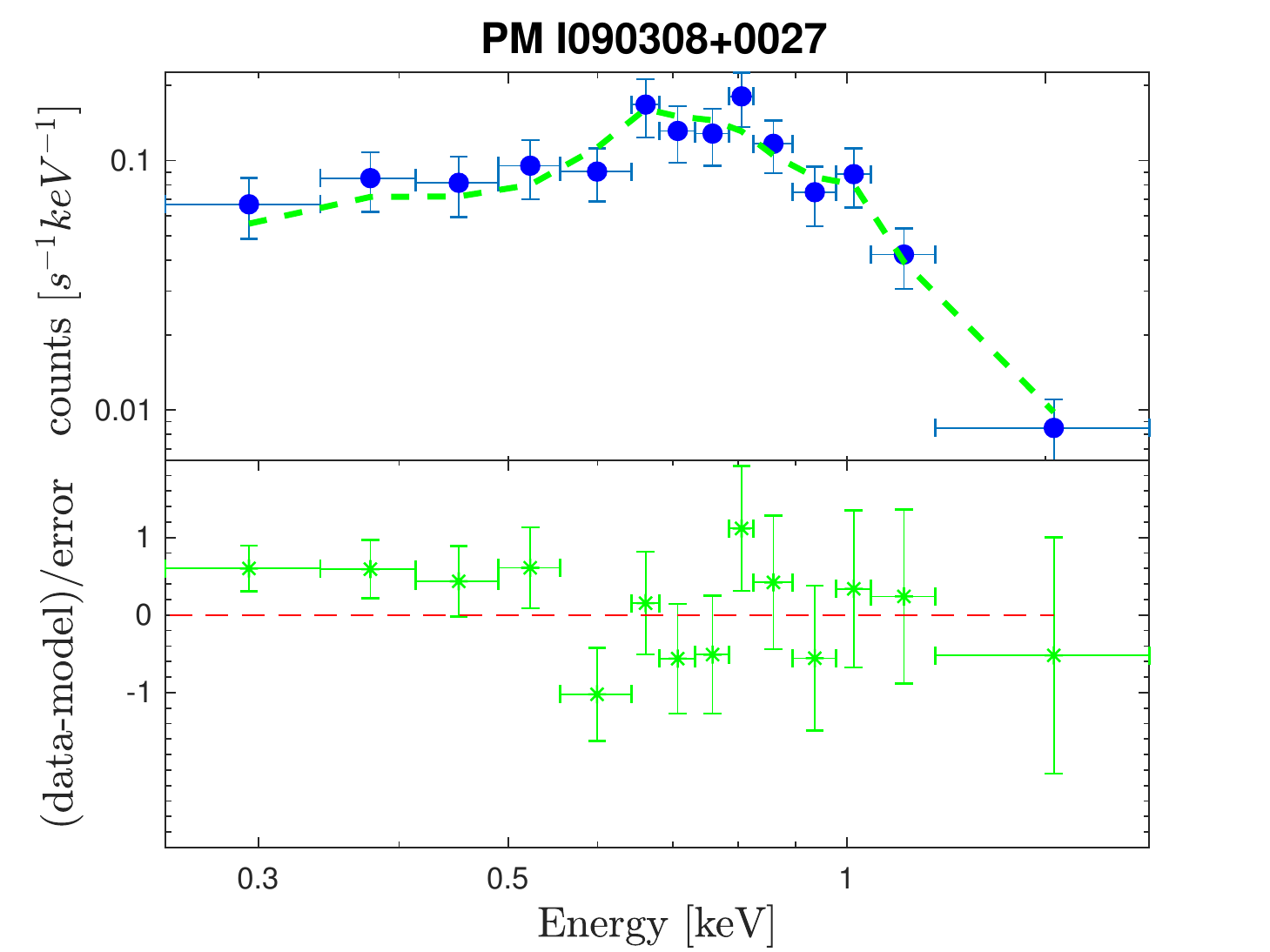}}
 	\end{multicols}
 	\caption{{\em eROSITA} spectrum, best-fit thermal model and residuals for the ten sources in the eFEDS fields with $>30$\,counts.}
 		\label{fig:eFEDS_spec}
 \end{figure*}

\section{eFEDS light curves}\label{app:lcs} 
 Fig.~\ref{fig:eFEDS_lcs} 
 comprises the {\em eROSITA} light curves from the eFEDS observation for the $14$ detected M dwarfs in these fields. The timing of the data intervals in these light curves is a consequence of the survey mode employed in the data acquisition for the eFEDS fields and is described in Sect.~\ref{subsubsect:analysis_efeds_xrays_lcs}.
 \renewcommand{\thetable}{B.\arabic{figure}} 
       \begin{figure*}[htbp]
 	    \parbox{18cm}{
 	    \parbox{6cm}{\includegraphics[width=0.33\textwidth]{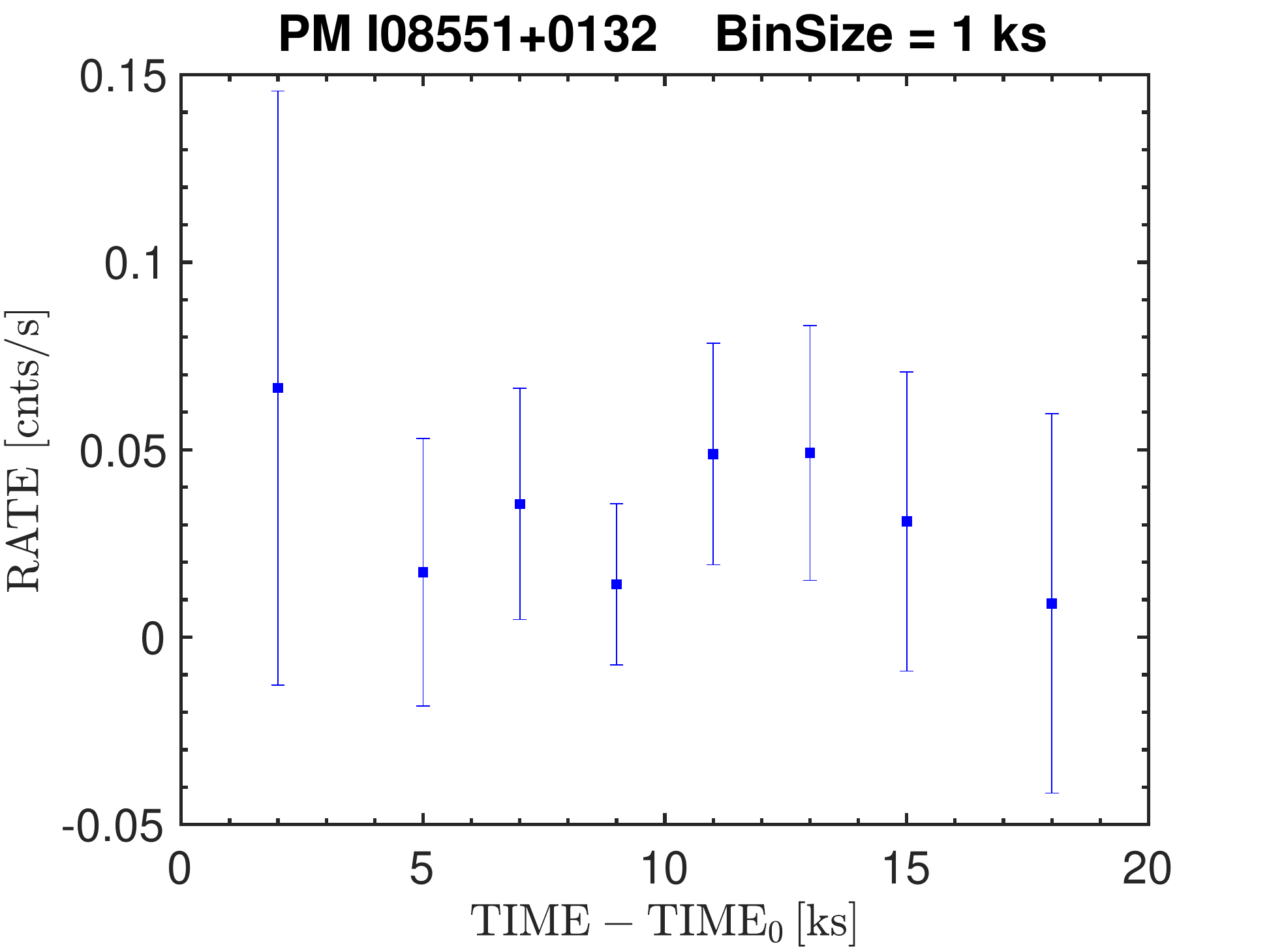}}
 	    \parbox{6cm}{\includegraphics[width=0.33\textwidth]{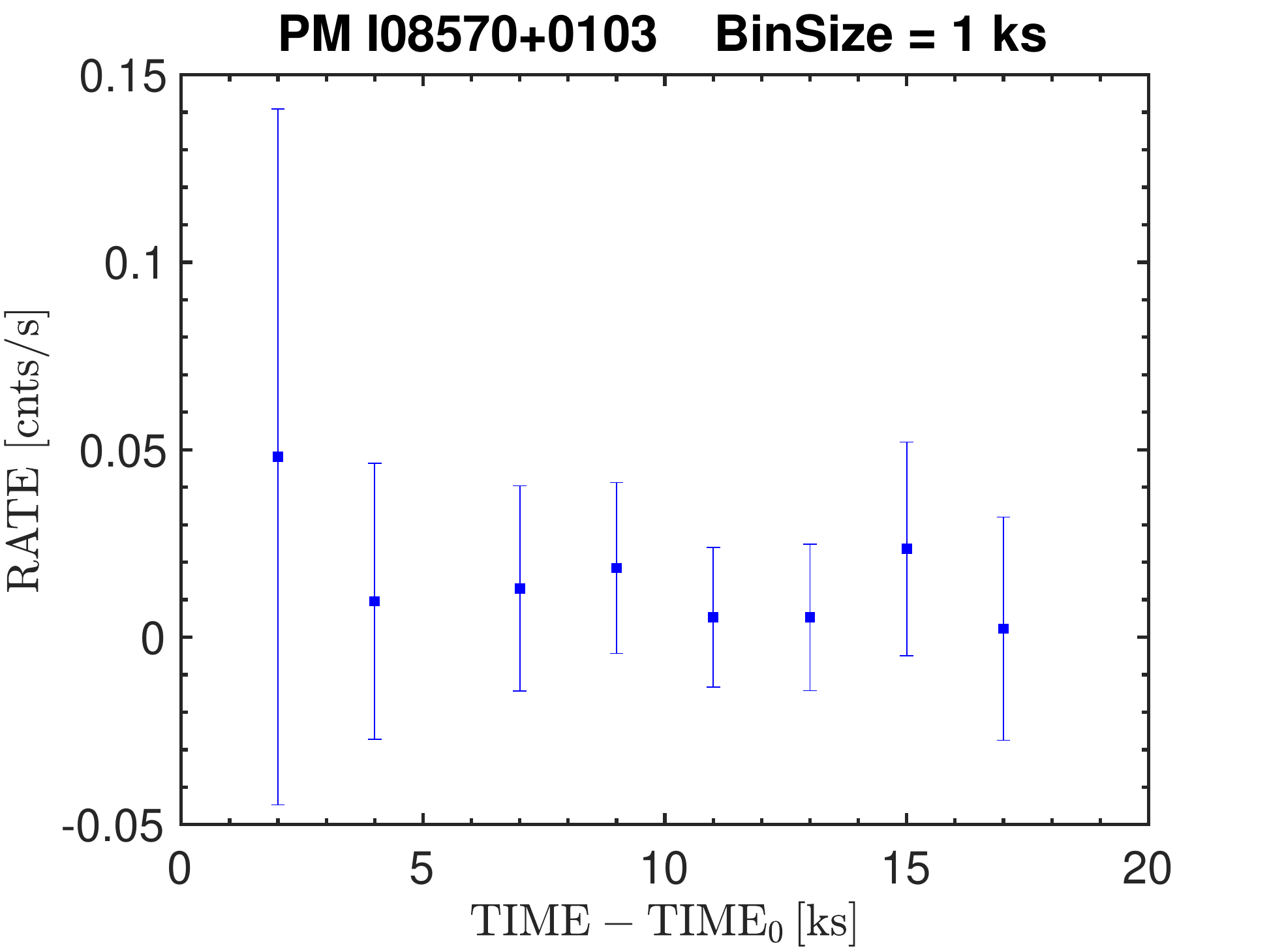}}	
 	    \parbox{6cm}{\includegraphics[width=0.33\textwidth]{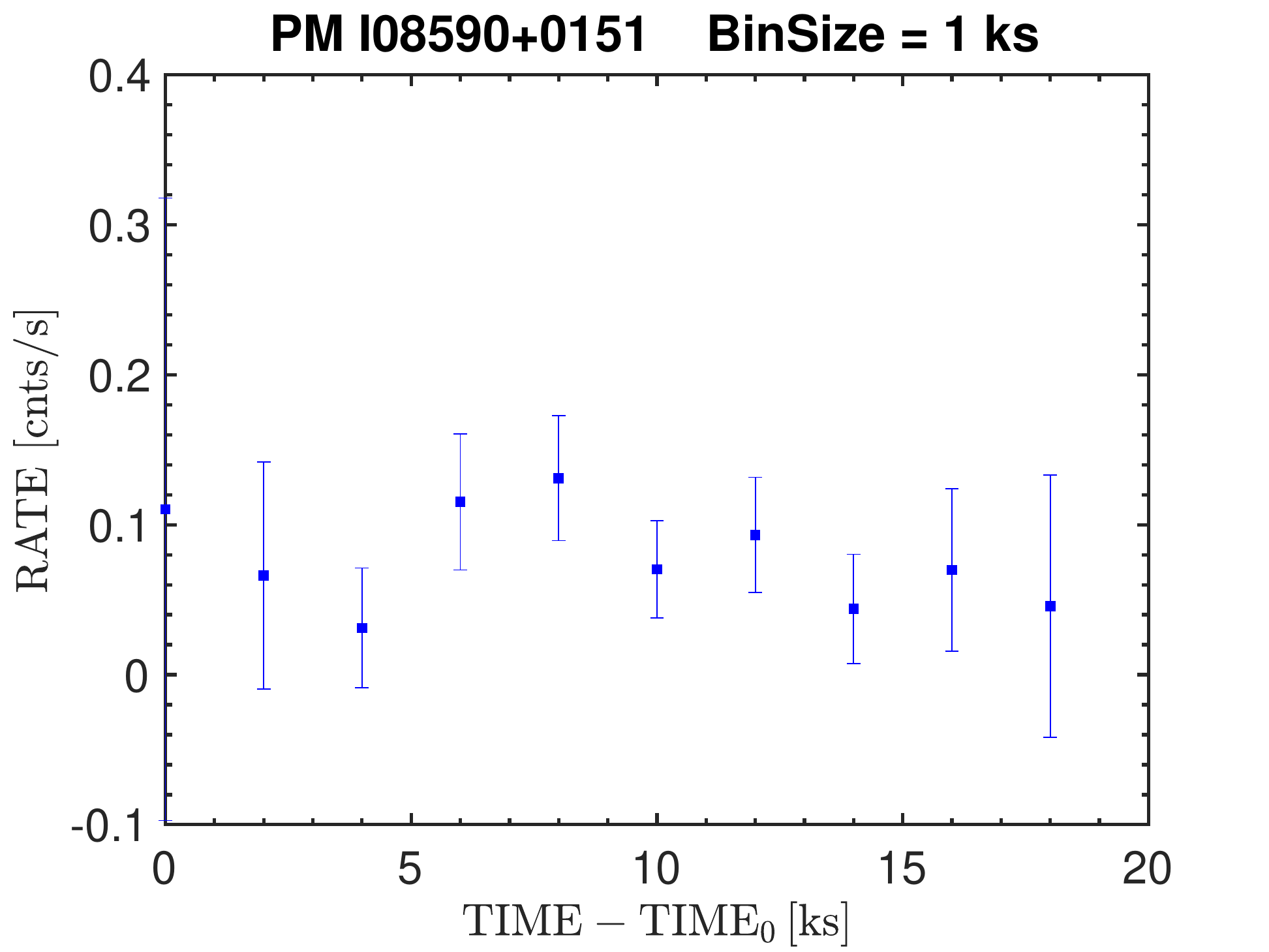}}	
 	    }
 	    
 	    \parbox{18cm}{
 	    \parbox{6cm}{\includegraphics[width=0.33\textwidth]{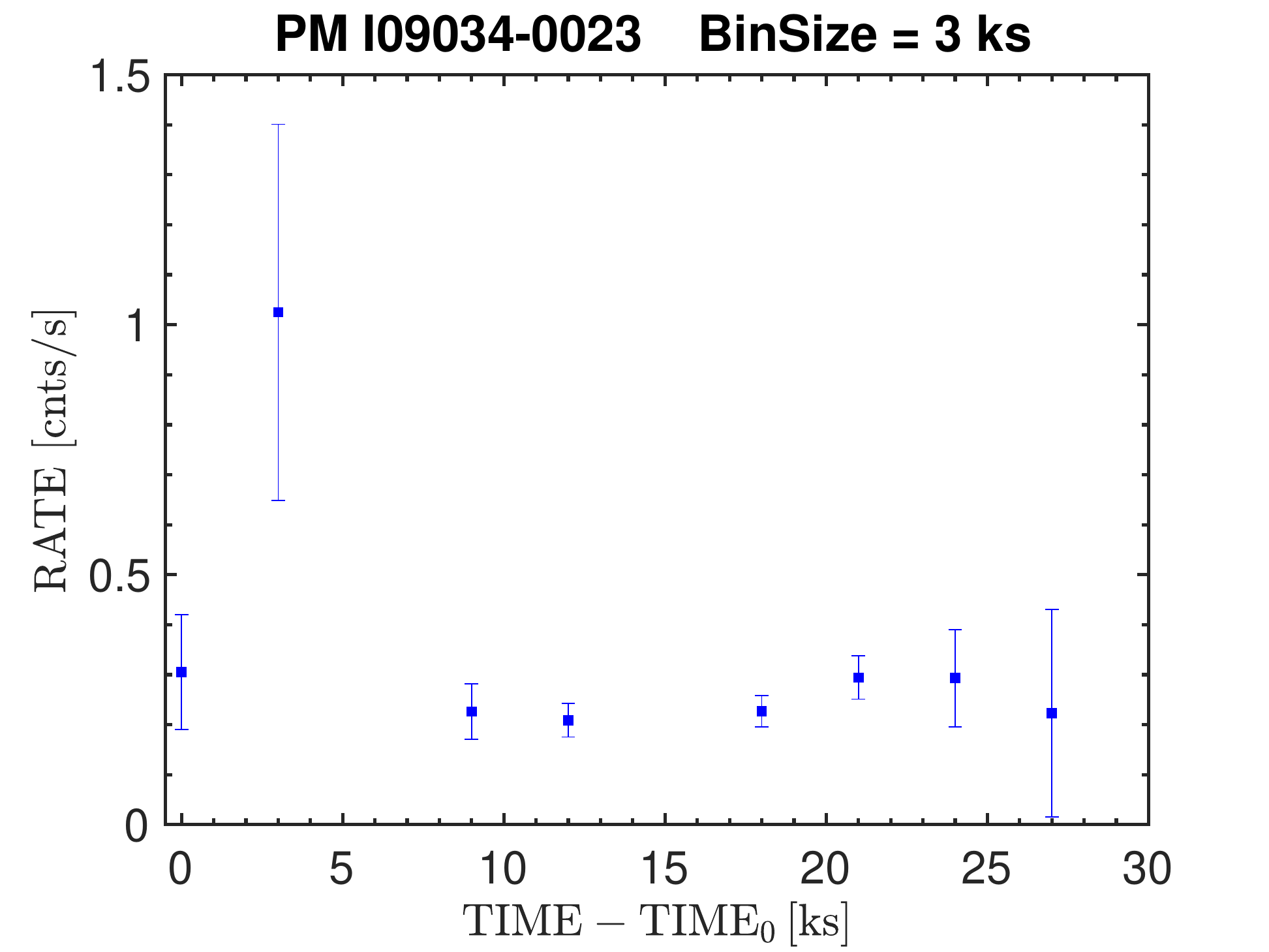}}
 	    \parbox{6cm}{\includegraphics[width=0.33\textwidth]{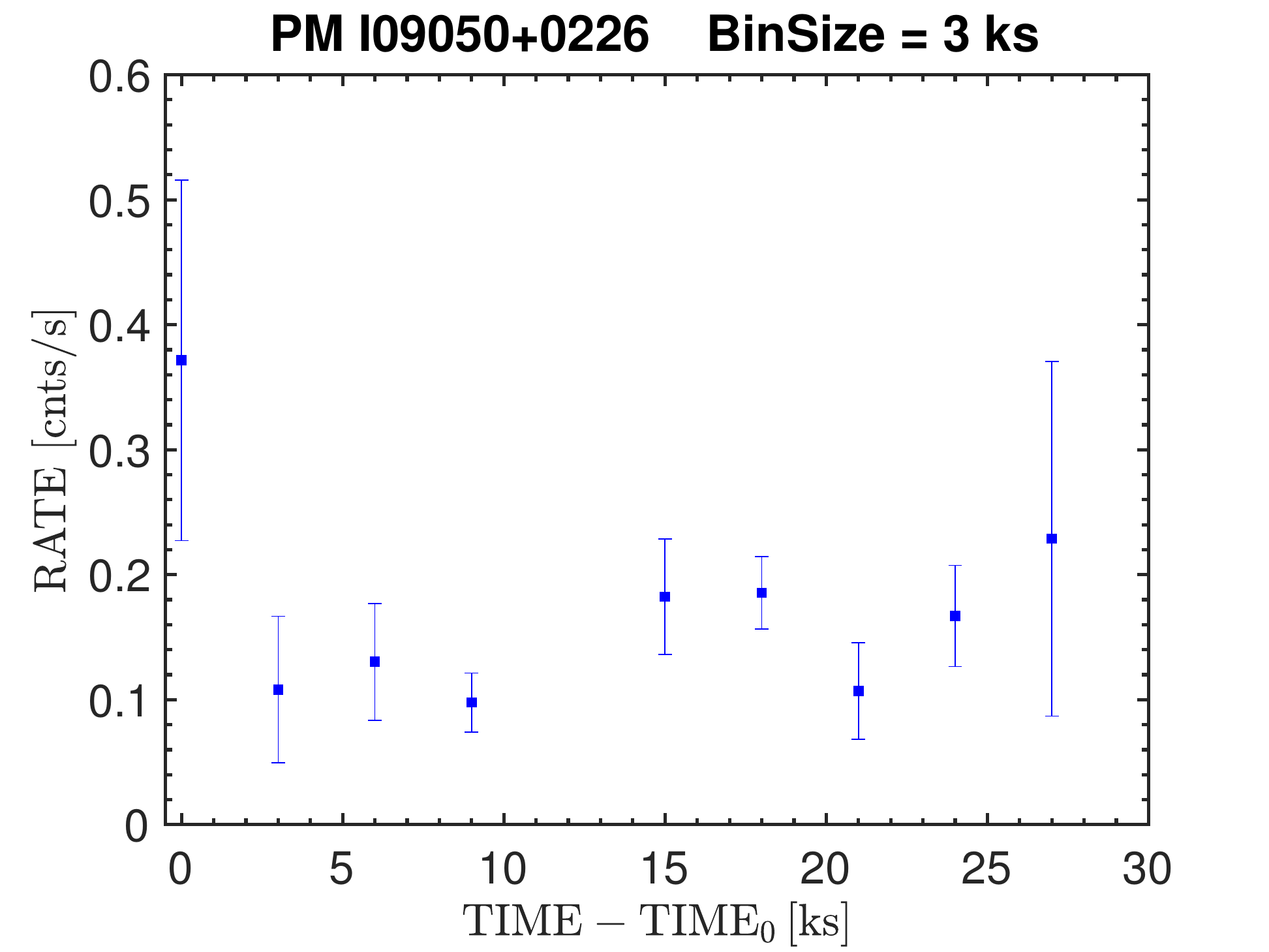}}	
 	    \parbox{6cm}{\includegraphics[width=0.33\textwidth]{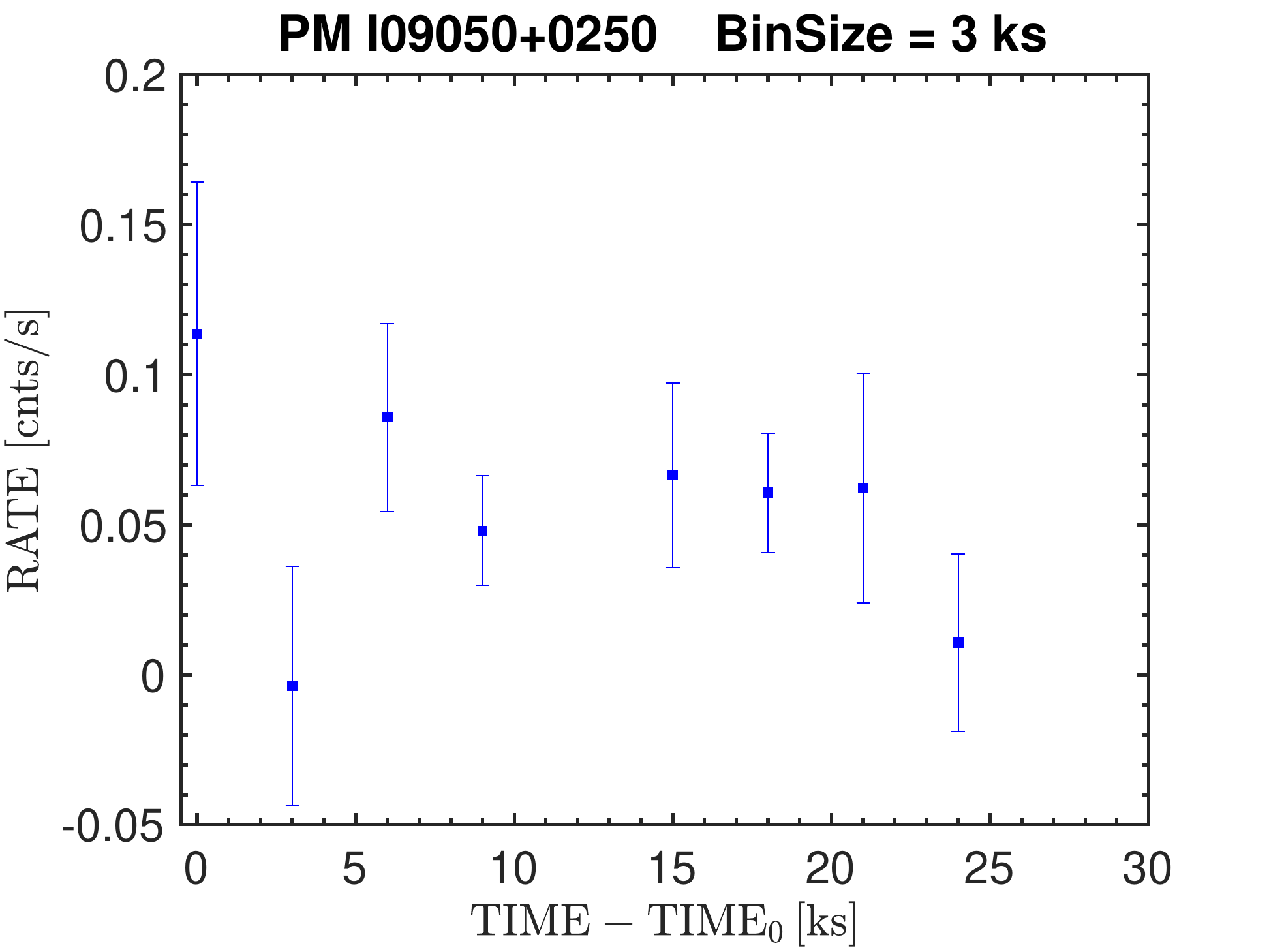}}	
 	    }
 		
 		\parbox{18cm}{
 	    \parbox{6cm}{\includegraphics[width=0.33\textwidth]{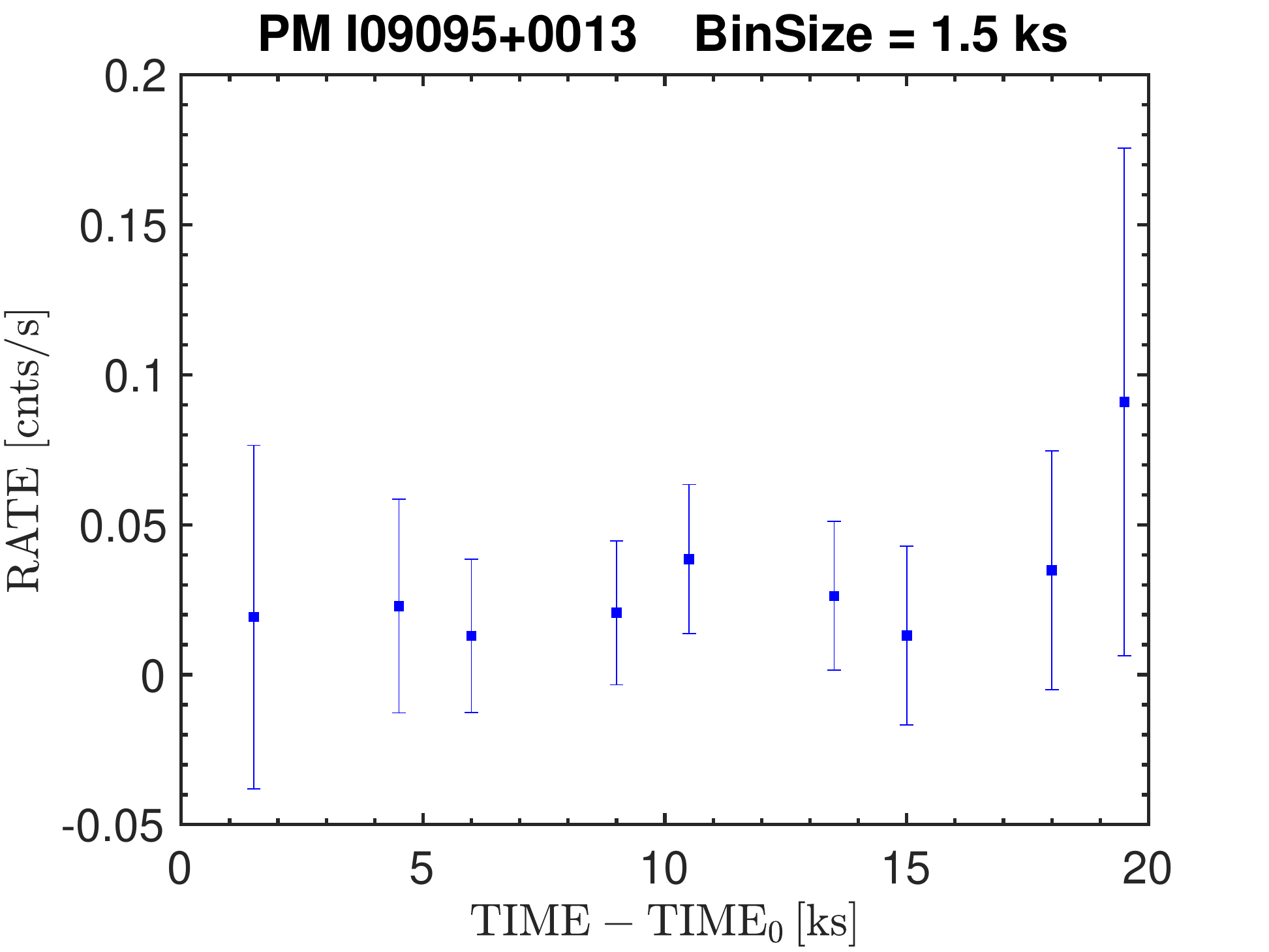}}
 	    \parbox{6cm}{\includegraphics[width=0.33\textwidth]{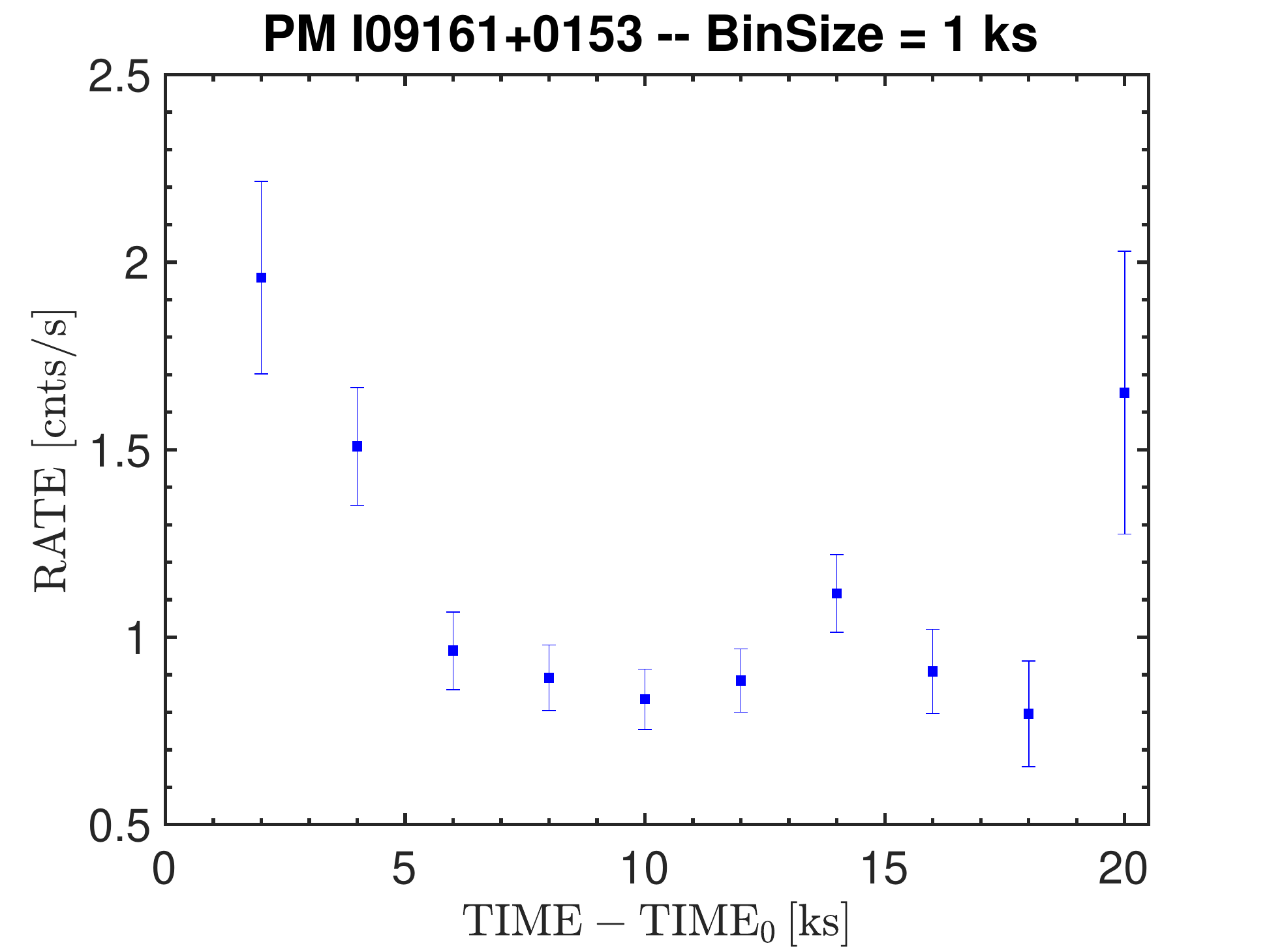}}	
 	    \parbox{6cm}{\includegraphics[width=0.33\textwidth]{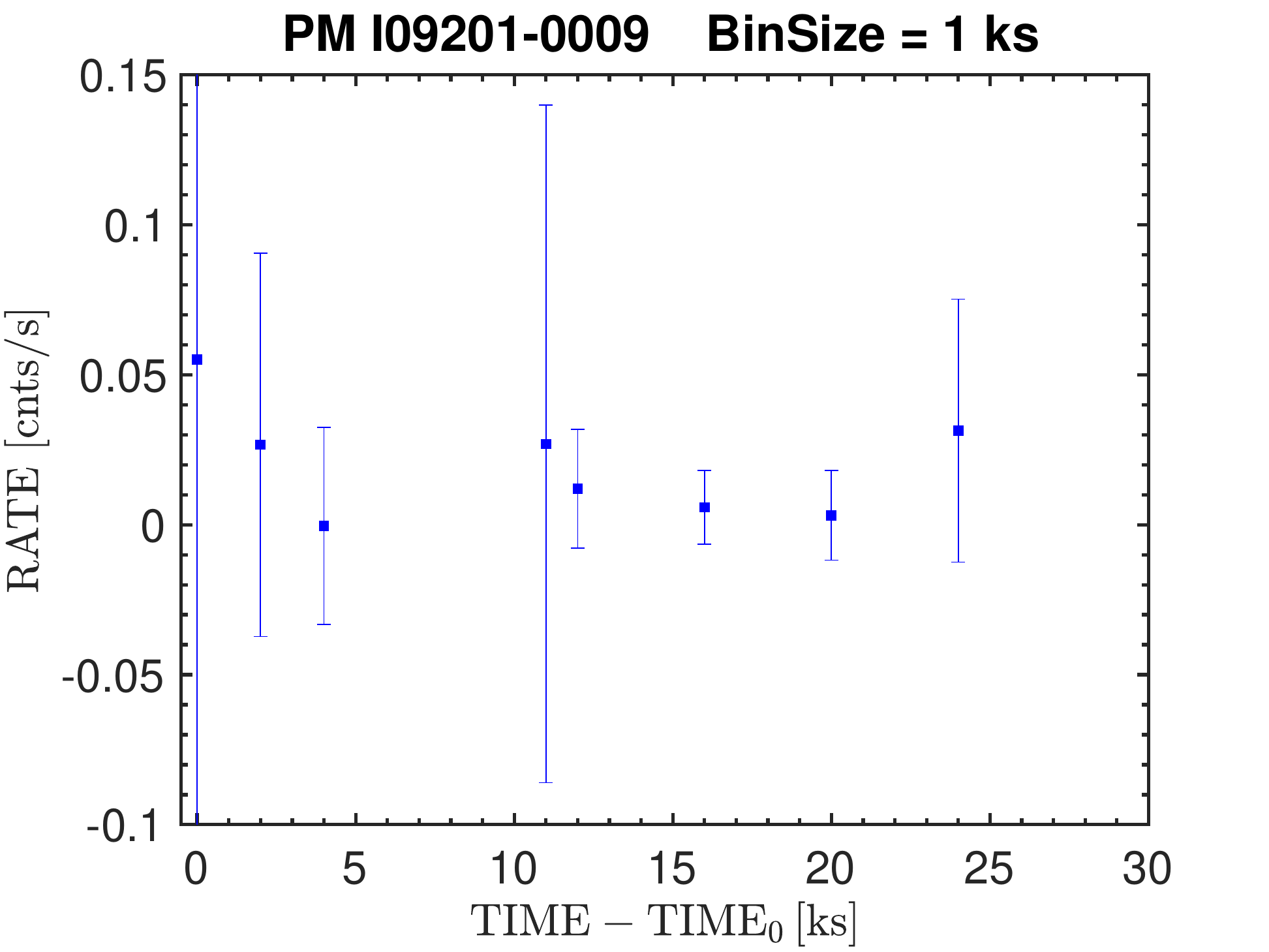}}	
 	    }
 	    \parbox{18cm}{
 	    \parbox{6cm}{\includegraphics[width=0.33\textwidth]{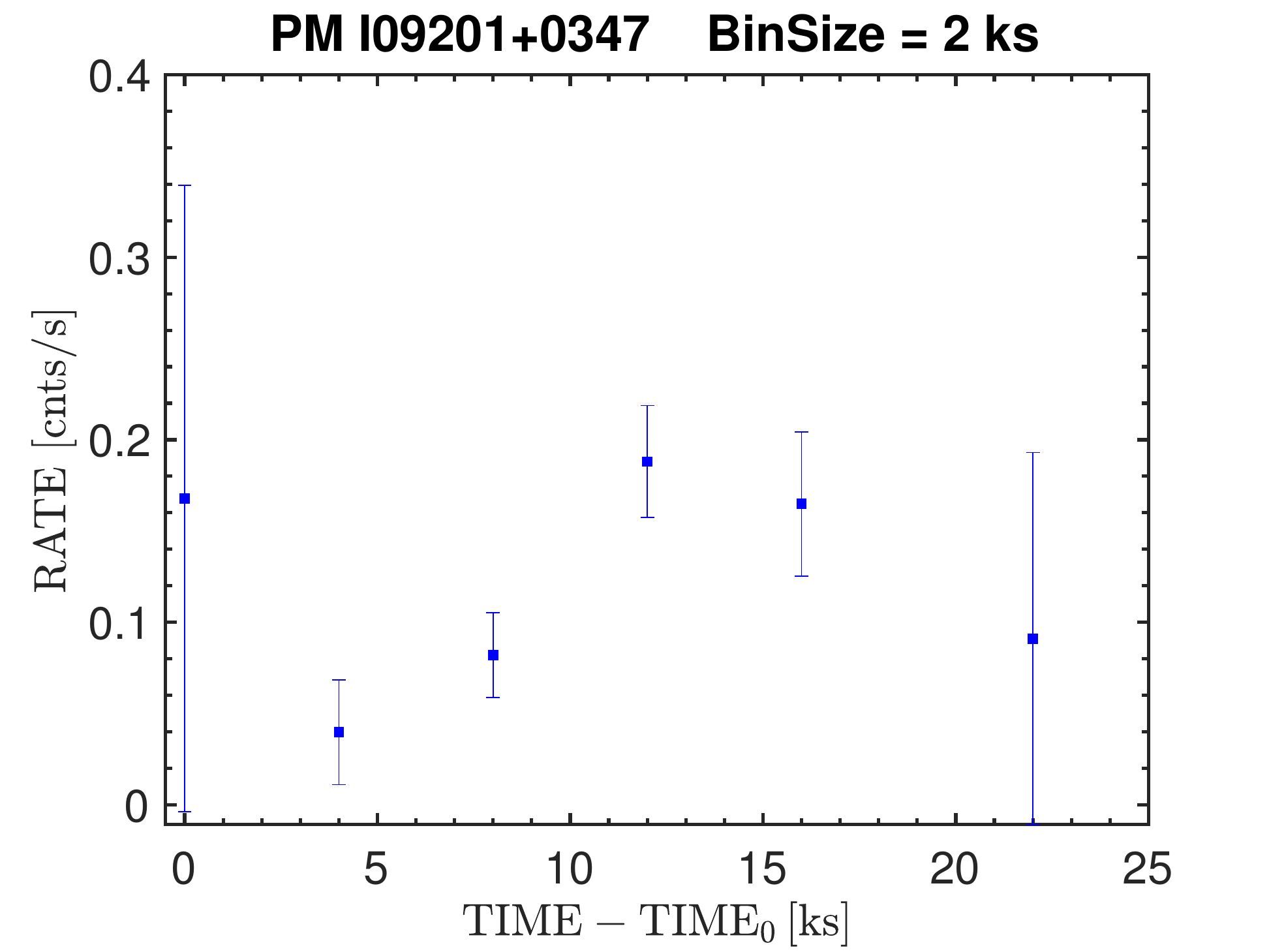}}
 	    \parbox{6cm}{\includegraphics[width=0.33\textwidth]{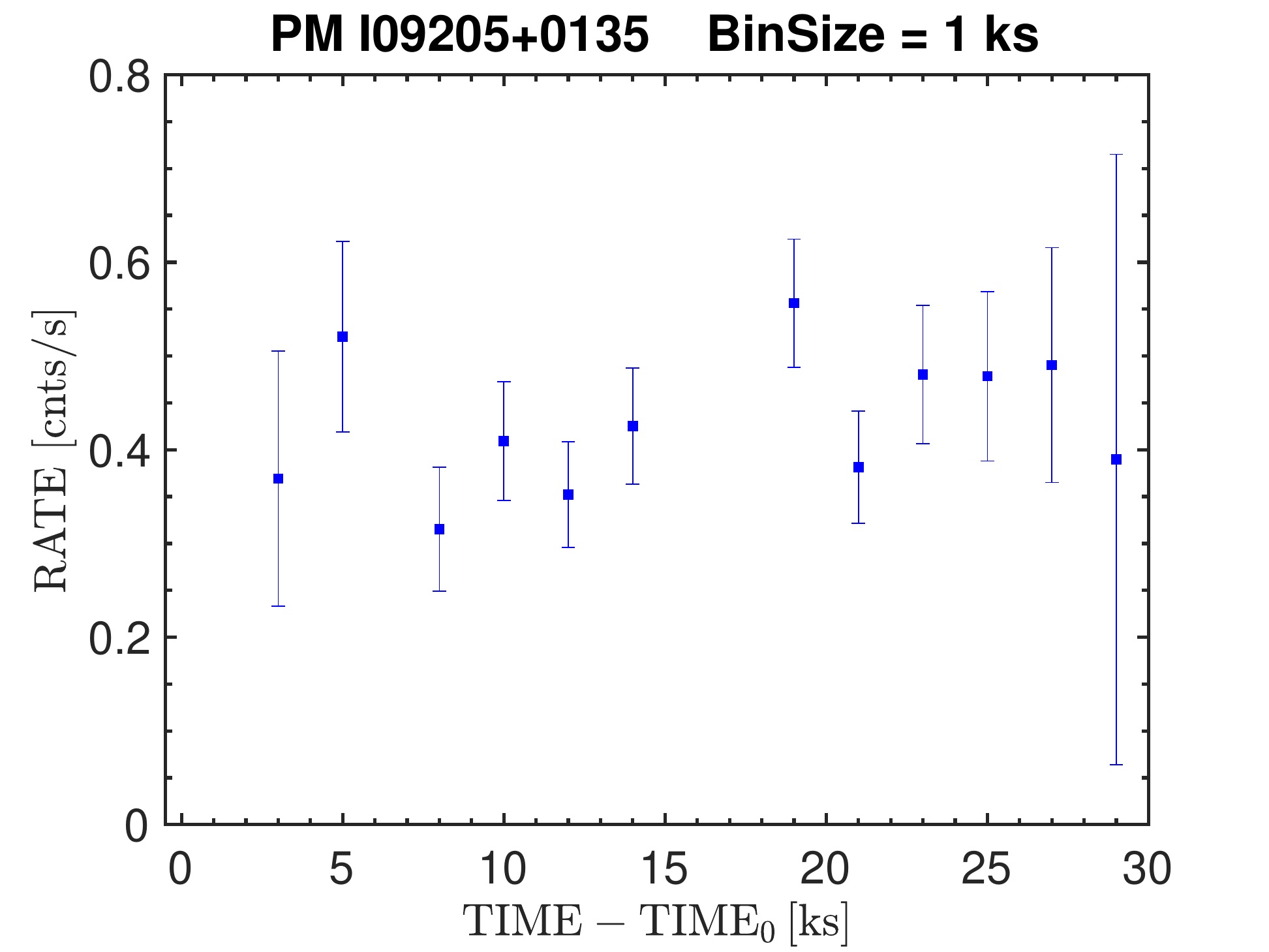}}	
 	    \parbox{6cm}{\includegraphics[width=0.33\textwidth]{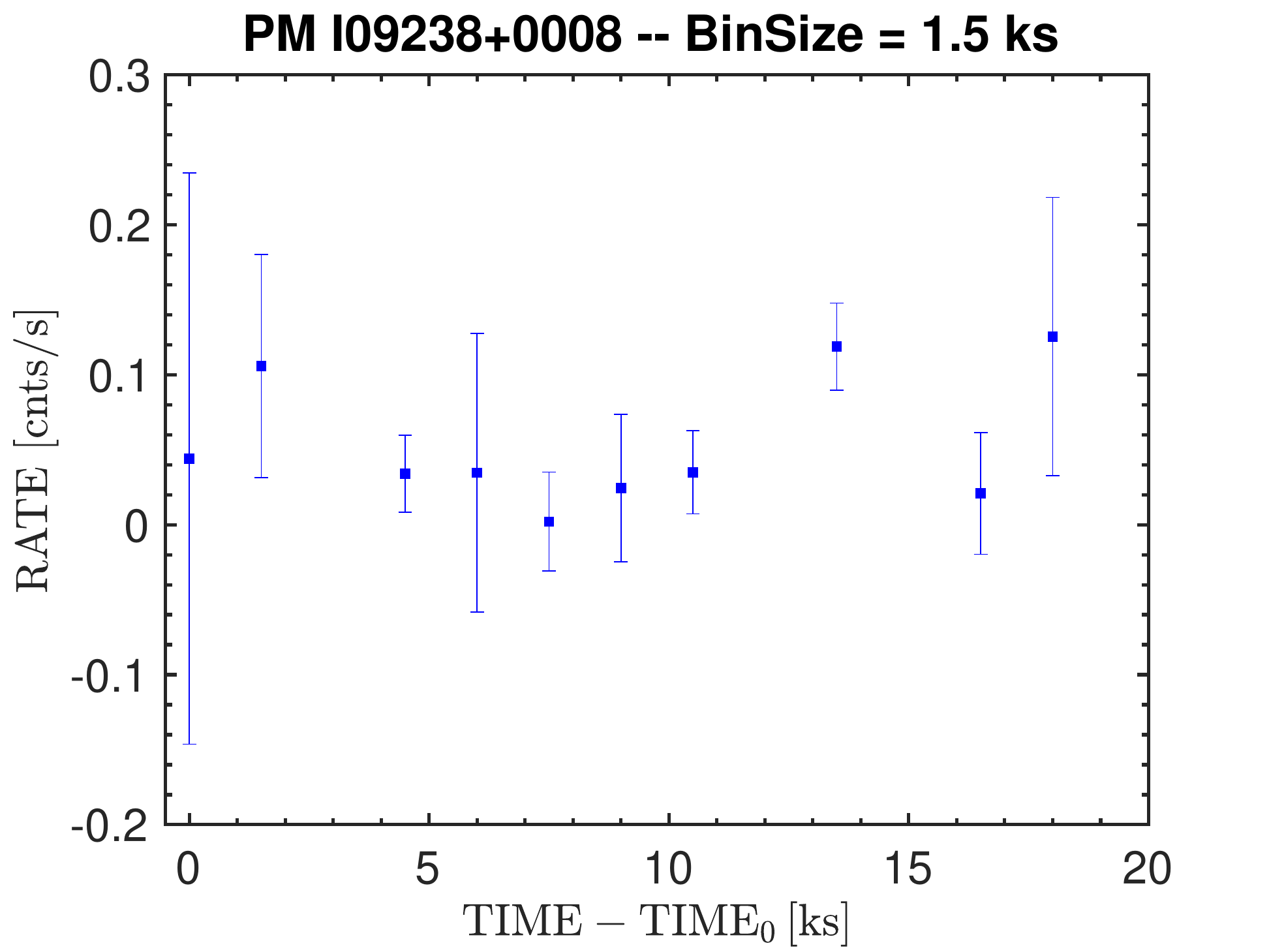}}	
 	    }
 		\parbox{18cm}{
 	    \parbox{6cm}{\includegraphics[width=0.33\textwidth]{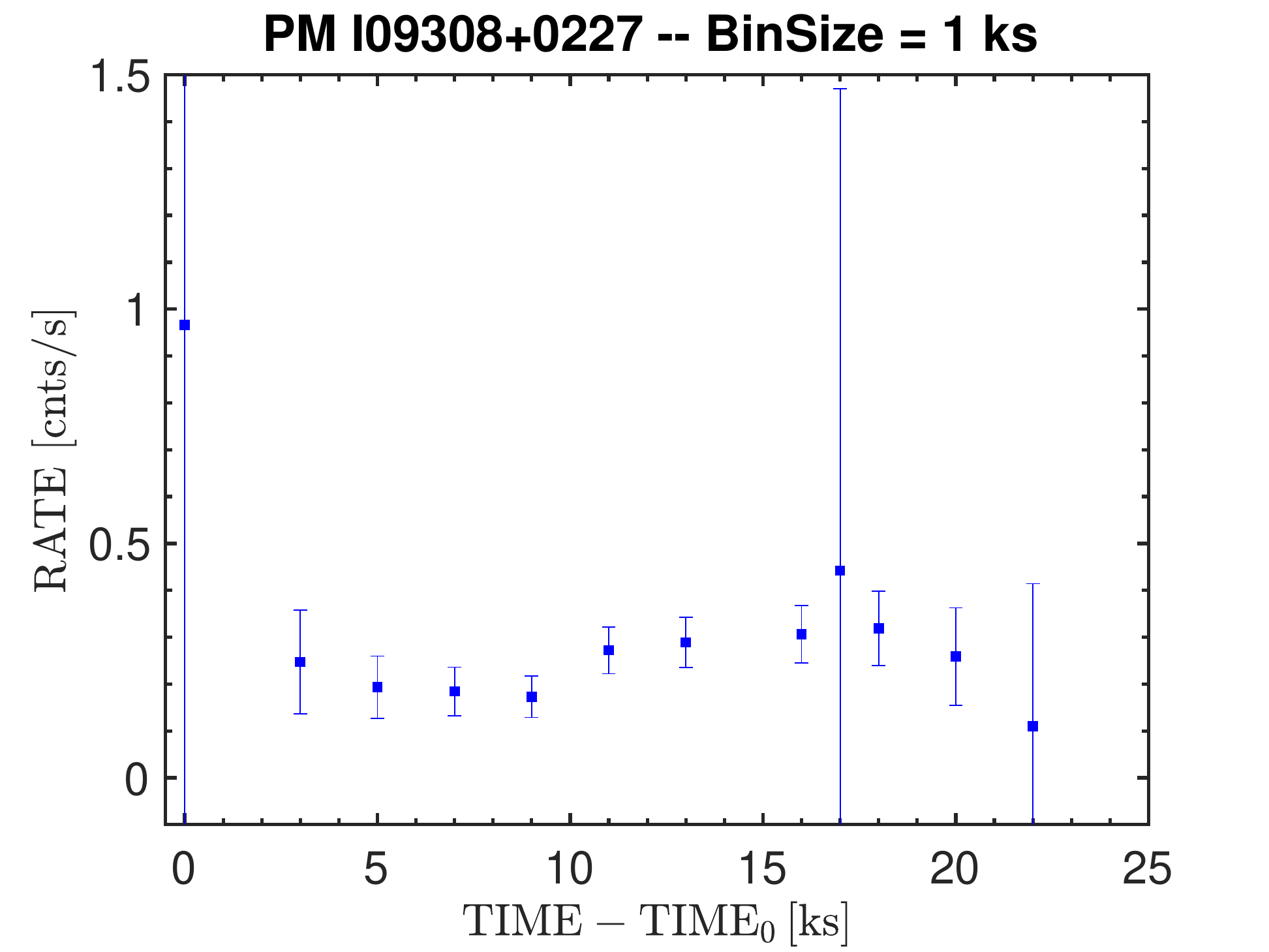}}
 	    \parbox{6cm}{\includegraphics[width=0.33\textwidth]{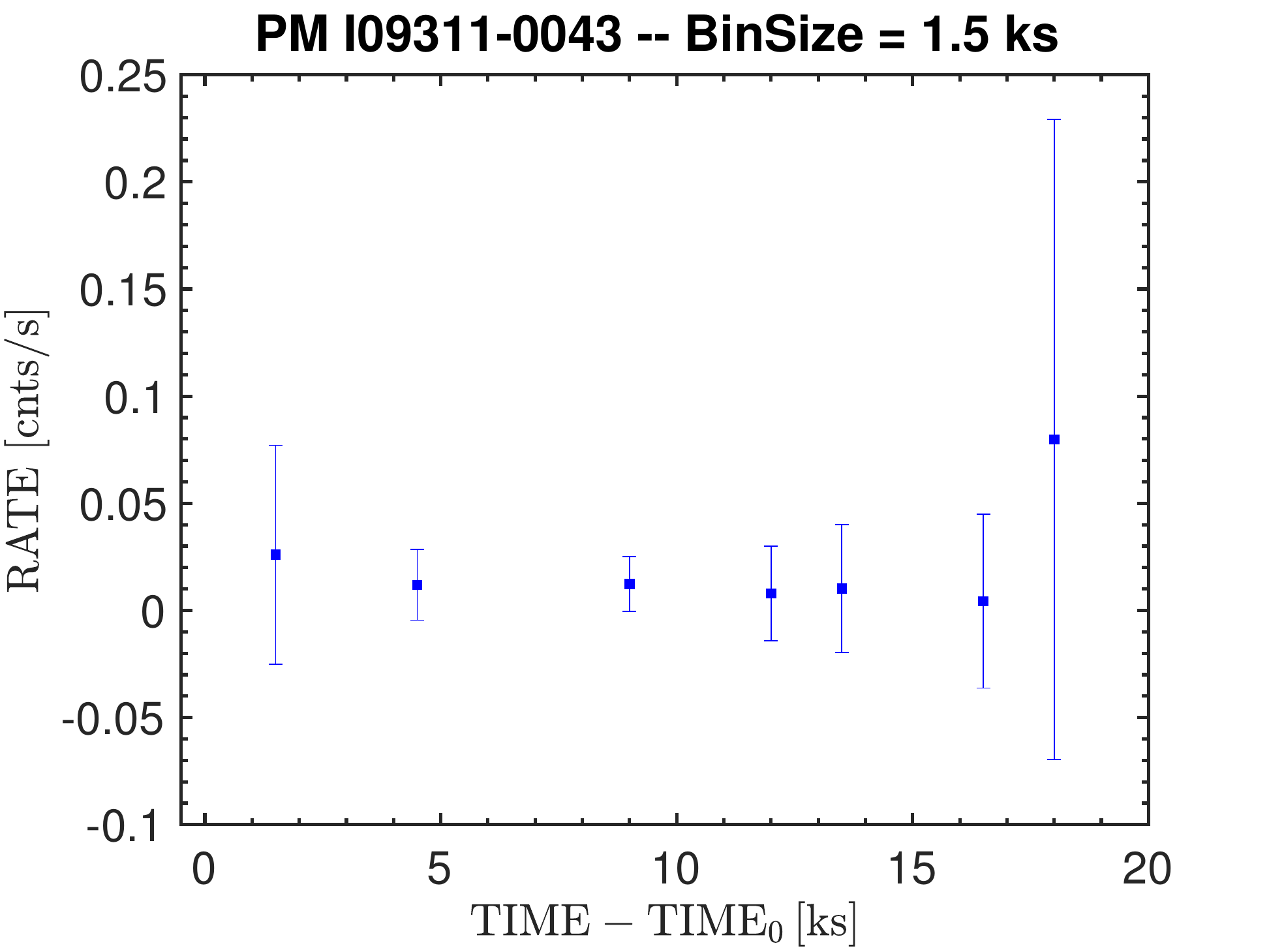}}	
 	    \parbox{6cm}{\caption{{\em eROSITA} light curves for the $14$ M dwarfs detected in the eFEDS fields; the bin size (given above each panel) was individually adapted taking into account the sampling of the light curve resulting from the scanning mode such as to produce the smallest error bars; see text in Sect.~\ref{subsubsect:analysis_efeds_xrays_lcs} for details.\label{fig:eFEDS_lcs}}}}
 	
\end{figure*}
\end{appendix}      

\end{document}